\documentclass[aps,prd,twocolumn,floatfix,nofootinbib,superscriptaddress]{revtex4-1}
\usepackage{graphicx}
\usepackage{amsmath,amssymb,siunitx}

\newcommand{\prx}{Phys. Rev. X}
\newcommand{\physrep}{Phys. Rep.}

\newcommand{\mnras}{Mon. Not. R. Astron. Soc.}
\newcommand{\apjl}{Astrophys. J. Lett.}

\begin{document}
\title{Reducing orbital eccentricity in initial data of black
hole--neutron star binaries in the puncture framework}
\author{Koutarou Kyutoku}
\affiliation{Department of Physics, Kyoto University, Kyoto 606-8502,
Japan}
\affiliation{Center for Gravitational Physics, Yukawa Institute for
Theoretical Physics, Kyoto University, Kyoto 606-8502, Japan}
\affiliation{Interdisciplinary Theoretical and Mathematical Sciences
Program (iTHEMS), RIKEN, Wako, Saitama 351-0198, Japan}
\author{Kyohei Kawaguchi}
\affiliation{Institute for Cosmic Ray Research, The University of Tokyo,
Chiba 277-8582, Japan}
\author{Kenta Kiuchi}
\affiliation{Max Planck Institute for Gravitational Physics (Albert
Einstein Institute), Am M{\"u}hlenberg 1, Potsdam-Golm 14476, Germany}
\affiliation{Center for Gravitational Physics, Yukawa Institute for
Theoretical Physics, Kyoto University, Kyoto 606-8502, Japan}
\author{Masaru Shibata}
\affiliation{Max Planck Institute for Gravitational Physics (Albert
Einstein Institute), Am M{\"u}hlenberg 1, Potsdam-Golm 14476, Germany}
\affiliation{Center for Gravitational Physics, Yukawa Institute for
Theoretical Physics, Kyoto University, Kyoto 606-8502, Japan}
\author{Keisuke Taniguchi}
\affiliation{Department of Physics, University of the Ryukyus,
Nishihara, Okinawa 903-0213, Japan}

\date{\today}

\begin{abstract}
 We develop a method to compute low-eccentricity initial data of black
 hole--neutron star binaries in the puncture framework extending
 previous work on other types of compact binaries. In addition to
 adjusting the orbital angular velocity of the binary, the approaching
 velocity of a neutron star is incorporated by modifying the helical
 Killing vector used to derive equations of the hydrostationary
 equilibrium. The approaching velocity of the black hole is then induced
 by requiring the vanishing of the total linear momentum of the system,
 differently from the case of binary black holes in the puncture
 framework where the linear momentum of each black hole is specified
 explicitly. We successfully reduce the orbital eccentricity to
 $\lesssim 0.001$ by modifying the parameters iteratively using
 simulations of $\approx 3$ orbits both for nonprecessing and precessing
 configurations. We find that empirical formulas for binary black holes
 derived in the excision framework do not reduce the orbital
 eccentricity to $\approx 0.001$ for black hole--neutron star binaries
 in the puncture framework, although they work for binary neutron stars.
\end{abstract}

\maketitle

\section{Introduction} \label{sec:intro}

One of the remaining and promising targets for ground-based
gravitational-wave detectors is the coalescence of black hole--neutron
star binaries (see Ref.~\cite{shibata_taniguchi2011} for
reviews). Indeed, we have already been informed of possible black
hole--neutron star binary coalescences in the LIGO-Virgo O3
\cite{ligovirgo-gwtc2}, including signals from sources whose identities
are not fully clear \cite{ligovirgo2020-2,ligovirgo2020-3}. If we would
have detected these events with a high signal-to-noise ratio, we could
infer finite-size properties of neutron stars such as the radius
\cite{vallisneri2000,shibata_kyt2009,kyutoku_st2010,kyutoku_ost2011,pannarale_bkls2015}
and tidal deformability \cite{lackey_ksbf2012,lackey_ksbf2014} as well
as the mass and the spin of each component. Because the finite-size
properties depend crucially on the underlying equation of state for
supranuclear-density matter (see
Refs.~\cite{lattimer_prakash2016,baym_hkpst2018} for reviews),
gravitational-wave observations of black hole--neutron star binaries
will provide invaluable information not only to astrophysics but also to
nuclear physics in a manner similar to the detections of binary neutron
stars \cite{ligovirgo2018,ligovirgo2019}.

Reliable theoretical templates of gravitational waveforms are the
prerequisite for accurate extraction of source properties
\cite{ligovirgo2016-5,ligovirgo2019-3}. Accordingly, theoretical
calculations of gravitational waveforms have been playing a central role
in gravitational-wave astronomy. Particularly high accuracy is required
to extract finite-size properties of neutron stars, namely the tidal
deformability, in a reliable manner
\cite{yagi_yunes2014,favata2014,wade_colflr2014}. Although the
systematic errors associated with waveform models are smaller than the
statistical errors for the first binary-neutron-star merger GW170817
\cite{ligovirgo2019,ligovirgo2020,narikawa_ukkkst2020} and largely
uninformative GW190425 \cite{ligovirgo2020-2}, improvement of the
detector sensitivity by an order of magnitude \cite{lvk2018} will make
the systematic error discernible for GW170817-equivalent sources
\cite{kawaguchi_kksst2018,dudi_pdhbob2018}.

Development of templates for black hole--neutron star binaries is
generally in its early stage \cite{chakravarti_etal2019,huang_hvvfb2020}
(see also Refs.~\cite{thompson_fknpdh2020,matas_etal2020} for recent
progress). One reason is attributed to the small number of long-term and
high-precision simulations of black hole--neutron star binaries in
numerical relativity, which is the unique tool to investigate
theoretically the late inspiral and merger phases and to derive
gravitational waveforms. While a lot of insight on dynamical mass
ejection and electromagnetic counterparts has been gained in the past
years
\cite{chawla_abllmn2010,kyutoku_ost2011,foucart_ddkmopsst2013,kyutoku_is2013,foucart_etal2014,kawaguchi_knost2015,kyutoku_iost2015,foucart_dbdkhkps2017,kyutoku_ksst2018,brege_dfdchkops2018,foucart_dknps2019},
we need to reignite numerical-relativity simulations of the inspiral and
merger phases to derive accurate gravitational waveforms
\cite{foucart_bdgkmmpss2013,foucart_etal2019}.

Realistic initial data of black hole--neutron star binaries are
necessary to compute realistic gravitational waves in numerical
relativity. In particular, the orbital eccentricity has to be low enough
because the majority of astrophysical compact binaries are circularized
nearly completely right before merger due to gravitational radiation
reaction during its long inspiral
\cite{peters_mathews1963,peters1964}. It has been pointed out that
seemingly tiny eccentricity, say $e \sim 0.01$, in theoretical templates
can significantly degrade the accuracy with which we can measure the
tidal deformability via gravitational-wave observation
\cite{favata2014}. Eccentricity reduction has already been performed for
and routinely applied to initial data of black hole--neutron star
binaries in the excision framework \cite{foucart_kpt2008}, in which the
physical singularity inside the horizon is removed from the
computational domain. However, the eccentricity reduction for initial
data of black hole--neutron star binaries in the puncture framework
\cite{shibata_uryu2006,shibata_uryu2007}, in which the singularity is
handled in an analytic manner, has not yet been reported.

In this paper, we present a method to reduce the orbital eccentricity in
initial data of black hole--neutron star binaries in the puncture
framework. Specifically, we adjust the orbital angular velocity and
incorporate the approaching velocity to obtain low-eccentricity initial
data,\footnote{Following our previous work \cite{kyutoku_st2014}, we
refer to initial data obtained by assuming helical symmetry as
``quasicircular'' and those with the approaching velocity as
``low-eccentricity.''} where their values are determined iteratively by
analyzing orbital evolution for a few orbits derived by dynamical
simulations. Differently from initial data of binary black holes in the
puncture framework, in which the approaching velocity of black holes is
incorporated by actively choosing the values of the linear-momentum
parameter \cite{husa_hgsb2008}, the translational motion of the black
hole cannot be specified freely in our formulation for black
hole--neutron star binaries. In this work, we first incorporate an
approaching velocity of the neutron star by modifying the
hydrostationary equations in the same manner as in the case of binary
neutron stars \cite{kyutoku_st2014}. Because the black hole is not
subject to hydrodynamics, the approaching velocity of the black hole is
passively induced by requiring the total linear momentum of the system
to vanish. To control the approaching velocity of the binary as a whole,
the velocity of the black hole is identified by the minus of the shift
vector at the puncture.

The paper is organized as follows. The formulation is described in
Sec.~\ref{sec:method}. To demonstrate the validity of our method, we
apply it to both nonprecessing and precessing configurations in
Sec.~\ref{sec:result}. Section \ref{sec:summary} is devoted to a
summary. Greek and Latin indices denote the spacetime and space
components, respectively. We adopt the geometrical unit in which
$G=c=1$, where $G$ and $c$ are the gravitational constant and the speed
of light, respectively.

\section{Numerical method} \label{sec:method}

\subsection{Formulation}

We describe our formulation for the low-eccentricity initial data of
black hole--neutron star binaries in the puncture framework. As a
concise summary, the update from quasicircular initial data
\cite{kyutoku_st2009} resides in the modified symmetry vector,
Eq.~\eqref{eq:symvec}, which we have adopted for binary neutron stars
\cite{kyutoku_st2014}. Identification of the approaching velocity of the
black hole, Eq.~\eqref{eq:vbh}, has not been adopted in previous related
work for compact binaries.

We compute initial data of black hole--neutron star binaries in the
puncture framework
\cite{brandt_brugmann1997,shibata_uryu2006,shibata_uryu2007,shibata_taniguchi2008,kyutoku_st2009}.
The singularity of gravitational fields associated with the black hole
is handled in an analytic manner by decomposing geometric quantities
into singular and regular parts. By adopting a mixture of extended
conformal-thin sandwich formulation \cite{york1999,pfeiffer_york2003}
and conformal transverse-traceless decomposition \cite{york1979}, only
the regular parts have to be computed numerically to satisfy Einstein
constraint equations and quasiequilibrium conditions. Hydrostationary
equations for the neutron-star matter are solved assuming the
zero-temperature and irrotational flow
\cite{bonazzola_gm1997,asada1998,shibata1998,teukolsky1998,gourgoulhon_gtmb2001},
and specifically we solve the continuity equation and integrated Euler's
equation in the same manner as described in
Ref.~\cite{kyutoku_st2014}. In this work, numerical computations are
performed using a public multidomain spectral method library, {\small
LORENE} \cite{LORENE}, and the details of the methods are presented in
Ref.~\cite{kyutoku_st2009}.

The puncture formulation for initial data of black hole--neutron star
binaries is summarized as follows. First, conformal transformation is
performed for the induced metric $\gamma_{ij}$ and the extrinsic
curvature $K_{ij}$ as
\begin{align}
 \gamma_{ij} & = \psi^4 \hat{\gamma}_{ij} , \\
 K_{ij} & = \psi^{-2} \hat{A}_{ij} + \frac{1}{3} K \psi^4
 \hat{\gamma}_{ij} ,
\end{align}
where $K := \gamma^{ij} K_{ij}$. In the puncture framework, the spatial
conformal flatness and maximal slicing conditions,
\begin{equation}
 \hat{\gamma}_{ij} = f_{ij} \; , \; K = 0 ,
\end{equation}
where $f_{ij}$ is the flat metric, are imposed. We also require them to
be preserved in time in the computation of initial data as usually done
in the extended conformal-thin sandwich formulation (see also
Refs.~\cite{wilson_mathews1995,wilson_mm1996}).

Next, we decompose the conformal factor $\psi$, a weighted lapse
function $\Phi := \alpha \psi$ with $\alpha$ being the lapse function,
and the conformally weighted traceless part of the extrinsic curvature
$\hat{A}_{ij}$ as
\begin{align}
 \psi & = 1 + \frac{M_\mathrm{P}}{2 r_\mathrm{BH}} + \phi , \\
 \Phi & = 1 - \frac{M_\Phi}{2 r_\mathrm{BH}} + \eta , \\
 \hat{A}_{ij} & = \hat{\nabla}_i W_j + \hat{\nabla}_j W_i - \frac{2}{3}
 f_{ij} \hat{\nabla}^k W_k + K_{ij}^\mathrm{P} .
\end{align}
Here, $r_\mathrm{BH}$ is the coordinate distance from the puncture,
$M_\mathrm{P}$ and $M_\Phi$ are constants of mass dimension, and
$\hat{\nabla}_i$ denotes the covariant derivative associated with
$f_{ij}$. The singular part of the extrinsic curvature
$K_{ij}^\mathrm{P}$ is determined by two sets of covariantly constant
vectorial parameters, namely the linear momentum $P^\mathrm{BH}_i$ and
the bare spin angular momentum $S_\mathrm{P}^i$ of the black hole, as
\begin{align}
 K_{ij}^\mathrm{P} & := \frac{3}{2 r_\mathrm{BH}^2} [ l_i
 P^\mathrm{BH}_j + l_j P^\mathrm{BH}_i - ( f_{ij} - l_i l_j ) l^k
 P^\mathrm{BH}_k ] \notag \\
 & + \frac{3}{r_\mathrm{BH}^3} [ \epsilon_{kil} S_\mathrm{P}^l l^k l_j +
 \epsilon_{kjl} S_\mathrm{P}^l l^k l_i ] ,
\end{align}
where $l^i := x_\mathrm{BH}^i / r_\mathrm{BH}$ is the unit radial
vector, $l_i = f_{ij} l^j$, and $\epsilon_{ijk}$ is the Levi-Civita
tensor associated with $f_{ij}$
\cite{bowen_york1980,brandt_brugmann1997}.

Finally, the regular parts of the geometric quantities $\phi$, $\eta$,
$W_i$, and the shift vector $\beta^i$ are obtained by solving elliptic
equations derived from a subset of the Einstein equation [see, e.g.,
Eqs.~(16)--(19) of Ref.~\cite{kyutoku_st2009} for the explicit form]. In
the puncture framework, we have no inner boundary at the horizon, and
the outer boundary condition is derived from the asymptotic flatness
condition.

One of the keys to obtain low-eccentricity initial data is incorporation
of the approaching velocity to the helical symmetry, which governs
quasicircular initial data of binaries. Following
Ref.~\cite{kyutoku_st2014}, we adopt a symmetry vector equipped with the
approaching velocity $v_\mathrm{app,NS}^i$ of the form
\begin{equation}
 \xi^\mu = ( \partial_t )^\mu + \Omega ( \partial_\varphi )^\mu +
  v_\mathrm{app,NS}^i ( \partial_i )^\mu \label{eq:symvec}
\end{equation}
around the neutron star. In our computation, the rotational axis is
taken to be the $z$ axis, and both the black hole and the neutron star
are chosen to be located on the $xz$ plane. That is, we have
$x_\mathrm{NS}^i = ( x_\mathrm{NS} , 0 , z_\mathrm{NS} )$ and
$x_\mathrm{BH}^i = ( x_\mathrm{BH} , 0 , z_\mathrm{BH} )$. We always
choose $z_\mathrm{NS} = 0$ without loss of generality because this
choice removes the need for moving fluid variables as a whole during the
iteration. For nonprecessing configurations, we also set $z_\mathrm{BH}
= 0$ and thus both members of the binary are located on the $x$
axis. Our symmetry vector with a translational approaching motion is
slightly different from the uniform contraction usually adopted in other
eccentricity reduction methods
\cite{pfeiffer_bklls2007,foucart_kpt2008,buonanno_kmpt2011}, and no
significant difference has been found for binary neutron stars
\cite{kyutoku_st2014}. We note that this modified symmetry vector is not
fully compatible with the spacetime symmetry as discussed in
Ref.~\cite{kyutoku_is2014}.

\subsection{Choice of free parameters}

\begin{table*}
 \caption{Free parameters and requirements for determining them in our
 formulation for initial data of black hole--neutron star binaries in
 the puncture framework. Because $\Omega$ is determined in a different
 manner for quasicircular and low-eccentricity initial data, we show two
 conditions separately by ``QC'' and ``low-$e$,'' respectively. The
 approaching velocity of the neutron star, $v_\mathrm{app,NS}^i$, is
 fixed to be zero for quasicircular initial data.}
 \begin{tabular}{ccc} \hline
  Symbol and meaning & Requirement \\
  \hline
  \multicolumn{2}{c}{Parameters of the black hole} \\
  \hline
  $M_\mathrm{P}$: bare mass parameter in $\psi$ & The mass of the black
      hole takes a desired value \\
  $S_\mathrm{P}^i$: bare spin parameter in $K^\mathrm{P}_{ij}$ & The
      spin vector of the black hole takes a desired value \\
  $P^\mathrm{BH}_i$: linear-momentum parameter in $K^\mathrm{P}_{ij}$ &
      The total linear momentum of the system vanishes \\
  $M_\Phi$: mass parameter in $\Phi$ & The Arnowitt-Deser-Misner and
      Komar masses agree \\
  \hline
  \multicolumn{2}{c}{Parameters of the neutron star} \\
  \hline
  $-h u_\mu \xi^\mu$: first integral of Euler's equation & The mass of
      the neutron star takes a desired value \\
  $v_\mathrm{app,NS}^i$: approaching velocity & The approaching velocity
      of the binary takes a desired value \\
  \hline
  \multicolumn{2}{c}{Parameters of the binary} \\
  \hline
  $d^x$: separation along the $x$ axis & This is fixed to specify a
      model \\
  $d^z$: separation along the $z$ axis & The force-balance condition
      along the $z$ direction is satisfied \\
  $\Omega$: orbital angular velocity & QC: The force-balance condition
      along the $x$ direction is satisfied \\
  & low-$e$: This is fixed to a desired value \\
  Location of the rotational axis & The azimuthal component of the shift
      vector at the puncture is equal to $- \Omega$ \\
  \hline
 \end{tabular}
 \label{table:free}
\end{table*}

Free parameters in the formulation must be determined by physical
requirements. In this subsection, we describe our method for determining
them. A concise summary is presented in Table
\ref{table:free}. Iterative procedures are described in Sec.~III B of
Ref.~\cite{kyutoku_st2009} except that we have included steps to adjust
the approaching velocity and the value of $z_\mathrm{BH}$.\footnote{In
the published version of Ref.~\cite{kyutoku_st2009}, ``Adjust the
maximum enthalpy of the NS, $h_c$, at the center of the NS, to fix the
baryon rest mass of the NS.''  should have been marked as step (5).}
The level of the convergence for our iterative solution is not affected
significantly by the eccentricity reduction procedure.

The bare mass, $M_\mathrm{P}$, and the bare spin parameter,
$S_\mathrm{P}^i$, are determined to obtain desired values of the mass
and spin of the black hole. The mass and spin magnitude of the black
hole are computed in the isolated-horizon framework (see
Ref.~\cite{gourgoulhon_jaramillo2006} for reviews) with an approximate
rotational Killing vector obtained by minimizing its shear on the
horizon \cite{cook_whiting2007}. In our computation, we restrict the
spin parameter to have only $x$ and $z$ components and define the
inclination angle $\iota$ as the angle between the coordinate components
of the orbital angular momentum and the spin angular momentum in a
gauge-dependent manner \cite{kawaguchi_knost2015}. The linear momentum
of the black hole, $P^\mathrm{BH}_i$, is determined by requiring the
total linear momentum of the system vanishes. This means that we cannot
choose values of $P^\mathrm{BH}_i$ to control the approaching velocity
of the black hole. This is the chief difference from the eccentricity
reduction of binary black holes in the puncture framework
\cite{husa_hgsb2008,purrer_hh2012,ramosbuades_hp2019} and is the reason
that we need to develop a method suitable for black hole--neutron star
binaries. A constant value of the first integral of the Euler equation,
$-hu_\mu \xi^\mu$ with $h$ and $u^\mu$ being the specific enthalpy and
the $4$-velocity of the fluid, respectively, is determined by requiring
the baryon rest mass of the neutron star to take a desired value.

The mass parameter in the weighted lapse function, $M_\Phi$, is
determined by the condition that the Arnowitt-Deser-Misner and Komar
masses agree, which holds for stationary and asymptotically flat
spacetimes \cite{beig1978,ashtekar_magnonashtekar1979}. This condition
also holds for quasicircular initial data computed in our formulation
\cite{friedman_us2002,shibata_uf2004}, and thus requiring this condition
is fully justified (see also
Refs.~\cite{gourgoulhon_gb2002,grandclement_gb2002,caudill_cgp2006} for
early work on binary black holes). However, this does not hold
rigorously for low-eccentricity initial data with an approaching
velocity. Despite this caveat, we still determine the value of $M_\Phi$
by the equality of the two masses even if the approaching velocity is
turned on. We expect that this condition is not very problematic because
we observe for initial data of binary neutron stars
\cite{kyutoku_st2014} that the differences between the
Arnowitt-Deser-Misner and Komar masses are of the same order for both
quasicircular and low-eccentricity cases. This situation is also
reported in black hole--neutron star initial data computed in the
excision framework \cite{foucart_kpt2008}. When more accurate numerical
computations become necessary, the condition for determining $M_\Phi$
should be elaborated. The Arnowitt-Deser-Misner and Komar masses are
computed both by surface and volume integrals, and these two variants
agree within the error of $O(\num{e-6})$.

As we will discuss in the next section, we would like to control the
orbital angular velocity $\Omega$ and approaching velocity of the binary
$v_\mathrm{app}$ to obtain low-eccentricity initial data.\footnote{The
parameter, $v_\mathrm{app}$, is denoted by $2v$ in
Ref.~\cite{kyutoku_st2014} for equal-mass binary neutron stars.} On one
hand, the orbital angular velocity, $\Omega$, appears explicitly in our
symmetry vector, Eq.~\eqref{eq:symvec}. On the other hand, the
approaching velocity of the binary is controlled implicitly via that of
the neutron star, $v_\mathrm{app,NS}^i$, in our formulation. The
approaching velocity vector of the black hole, $v_\mathrm{app,BH}^i$, in
the initial data is identified as the minus of the $x$ and $z$
components (see below for the $y$ component) of the shift vector at the
puncture as
\begin{equation}
 v_\mathrm{app,BH}^i = \left. ( - \beta^x , 0 , - \beta^z )
                                            \right|_\mathrm{BH} .
 \label{eq:vbh}
\end{equation}
Then, we adjust the value of $| v_\mathrm{app,NS}^i |$, where the usual
Euclidean norm is assumed, so that $| v_\mathrm{app,NS}^i | + |
v_\mathrm{app,BH}^i |$ agrees with the desired value of $|
v_\mathrm{app} |$. Although we could have adjusted the value of
$|v_\mathrm{app,NS}^i - v_\mathrm{app,BH}^i|$ instead, this is not
necessarily preferable in a curved spacetime. Because we always require
the vanishing of the total linear momentum of the system, the
approaching velocity of the neutron star automatically induces that of
the black hole.

When we compute quasicircular initial data for a given separation along
the $x$ axis, $d^x$, the value of $\Omega$ is determined by requiring
the force balance at the neutron-star center (see Sec.~IV D 2 of
Ref.~\cite{gourgoulhon_gtmb2001}),
\begin{equation}
 \left. \frac{\partial h}{\partial x} \right|_\mathrm{NS} = 0 .
  \label{eq:balancex}
\end{equation}
Specifically, we insert constancy of the first integral of the Euler
equation to Eq.~\eqref{eq:balancex}. Because the first integral of the
Euler equation includes $\Omega$ through the shift vector, the force
balance condition, Eq.~\eqref{eq:balancex}, can be rewritten as an
equation to determine the orbital angular velocity. The approaching
velocity is set to be zero by choosing $v_\mathrm{app,NS}^i = 0$. This
results in $v_\mathrm{app,BH}^i = 0$ within the numerical error.

When we compute low-eccentricity initial data for a given value of
$d^x$, we prescribe desired values of $\Omega$ and $v_\mathrm{app}$
according to the estimates from dynamical simulations (see
Sec.~\ref{sec:method_iter}). The value of $\Omega$ in
Eq.~\eqref{eq:symvec} is fixed to this prescribed value for each
computation of initial data. For the approaching velocity vector of the
neutron star, $v_\mathrm{app,NS}^i$, in Eq.~\eqref{eq:symvec}, we need
to determine the magnitude and the direction. The magnitude is
determined so that $|v_\mathrm{app,NS}^i| + |v_\mathrm{app,BH}^i|$ takes
the prescribed value, $|v_\mathrm{app}|$. The direction is set to point
toward the black hole. For nonprecessing binaries, $v_\mathrm{app,NS}^i$
has only the $x$ component and $v_\mathrm{app,BH}^i$ points exactly
opposite to $v_\mathrm{app,NS}^i$. For precessing binaries, we find that
the induced approaching velocity of the black hole,
$v_\mathrm{app,BH}^i$, does not point exactly toward the neutron star in
our coordinates, but the deviation is smaller than \ang{0.1} for the
case studied here. Because the direction is inherently gauge dependent,
we regard this deviation as acceptable.

We have to determine the location of the rotational axis, or the
positions of the black hole and neutron star relative to the rotational
axis, in our computations. In the excision framework, this location is
fixed by the condition that the total linear momentum of the system
vanishes
\cite{taniguchi_bfs2006,taniguchi_bfs2007,taniguchi_bfs2008,foucart_kpt2008}.
However, this condition has already been used to determine
$P^\mathrm{BH}_i$ in the puncture framework. In this work, following
Ref.~\cite{shibata_taniguchi2008}, we determine the location of the
rotational axis by requiring that the azimuthal component of the shift
vector at the puncture is equal to the minus of the angular velocity,
\begin{equation}
 \left. \beta^\varphi \right|_\mathrm{BH} = - \Omega
  . \label{eq:betaphi}
\end{equation}
This states that the puncture moves along the symmetry vector,
Eq.~\eqref{eq:symvec}, and is consistent with our definition of
$v_\mathrm{app,BH}^i$, where the unused $y$ component of the shift
vector is interpreted as the orbital velocity.

We also have to determine the separation between the black hole and the
neutron star along the rotational axis, $d^z$, when we compute
precessing configurations. It is determined by requiring the
force-balance condition like Eq.~\eqref{eq:balancex} but along the $z$
direction \cite{foucart_dkt2011,kawaguchi_knost2015},
\begin{equation}
 \left. \frac{\partial h}{\partial z} \right|_\mathrm{NS} = 0 .
\end{equation}
The position of the neutron star is fixed to $z_\mathrm{NS} = 0$
throughout.

\subsection{Iterative correction} \label{sec:method_iter}

We seek the optimal choice of $\Omega$ and $v_\mathrm{app}$ for a given
value of the separation along the $x$ axis, $d^x$, via iterative
corrections estimated from the orbital evolution derived by dynamical
simulations. This strategy is originally developed for binary-black-hole
initial data \cite{pfeiffer_bklls2007} and later applied successfully to
black hole--neutron star binaries in the excision framework
\cite{foucart_kpt2008} and to binary neutron stars
\cite{kyutoku_st2014,moldenhauer_mjtb2014,haas_etal2016}. Our dynamical
simulations are performed with an adaptive-mesh-refinement code, {\small
SACRA} \cite{yamamoto_st2008}, and the formulation adopted in the
current version is explained in Ref.~\cite{kyutoku_st2014}. We do not
use the MPI-parallelized version of {\small SACRA}
\cite{kiuchi_kksst2017} because the eccentricity reduction does not
require orbital evolution with a very high precision.

The position and velocity\footnote{The quantities denoted by $\dot{x}^i$
refer to the velocity as a sum of the orbital and approaching
velocities.} of the neutron star are identified with the integration
over the fluid as
\begin{align}
 x_\mathrm{NS}^i & = \frac{\int \rho_* x^i d^3 x}{\int \rho_* d^3 x}
 , \label{eq:locNS} \\
 \dot{x}_\mathrm{NS}^i & = \frac{\int \rho_* \dot{x}^i d^3 x}{\int
 \rho_* d^3 x} ,
\end{align}
where $\rho_* := \rho \alpha u^t \sqrt{\gamma}$ and $\dot{x}^i := u^i /
u^t$, with $\rho$ being the rest-mass density. The velocity of the black
hole is identified as the minus of the shift vector at the puncture in a
manner similar to Eq.~\eqref{eq:vbh} as
\begin{equation}
 \dot{x}_\mathrm{BH}^i = \left. - \beta^i \right|_\mathrm{BH} ,
\end{equation}
and the position is obtained by integrating this in time
\cite{campanelli_lmz2006,brugmann_ghhst2008}. In {\small SACRA}, the
shift vector at the puncture is determined by trilinear interpolation
from surrounding eight grid points, and this limits the accuracy with
which we can determine the orbital evolution for given data of
gravitational fields. The orbital angular velocity of the binary is
computed from the Euclidean outer product of $x^i := x_\mathrm{NS}^i -
x_\mathrm{BH}^i$ and $\dot{x}^i := \dot{x}_\mathrm{NS}^i -
\dot{x}_\mathrm{BH}^i$ as \cite{buonanno_kmpt2011,Boyle2013}
\begin{equation}
 \mathbf{\Omega} = \frac{\mathbf{x} \times \dot{\mathbf{x}}}{|
  \mathbf{x} |^2} .
\end{equation}

We estimate appropriate values of the correction to $\Omega$ and
$v_\mathrm{app}$ by fitting the time evolution of orbital angular
velocity, $\dot{\Omega} (t)$, by a function
\cite{pfeiffer_bklls2007,boyle_bkmpsct2007,buonanno_kmpt2011}
\begin{equation}
 \dot{\Omega} (t) = A_0 + A_1 t + B \cos ( \omega t + \phi_0 ) ,
\end{equation}
where $\{ A_0 , A_1 , B , \omega , \phi_0 \}$ are parameters determined
by the fitting. We average numerical data over $\sim 100$ time steps in
deriving $\dot{\Omega} (t)$ to remove high-frequency noise
\cite{kyutoku_st2014}. The fitting is performed using $\dot{\Omega} (t)$
during $0.5P_0 < t < 3P_0$, where $P_0$ is the initial orbital period of
the binary. Aiming at removing the modulation term, $B \cos ( \omega t +
\phi_0 )$, we modify $\Omega$ and $v_\mathrm{app}$ in the initial-data
computation by
\begin{align}
 \delta \Omega & = - \frac{B \omega \sin \phi_0}{4 \Omega^2} ,
 \label{eq:coromega} \\
 \delta v_\mathrm{app} & = \frac{B d \cos \phi_0}{2 \Omega}
 \label{eq:corvapp} ,
\end{align}
according to Newtonian expressions
\cite{pfeiffer_bklls2007,buonanno_kmpt2011,kyutoku_st2014}. Here, $d =
\sqrt{(d^x)^2 + (d^z)^2} = |x^i(t=0)|$ is the initial orbital
separation. We also estimate the orbital eccentricity by
\begin{equation}
 e \approx \frac{|B|}{2 \omega \Omega} \label{eq:ecc}
\end{equation}
in this fitting procedure. Exceptionally when $e \lesssim 0.001$, we
find that the beginning of the fitting interval has to be delayed until
$t = 0.75P_0$ to obtain meaningful estimates of the residual
eccentricity.

\section{Demonstration} \label{sec:result}

\begin{table*}
 \caption{Key quantities of the models of black hole--neutron star
 binaries constructed in this work. Names of models represent the spin
 configuration and the stage of eccentricity reduction. Specifically, QC
 and IterX stand for quasicircular and the Xth iteration,
 respectively. The total mass of the binary at infinite separation,
 $m_0$, is $5.4 M_\odot$ for all the models. The normalized orbital
 angular velocity $m_0 \Omega$ and approaching velocity $v_\mathrm{app}$
 of the binary characterize the initial data. The Arnowitt-Deser-Misner
 mass is denoted by $M_0$, where we show it as the gravitational binding
 energy, $| M_0 - m_0 |$. The magnitude of the orbital angular momentum
 of the system is given by $L_0$, which does not include the spin
 angular momentum of the black hole. The eccentricity $e$ is estimated
 by fitting the time derivative of the orbital angular velocity obtained
 in dynamical simulations. The initial orbital period,
 gravitational-wave frequency, and wavelength are $\approx
 \SI{6.7}{\ms}, \approx \SI{300}{\hertz}$, and $\approx \SI{1000}{\km}$,
 respectively.}
 \begin{tabular}{cccccc} \hline
  Model & $m_0 \Omega$ & $v_\mathrm{app}$ & $| M_0 - m_0 | ( M_\odot )$
  & $L_0 ( M_\odot^2 )$ & $e$ \\
  \hline
  \multicolumn{6}{c}{Zero Spin: $\chi = 0, d = \SI{86.1}{\km}$} \\
  \hline
  ZS-QC & $0.0250001$ & $0$ & $0.0399$ & $21.29$ & $0.04$ \\
  ZS-Iter1 & $0.0252757$ & $-0.00723672$ & $0.0388$ & $21.54$ &
                      $0.02$ \\
  ZS-Iter2 & $0.0254085$ & $-0.00233390$ & $0.0382$ & $21.66$ &
                      $0.006$ \\
  ZS-Iter3 & $0.0253783$ & $-0.00138723$ & $0.0384$ & $21.63$ &
                      $0.003$ \\
  ZS-Iter4 & $0.0253797$ & $-0.00207750$ & $0.0384$ & $21.63$ &
                      $0.003$ \\
  ZS-Iter5 & $0.0253657$ & $-0.00173550$ & $0.0384$ & $21.62$ &
                      $0.0005$ \\
  \hline
  \multicolumn{6}{c}{Aligned Spin: $\chi = 0.75, \iota = \ang{0}, d =
  \SI{85.5}{\km}$} \\ \hline
  AS-QC & $0.0249999$ & $0$ & $0.0416$ & $20.36$ & $0.04$ \\
  AS-Iter1 & $0.0253334$ & $-0.00713689$ & $0.0403$ & $20.65$ & $0.02$\\
  AS-Iter2 & $0.0254891$ & $-0.00264875$ & $0.0396$ & $20.79$ &
                      $0.007$ \\
  AS-Iter3 & $0.0254545$ & $-0.00097836$ & $0.0398$ & $20.75$ &
                      $0.003$ \\
  AS-Iter4 & $0.0254355$ & $-0.00160145$ & $0.0399$ & $20.74$ &
                      $0.002$ \\
  AS-Iter5 & $0.0254308$ & $-0.00194680$ & $0.0399$ & $20.73$ &
                      $0.002$ \\
  AS-Iter6 & $0.0254454$ & $-0.00164779$ & $0.0398$ & $20.75$ &
                      $0.002$ \\
  AS-Iter7 & $0.0254302$ & $-0.00177446$ & $0.0399$ & $20.73$ &
                      $0.001$ \\
  AS-Iter8 & $0.0254431$ & $-0.00174272$ & $0.0398$ & $20.74$ &
                      $0.002$ \\
  AS-Iter9 & $0.0254245$ & $-0.00165976$ & $0.0399$ & $20.73$ &
                      $0.0008$ \\
  \hline
  \multicolumn{6}{c}{Inclined Spin: $\chi = 0.75$, $\iota \approx
  \ang{92}, d^x = \SI{86.0}{\km}, d^z \approx \SI{3.5}{\km}$}
  \\ \hline
  IS-QC & $0.0250000$ & $0$ & $0.0399$ & $21.27$ & $0.04$ \\
  IS-Iter1 & $0.0252851$ & $-0.00742954$ & $0.0387$ & $21.53$ &
                      $0.03$ \\
  IS-Iter2 & $0.0254223$ & $-0.00246477$ & $0.0382$ & $21.65$ &
                      $0.006$ \\
  IS-Iter3 & $0.0253949$ & $-0.00139129$ & $0.0383$ & $21.62$ &
                      $0.003$ \\
  IS-Iter4 & $0.0253965$ & $-0.00217036$ & $0.0383$ & $21.63$ &
                      $0.003$ \\
  IS-Iter5 & $0.0253833$ & $-0.00166181$ & $0.0383$ & $21.61$ &
                      $0.0008$\\
  \hline
 \end{tabular}
 \label{table:model}
\end{table*}

We present results of our eccentricity reduction for a few models of
black hole--neutron star binaries. Key quantities of the initial data
constructed in this work are summarized in Table~\ref{table:model}. The
neutron stars are modeled by a piecewise polytropic approximation
\cite{read_lof2009} of the APR4 equation of state \cite{akmal_pr1998}
with the gravitational mass in isolation of $M_\mathrm{NS} = 1.35
M_\odot$. The APR4 equation of state is consistent with GW170817
\cite{de_flbbb2018,ligovirgo2018,ligovirgo2019,ligovirgo2020,narikawa_ukkkst2020}
and gives the dimensionless tidal deformability of $323$ for this
$1.35M_\odot$ neutron star. The gravitational mass in isolation of the
black hole is fixed to be $M_\mathrm{BH} = 4.05M_\odot$, giving the mass
ratio $Q := M_\mathrm{BH}/M_\mathrm{NS} = 3$, which has been studied
vigorously in the literature. Accordingly, the total mass at infinite
separation $m_0 := M_\mathrm{BH} + M_\mathrm{NS}$ is $5.4M_\odot$. We
have checked that our eccentricity reduction method works similarly for
other equations of state and/or binary parameters at least for the range
considered in our previous work \cite{kyutoku_iost2015} (see
Ref.~\cite{kyutoku_st2014} for binary neutron stars). We plan to present
results of systematic long-term simulations of low-eccentricity black
hole--neutron star binary coalescences elsewhere.

All the simulations are performed with computational domains consisting
of five coarser boxes fixed around an approximate center of mass and
four pairs of finer boxes comoving with each binary component. The edge
length of the largest computational domain is $\approx \SI{2500}{\km}$
and the grid resolution at the finest domain is $\approx
\SI{240}{\meter}$. With this resolution, the coordinate radius of the
neutron star is covered by $\approx 35$ points, and that of the apparent
horizon is covered by $\approx 25$ and $18$ points for $\chi = 0$ and
$0.75$, respectively. The grid resolution is intentionally kept moderate
for demonstrating that the eccentricity reduction can be performed with
a reasonable computational cost. We checked for selected models, both
before and after the eccentricity reduction, that the eccentricity does
not depend on the grid resolution of the dynamical simulations.

\subsection{Nonprecessing case}

First, we apply the eccentricity reduction described in
Sec.~\ref{sec:method} to two nonprecessing binaries. One is the ZS (zero
spin) model, for which the black hole is nonspinning. The other is the
AS (aligned spin) model, for which the black hole is spinning in a
prograde sense with respect to the orbital angular momentum with its
magnitude being $\chi = 0.75$. (See the next subsection for IS.) The
orbital modulation is induced only by the residual eccentricity and
possible gauge artifacts (see, e.g.,
Refs.~\cite{purrer_hh2012,kyutoku_st2014}) for these models. Thus, the
eccentricity reduction should be straightforward.

\subsubsection{Time evolution}

\begin{figure*}
 \begin{tabular}{cc}
  \includegraphics[width=.48\linewidth]{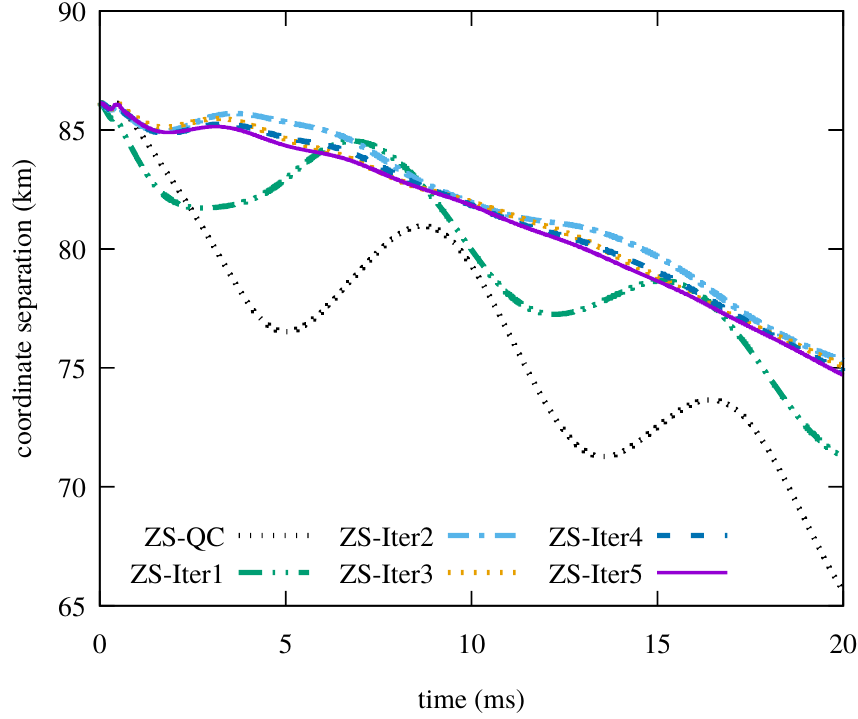} &
  \includegraphics[width=.48\linewidth]{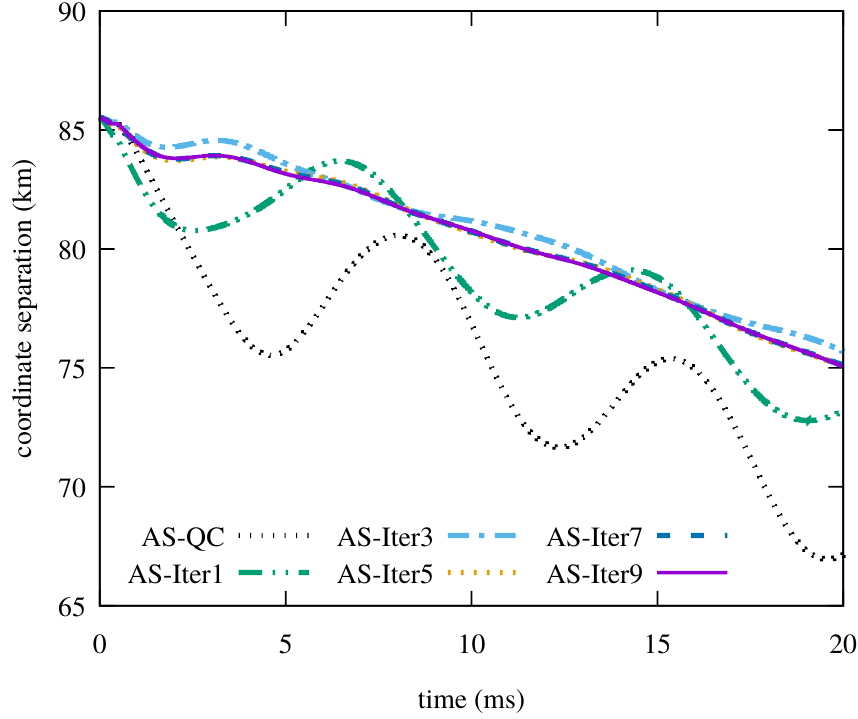}
 \end{tabular}
 \caption{Time evolution of the orbital separation for the ZS (left) and
 AS (right) families during initial $\approx 3$ orbital cycles. The
 eccentricity-induced modulation has a longer period than the orbital
 period due to the periastron advance. Differently from
 Ref.~\cite{kyutoku_st2014}, we do not need to perform Bezier smoothing
 to eliminate high-frequency noises because the location of the neutron
 star is defined not by the maximum density on discrete grid points but
 by the integral, Eq.~\eqref{eq:locNS}. Dips seen during the first
 $2$--\SI{3}{\ms} reflect initial gauge transition.}
 \label{fig:orbit_np}
\end{figure*}

\begin{figure}
 \includegraphics[width=0.95\linewidth]{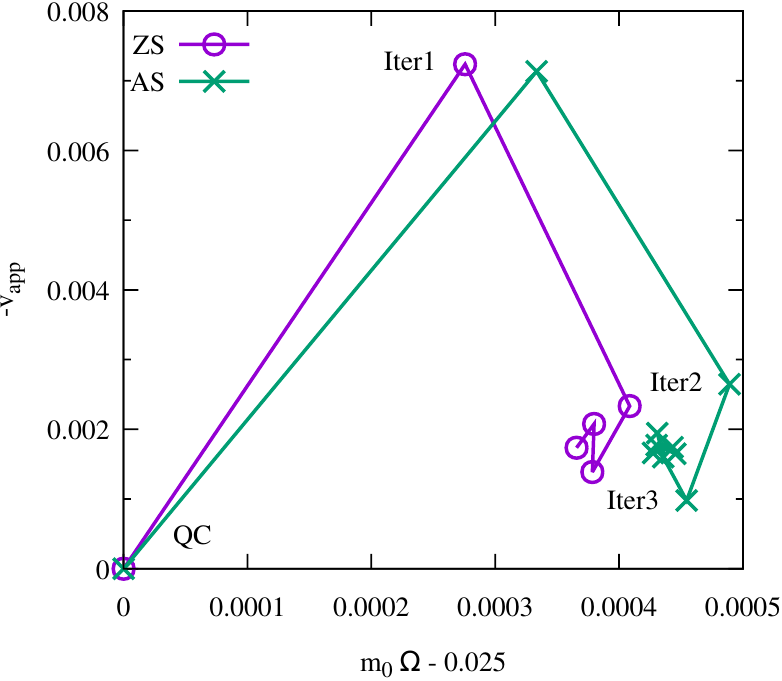} \caption{Orbital
 angular velocity and approaching velocity of initial data for the ZS
 (circle) and AS (cross) families. The former is plotted as the
 difference from $0.025$ of a value normalized by the total mass of the
 binary, $m_0$, and the latter is given by its negative. The origin
 represents the QC data, and successive points correspond to the data of
 subsequent iteration stages. We do not attach labels to data beyond
 Iter3 to avoid overcrowding.} \label{fig:cor_np}
\end{figure}

Figure~\ref{fig:orbit_np} shows the time evolution of the orbital
separation for selected models in the sequence of eccentricity
reduction. The orbital eccentricity is reduced from $~0.04$ to $<0.001$
(see Table \ref{table:model}) by several iterative corrections. To
achieve this eccentricity with current models, for which $m_0 \Omega
\approx 0.025$, we need to modify the orbital angular velocity by
$\approx 1.5\%$--1.7\% and to add the approaching velocity of $\approx
0.17\%$ of the speed of light. While the required approaching velocity
is similar to the value found for binary neutron stars considered in
Ref.~\cite{kyutoku_st2014}, the fractional amount of the required
correction to the orbital angular velocity, $\delta \Omega / \Omega$, is
larger by a factor of $\approx 4$. This is consistent with the degree of
the residual eccentricity, which is also higher by a factor of $\approx
4$ than that of binary neutron stars, $e \approx 0.01$, for
quasicircular initial data considered in Ref.~\cite{kyutoku_st2014},
because Eqs.~\eqref{eq:coromega} and \eqref{eq:ecc} indicate that these
quantities are approximately proportional to each other. The approaching
velocity may play a subdominant role for determining the residual
eccentricity.

The number of required iterations is typically larger than that for
binary neutron stars with our method \cite{kyutoku_st2014}. Changes of
the relevant parameters during the eccentricity reduction are presented
graphically in Fig.~\ref{fig:cor_np}, which should be compared with
Fig.~1 of Ref.~\cite{kyutoku_st2014}. This difference means that the
eccentricity reduction is cumbersome for black hole--neutron star
binaries, which are highly asymmetric and involve singularities
associated with the puncture. The necessity of many iterations is partly
ascribed to larger eccentricities in quasicircular initial data of black
hole--neutron star binaries in the puncture framework described
above. Another reason may be that our current correction formulas,
Eqs.~\eqref{eq:coromega} and \eqref{eq:corvapp}, are not efficient at $e
\lesssim 0.2\%$--0.3\%. The situation is visualized as wandering of
points beyond Iter3 in Fig.~\ref{fig:cor_np}. This low efficiency may be
ascribed to the fact that initial transition of the gauge condition in
the simulations (see Fig.~\ref{fig:orbit_np}) limits the accuracy with
which we can extrapolate results of the fitting to $t=0$
\cite{purrer_hh2012}. Our results also suggest that the presence of a
black hole spin increases the computational cost for the eccentricity
reduction, and indeed our experience supports this observation.

\begin{figure}
 \includegraphics[width=0.95\linewidth]{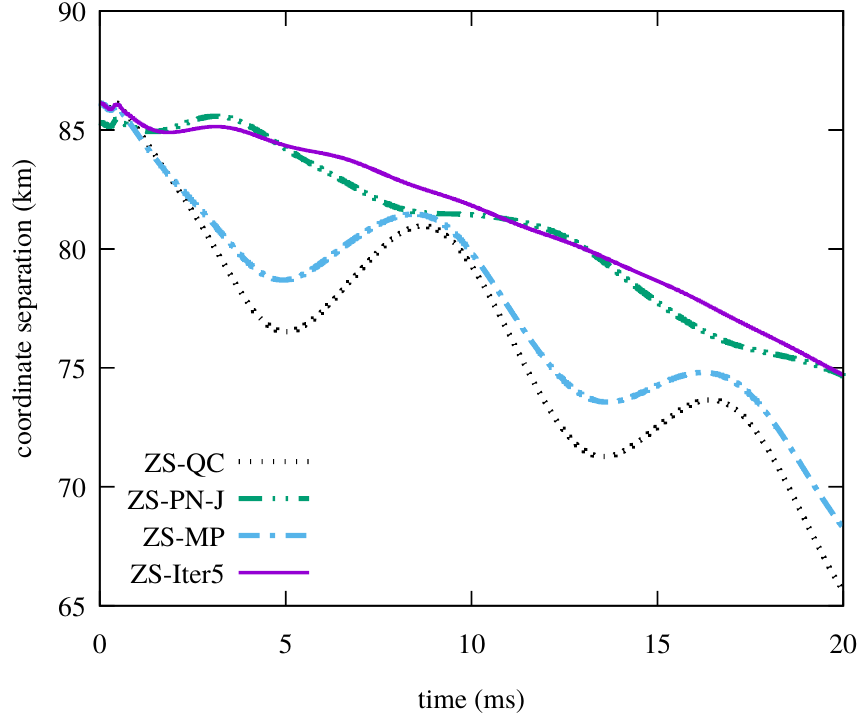} \caption{Same as
 Fig.~\ref{fig:orbit_np} but for different initial data of the ZS
 model. The curve labeled by ZS-PN-J shows the result for quasicircular
 initial data for which the location of the rotational axis is
 determined by the PN-J condition (see the body text). That labeled by
 ZS-MP shows the result for low-eccentricity initial data whose
 parameters are given by the fitting formulas due to
 Ref.~\cite{mroue_pfeiffer2012}.} \label{fig:orbit_compare}
\end{figure}

We recall that the orbital eccentricity of quasicircular initial data,
i.e., those without the approaching velocity, can be reduced by a factor
of 2--3 compared to the QC data adopted in this study simply by changing
the condition to determine the location of the rotational axis
\cite{shibata_kyt2009,kyutoku_st2009}. We compare orbital evolution
derived with various quasicircular initial data as well as
low-eccentricity ones for ZS models in
Fig.~\ref{fig:orbit_compare}. Here, the PN-J model refers to
quasicircular initial data in which the location of the rotational axis
is determined by requiring that the total angular momentum of the system
agrees with post-Newtonian predictions
\cite{kyutoku_st2009},\footnote{Many of our previous numerical
simulations have been performed using these PN-J data (e.g.,
Refs.~\cite{shibata_kyt2009,kyutoku_st2010,kyutoku_ost2011,kawaguchi_knost2015})
because of their moderately low eccentricity.} while the QC model is
derived using Eq.~\eqref{eq:betaphi}. No iterative eccentricity
reduction is applied to either initial data. This figure shows that the
PN-J data are superior to the QC data but cannot be a substitute for
low-eccentricity initial data, ZS-Iter5. At the same time, this figure
suggests that the eccentricity reduction described in this work may
further be improved by modifying the method to determine the location of
the rotational axis. In addition, the number of required iterations may
be reduced by changing the method appropriately. We leave this topic as
a future task.

We also find that phenomenological formulas for $\Omega$ and
$v_\mathrm{app}$ derived by simulations of binary-black-hole mergers in
the excision framework \cite{mroue_pfeiffer2012} do not reduce the
orbital eccentricity of black hole--neutron star binaries in the
puncture framework to $\lesssim 0.001$, although they work successfully
for binary neutron stars
\cite{hotokezaka_kos2015,hotokezaka_kss2016,kiuchi_kksst2017,kiuchi_kkss2020}. The
orbital evolution is shown in Fig.~\ref{fig:orbit_compare} as ZS-MP. The
eccentricity of this model is only smaller by a factor of $\lesssim 2$
than that of ZS-QC. This inefficiency may be ascribed to different
approaches for handling black holes, i.e., the puncture or
excision. Actually, it has been shown in the study of quasiequilibrium
sequences that the puncture initial data are characterized by
insufficient orbital angular momenta compared to the excision initial
data and post-Newtonian predictions \cite{kyutoku_st2009}. Thus, it is
naturally expected that the parameters to achieve low eccentricity
depend significantly on the framework to handle the black hole.

\begin{figure*}
 \begin{tabular}{cc}
  \includegraphics[width=.48\linewidth]{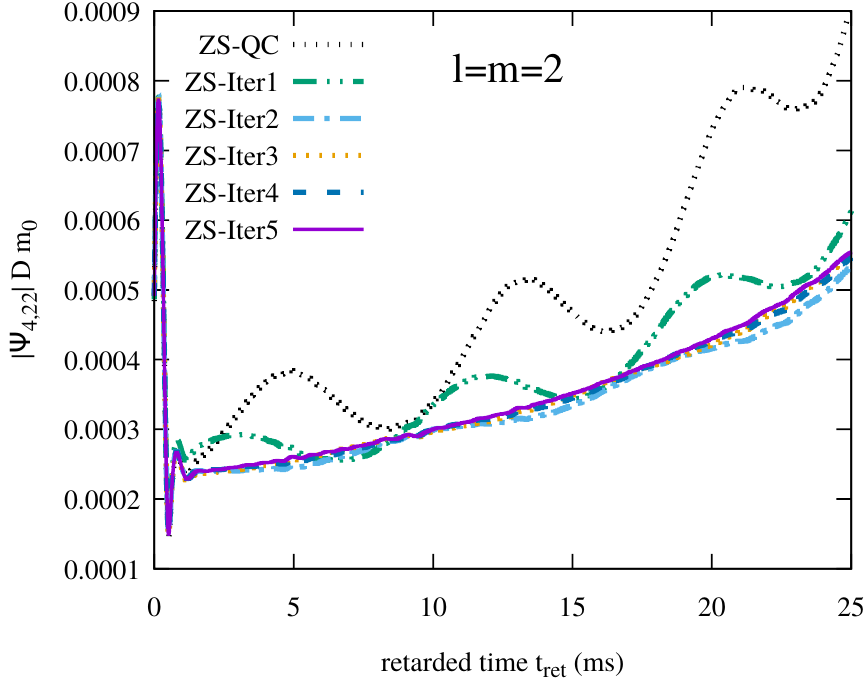} &
  \includegraphics[width=.48\linewidth]{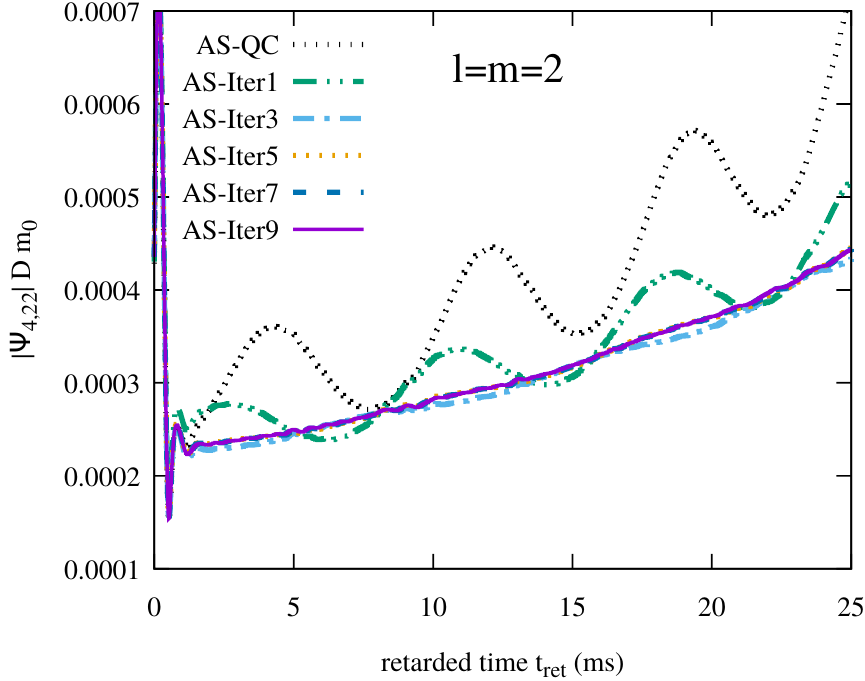} \\
  \includegraphics[width=.48\linewidth]{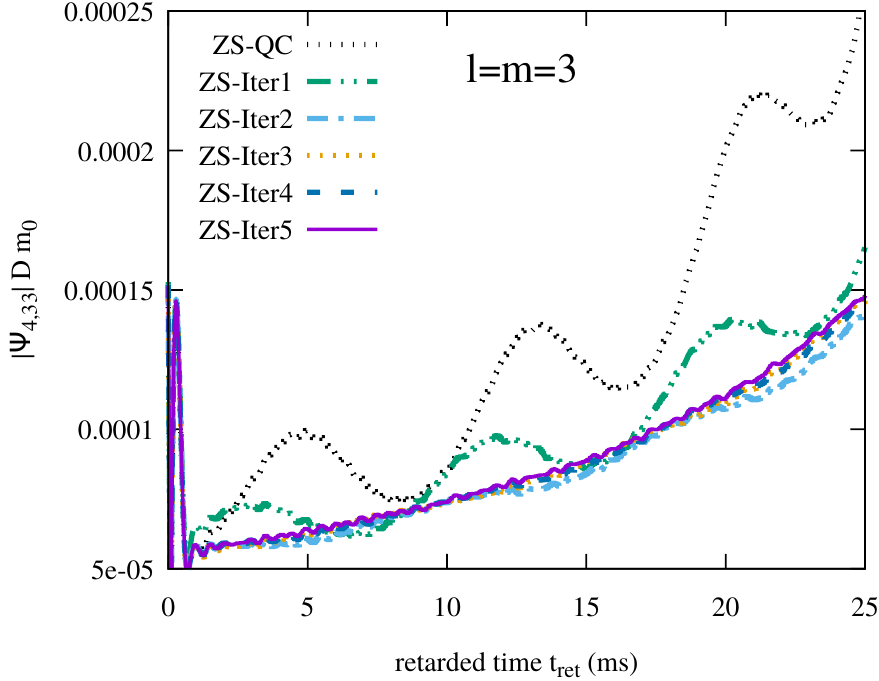} &
  \includegraphics[width=.48\linewidth]{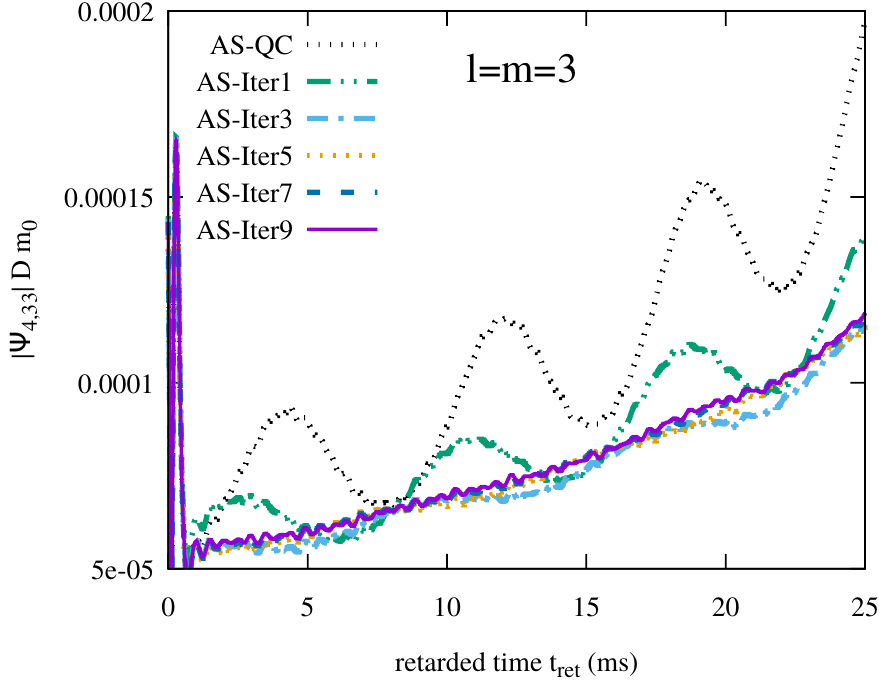} \\
  \includegraphics[width=.48\linewidth]{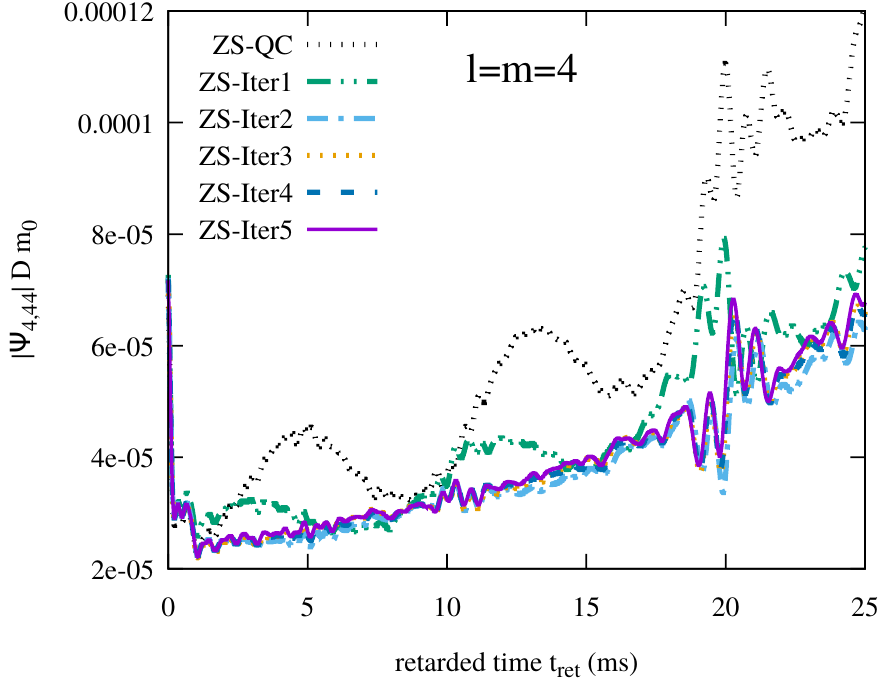} &
  \includegraphics[width=.48\linewidth]{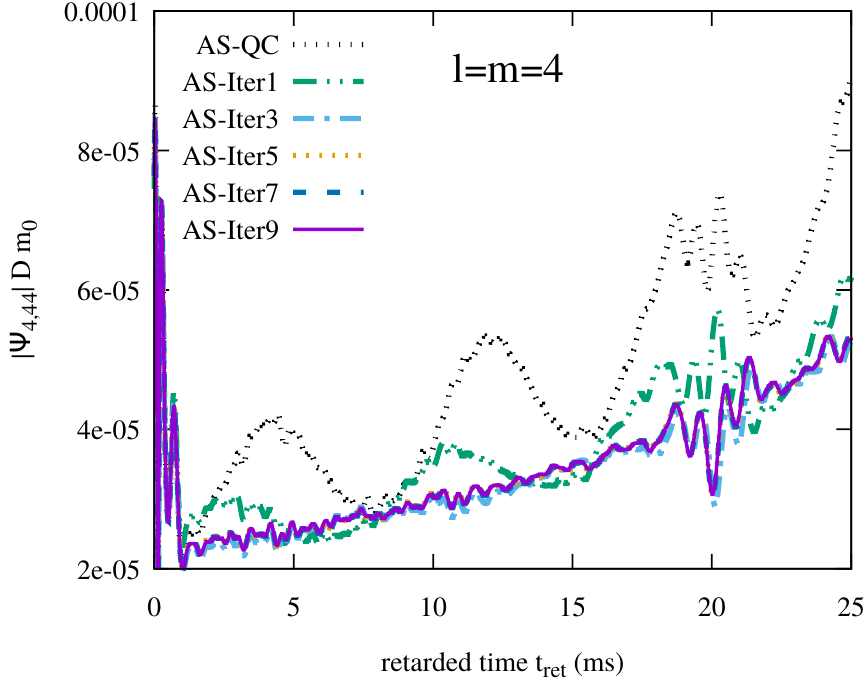} \\
 \end{tabular}
 \caption{Time evolution of the mode amplitude of $\Psi_4$ normalized by
 $D m_0$ for the ZS (left) and AS (right) families. The top, middle, and
 bottom rows show the $l=m=2, 3,$ and $4$ modes, respectively. The
 initial $\approx \SI{1}{\ms}$ suffers from junk radiation in the
 initial data. Wiggles observed in our previous work
 \cite{kyutoku_st2014} are not significant for the $l=m=2$ mode because
 the extraction radii are larger by a factor of $\approx 4$ here. Still,
 they are discernible in higher harmonics.} \label{fig:psi4_np}
\end{figure*}

To demonstrate that the successful reduction of the eccentricity is not
a gauge artifact, we also show the time evolution of the mode amplitude
of a Weyl scalar $\Psi_4$ as a gauge-invariant quantity in
Fig.~\ref{fig:psi4_np}. The time coordinate is taken to be a retarded
time defined by
\begin{align}
 t_\mathrm{ret} & := t - D - 2 m_0 \ln ( D / m_0 ) , \\
 D & := r_\mathrm{ext} \left( 1 + \frac{m_0}{2r_\mathrm{ext}} \right)^2
 ,
\end{align}
where $r_\mathrm{ext} = 800M_\odot = \SI{1181}{\km}$ is the extraction
radius. The suppression of the modulation in $\Psi_4$ confirms that the
reduced orbital modulation shown in Fig.~\ref{fig:orbit_np} is a
physical outcome caused by the reduced eccentricity. This figure also
shows that the modulation is eliminated irrespective of the harmonic
modes. This is particularly important for asymmetric systems like black
hole--neutron star binaries, for which higher multipole modes are
prominent in the actual gravitational-wave signals
\cite{ligovirgo2020-3,ligovirgo2020-4}.

\subsubsection{Properties of initial data}

We briefly comment on notable features of initial data. The eccentricity
reduction decreases the absolute value of the binding energy and
increases the orbital angular momentum of the binary (e.g., compare
ZC-QC and ZC-Iter5). Although it is not as good as in the case of binary
neutron stars \cite{kyutoku_st2014}, the agreement of the angular
momentum between our numerical initial data and post-Newtonian
approximations is improved by a factor of $\approx 3$ by the
eccentricity reduction. Quantitatively, for $\chi = 0$ associated with
the ZS model, fourth post-Newtonian approximations give $|M_0 - m_0| =
0.0394 M_\odot$ and $L_0 = 21.59 M_\odot^2$ for $m_0 \Omega = 0.025$
\cite{blanchet2014}, where the tidal effect does not significantly
affect these values \cite{vines_flanagan2013}. For an updated value of
$m_0 \Omega = 0.0253657$ corresponding to ZS-Iter5, post-Newtonian
approximations give $|M_0 - m_0| = 0.0397 M_\odot$ and $L_0 = 21.52
M_\odot^2$.

The increase of the angular momentum for low-eccentricity initial data
is consistent with our previous finding that the eccentricity in
quasicircular initial data can be reduced by enhancing the angular
momentum via the condition to determine the location of the rotational
axis \cite{shibata_kyt2009,kyutoku_st2009}. However, the binding energy
for low-eccentricity initial data tends to deviate more from the
post-Newtonian predictions than that for quasicircular initial data. The
insufficient binding of the low-eccentricity initial data is consistent
with our previous work \cite{kyutoku_st2009} and may reflect an inherent
limitation of initial data for black hole--neutron star binaries in the
puncture framework.

We find that junk radiation may account for a significant fraction of
the differences in the binding energy and the orbital angular
momentum. Specifically, the junk radiation\footnote{We estimate
quantities of junk radiation at $r_\mathrm{ext} = 200 M_\odot$ to
account for the short wavelength.} is found to carry away the energy
$\Delta E \approx 4$--$\num{5e-4} M_\odot$ and the angular momentum
$\Delta L \approx 5$--$\num{6e-2} M_\odot^2$ irrespective of the
residual eccentricity for the ZS family. They correspond to $\approx
30\%$--40\% for the binding energy and $\approx 50\%$--60\% for the
orbital angular momentum of the differences between ZS-Iter5 and the
post-Newtonian predictions. For ZC-QC, taking into account the junk
radiation only increases the differences.

The agreement of the angular momentum is improved only moderately for
spinning black hole--neutron star binaries. For $\chi = 0.75$ associated
with the AS model and $m_0 \Omega = 0.025$, the post-Newtonian
prediction gives $|M_0 - m_0| = 0.0413M_\odot$ and $L_0 = 20.63
M_\odot^2$, where the spin effect is included up to 3.5th order
\cite{bohe_mfb2013,blanchet2014}. These values change to $|M_0 - m_0| =
0.0417 M_\odot$ and $L_0 = 20.54 M_\odot^2$ for $m_0 \Omega = 0.0254245$
corresponding to AS-Iter9. In addition, the binding energy of
quasicircular initial data appears closer to the post-Newtonian
prediction than that of low-eccentricity initial data is. Still, our
results indicate that the key to reduce the eccentricity is to enhance
the orbital angular momentum and the binding energy from quasicircular
initial data derived with the helical symmetry.

Again, junk radiation may account for a significant fraction of the
difference between AS-Iter9 and the post-Newtonian predictions. In
particular, the energy carried away by the junk radiation is $\Delta E
\approx \num{2e-3} M_\odot$ and agrees approximately with the difference
in the binding energy. On another front, the angular momentum carried
away by the junk radiation is $\Delta L \approx 5$--$\num{6e-2}
M_\odot^2$, approximately coinciding with the value for the ZS
family. This account for $\approx 30\%$ of the difference.

\begin{figure*}
 \begin{tabular}{cc}
  \includegraphics[width=.48\linewidth]{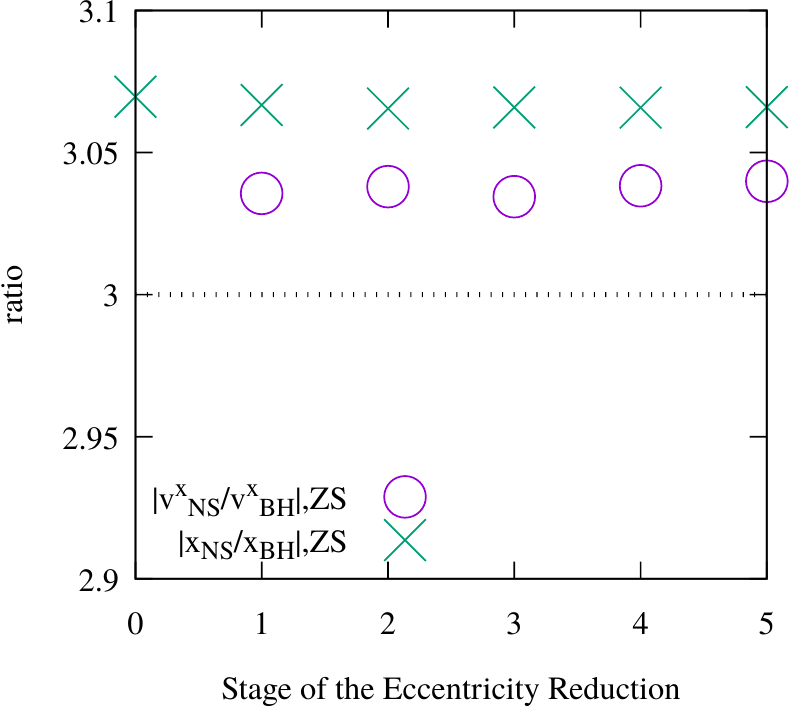} &
  \includegraphics[width=.48\linewidth]{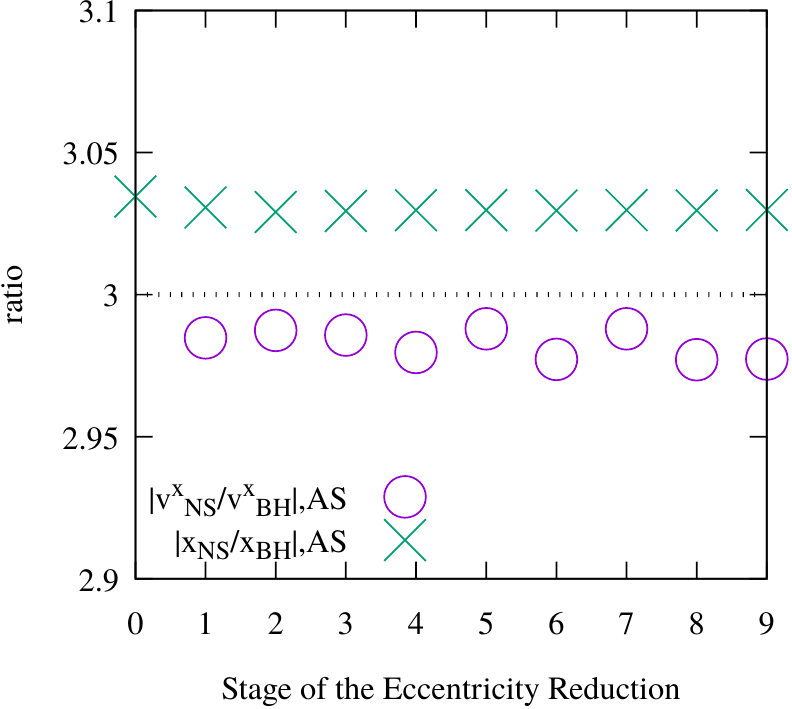}
 \end{tabular}
 \caption{Ratio of the approaching velocity (circle) and position
 (cross) between the neutron star and the black hole for the ZS (left)
 and AS (right) families. The horizontal axis denote the stage of
 iterative corrections with $0$ being the QC data. The dotted lines
 denote $Q=3$.}  \label{fig:ratio_np}
\end{figure*}

Finally, as we do not have a direct control of $v_\mathrm{app,BH}^i$, it
is worthwhile to check whether the approaching velocity of the binary in
the initial data is reasonably distributed to the black hole and the
neutron star with the ratio around an expected value of $Q=3$. Figure
\ref{fig:ratio_np} shows the ratio of the $x$ components of the velocity
as well as the $x$ coordinates for ZS and AS families. This figure
indicates that the reasonable distribution is automatically
achieved. Because both the velocity and the position are gauge-dependent
quantities, we believe that the deviation of 1\%--2\% from $Q=3$ found
here is acceptable.

\subsection{Precessing case}

Next, we perform eccentricity reduction on precessing binaries, for
which the black hole spin angular momentum is inclined with respect to
the orbital angular momentum of the binary. The spin-orbit, spin-spin,
and quadrupole-monopole couplings induce orbital precessions for
inclined spins \cite{barker_oconnel1975,apostolatos_cst1994,racine2008},
and precession-induced modulation appears in the orbital evolution and
gravitational waves \cite{buonanno_kmpt2011}. This feature could make
the eccentricity reduction more complicated than in the cases of
nonprecessing binaries.

We specifically consider the IS (inclined spin) model, for which the
black hole spin of $\chi = 0.75$ is inclined toward the neutron star by
$\approx \ang{92}$ from the initial orbital angular momentum in the
initial data.\footnote{The angle between the rotational axis and the
spin angular momentum is taken to be \ang{90}.} This model is expected
to exhibit precession of the orbital plane with the period $\approx
\SI{130}{\ms}$, which decreases during the orbital evolution. Following
Ref.~\cite{kawaguchi_knost2015}, the $z$ axis of the dynamical
simulation is taken to be the direction of the total angular momentum,
which is approximately fixed throughout the evolution.

\subsubsection{Time evolution}

\begin{figure}
 \includegraphics[width=.95\linewidth]{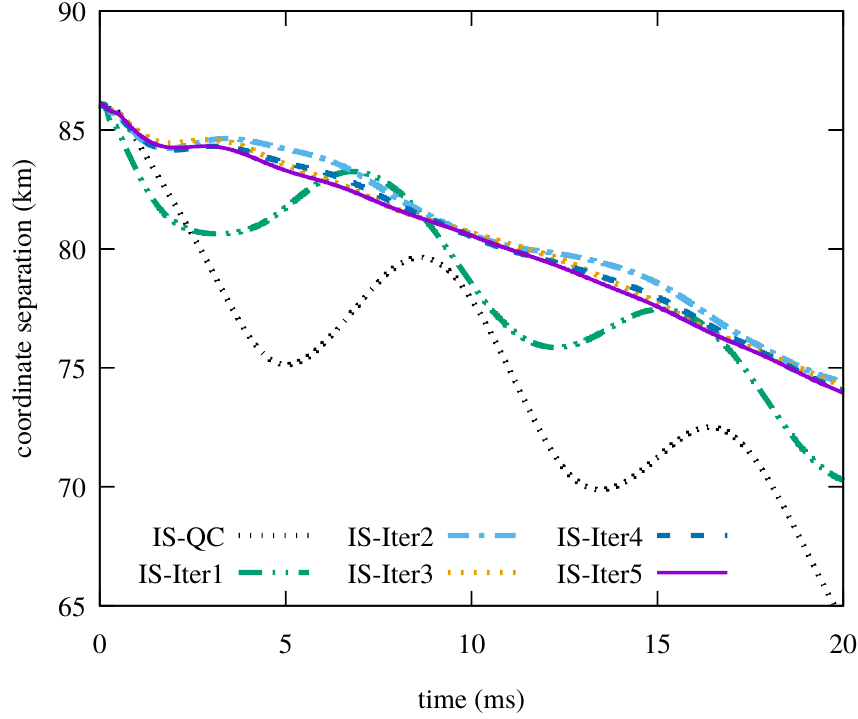} \caption{Same as
 Fig.~\ref{fig:orbit_np} but for the IS family. The orbital plane is
 expected to exhibit $\sim 20\%$ of the precession cycle on the time
 scale of this plot.} \label{fig:orbit_is}
\end{figure}

Figure \ref{fig:orbit_is} shows the time evolution of the orbital
separation for the IS family. This indicates that the performance of our
eccentricity reduction is similar to that for the ZS family. The
evolution of the orbital separation itself is also similar to that of
the ZS model because the black hole spin is approximately confined in
the orbital plane so that the inspiral is not decelerated or accelerated
by the spin-orbit interaction \cite{kidder1995}. Because we aim only at
reducing the orbital eccentricity to a moderately low value of $\lesssim
0.001$, the precession-induced modulation does not appear even in our
low-eccentricity evolution. We have checked that we did not incorrectly
subtract the precession-induced modulation during the iterative
eccentricity reduction. Specifically, frequency of the
precession-induced modulation is expected to be twice the orbital
frequency. Because the eccentricity-induced modulation should have
frequency lower than the orbital frequency, it is straightforward to
distinguish these two modulation effects.

\begin{figure}
 \includegraphics[width=0.95\linewidth]{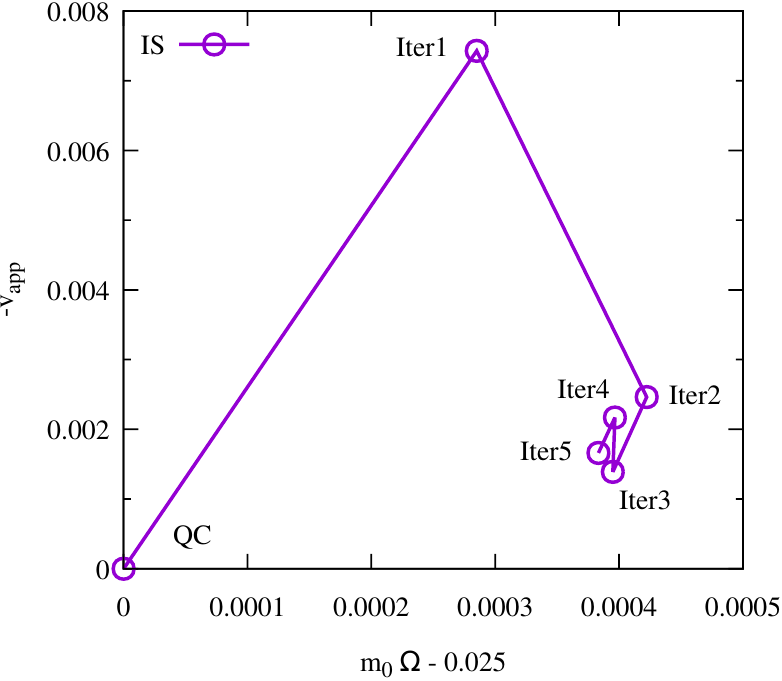} \caption{Same as
 Fig.~\ref{fig:cor_np} but for the IS family. All the models are
 labeled.} \label{fig:cor_is}
\end{figure}

Required corrections to the orbital parameters are presented in
Fig.~\ref{fig:cor_is}. They also exhibit similarity with those for the
ZS family shown in Fig.~\ref{fig:cor_np}. The situation should change,
however, if we further reduce the eccentricity to, say, $e \lesssim
\num{e-4}$, where precession-induced modulation is likely to play a role
\cite{buonanno_kmpt2011}.

To monitor the eccentricity using $\Psi_4$, the mode mixing due to the
precession has to be removed appropriately. Because our simulations
adopt the direction of the total angular momentum as the $z$ axis, the
precession-induced modulation is likely minimal but definitely
non-negligible \cite{oshaughnessy_vhms2011,schmidt_hh2012}. In this
paper, we transform $\Psi_4$ obtained in our simulations to components
in the coprecessing frame defined according to
Refs.~\cite{oshaughnessy_vhms2011,boyle_op2011,ochsner_oshaughnessy2012}
restricting ourselves only to $l=2$ modes. To our knowledge, this is the
first application of the radiation axis defined in
Ref.~\cite{oshaughnessy_vhms2011} to black hole--neutron star binaries
in numerical relativity (see also Ref.~\cite{kawaguchi_kns2017} for
another choice of the radiation axis \cite{schmidt_hha2011}).

\begin{figure}
 \includegraphics[width=.95\linewidth]{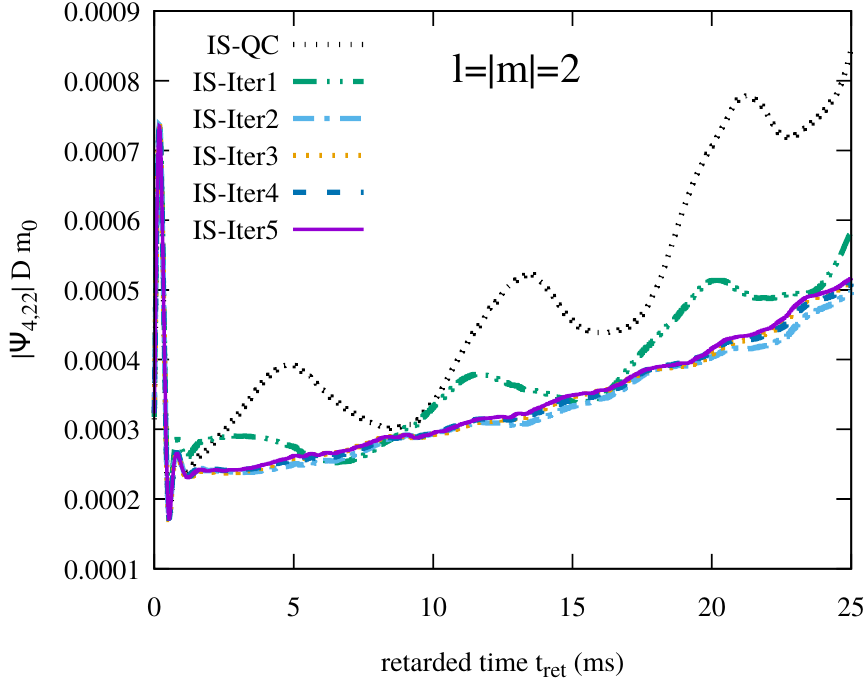} \caption{Same
 as Fig.~\ref{fig:psi4_np} but for the average of the amplitudes of
 $l=m=2$ and $l=-m=2$ modes of the IS family.} \label{fig:psi4_is}
\end{figure}

Time evolution of the $l=|m|=2$ mode amplitude of $\Psi_4$ is shown in
Fig.~\ref{fig:psi4_is}. Specifically, we average the amplitudes of $m=2$
and $m=-2$ modes to remove reflection-asymmetric components
\cite{pekowsky_ohs2013,boyle_kop2014}. Overall, we confirm that the
reduction of eccentricity is not a gauge artifact. Although the
evolution appears similar to that of the ZS model as expected from the
approximate absence of the spin component along the orbital angular
momentum, definite comparisons are difficult because of the presence of
additional structures in $\Psi_4$ of the IS model. In particular,
modulation with approximately twice the orbital frequency, which is
consistent with the precession-induced one, appears to be present in our
coprecessing-frame data. We defer detailed investigations of this
feature to a future task.

\subsubsection{Properties of initial data}

\begin{figure}
 \includegraphics[width=.95\linewidth]{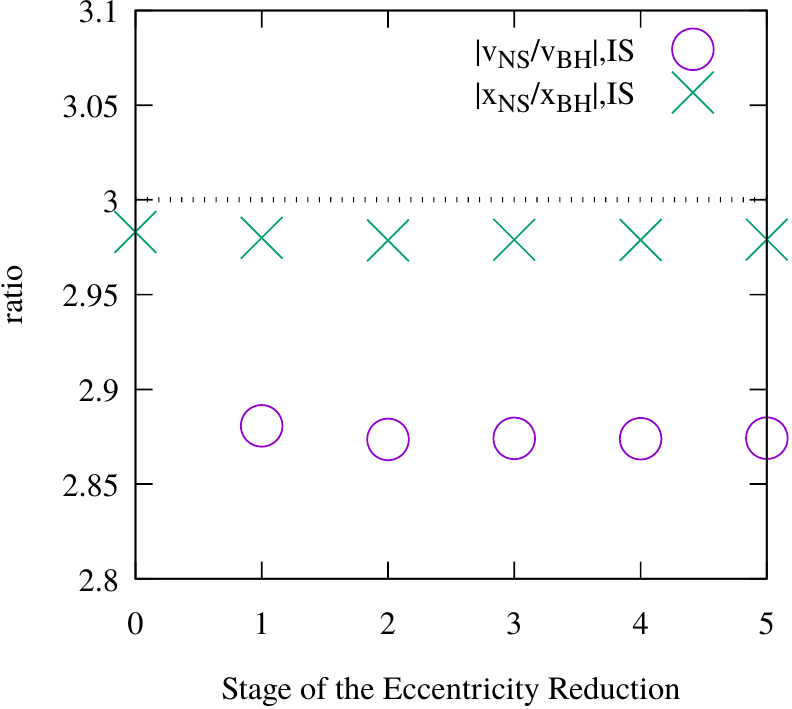} \caption{Same as
 Fig.~\ref{fig:ratio_np} but for the IS family. Note the different
 vertical scale.} \label{fig:ratio_is}
\end{figure}

The binding energy and the orbital angular momentum of the IS family are
shown in Table \ref{table:model}. They again behave similarly to those
of the ZS family. Post-Newtonian predictions are $|M_0 - m_0| = 0.0393
M_\odot$ and $L_0 = 21.63 M_\odot^2$ for $m_0 \Omega = 0.025$, and they
change to $|M_0 - m_0| = 0.0397M_\odot$ and $L_0 = 21.56 M_\odot^2$ for
$m_0 \Omega = 0.0253833$ corresponding to IS-Iter5. The level of
agreement and its improvement are similar to those for the ZS model.

The amount of junk radiation is different between the ZS and IS
models. The junk radiation carries away the energy $\Delta E \sim
\num{2e-3} M_\odot$ and the angular momentum $\Delta L \sim
5$--$\num{6e-2} M_\odot^2$ for the IS model. These values are closer to
those for the AS model than for the ZS model because the property of
junk radiation is influenced strongly by the magnitude of the black hole
spin. While $\Delta E$ corresponds to $\approx 140\%$ of the difference
of the binding energy between IS-Iter5 and the post-Newtonian
prediction, $\Delta L$ approximately coincides with the difference of
the orbital angular momentum. What is common among all the models
considered in this study is that a significant fraction of the
difference between low-eccentricity initial data and the post-Newtonian
prediction may be ascribed to the junk radiation.

We also plot the ratio of the approaching velocity and the $x$
coordinate of the black hole and the neutron star in
Fig.~\ref{fig:ratio_is}. They are reasonably close to the expected
values of $Q=3$. To sum up, our initial-data computations including the
eccentricity reduction perform similarly for both nonprecessing and
precessing configurations.

\section{Summary} \label{sec:summary}

We demonstrated that the orbital eccentricity in black hole--neutron
star binaries prepared in the puncture framework can be reduced to $e
\lesssim 0.001$ irrespective of the spin configuration. Following
previous work
\cite{pfeiffer_bklls2007,buonanno_kmpt2011,kyutoku_st2014}, we
iteratively modify parameters specifying the initial data, namely the
orbital angular velocity and the approaching velocity of the binary, by
analyzing orbital evolution derived by dynamical simulations for a few
orbits. Because the linear momentum of the black hole cannot be
specified freely in our puncture-based formulation for initial data of
black hole--neutron star binaries, we cannot adopt methods developed for
binary black holes in a straightforward manner
\cite{husa_hgsb2008,boyle_bkmpps2008,purrer_hh2012,ramosbuades_hp2019}.\footnote{We
could potentially rely on post-Newtonian approximation of the coordinate
velocity to specify the approaching velocity of the neutron star.} In
this work, we instead control the approaching velocity of the neutron
star by modifying the helical Killing vector required to integrate
Euler's equation \cite{kyutoku_st2014} (see also
Ref.~\cite{moldenhauer_mjtb2014}). The approaching velocity of the black
hole is induced automatically by the requirement that the total linear
momentum of the system vanishes. To control the approaching velocity of
the binary, the velocity of the black hole is defined by the minus of
the shift vector at the puncture. Accordingly, we also determine the
location of the rotational axis by requiring that the puncture moves in
the azimuthal direction with the orbital angular velocity
\cite{shibata_taniguchi2008}.

Our work completes, at least as a first serious attempt, the
eccentricity reduction for initial data of compact binaries required for
fully exploiting ground-based gravitational-wave detectors
\cite{ligovirgo2019-3}. For binary neutron stars, essentially the same
formulation has been adopted by various authors to reduce the orbital
eccentricity
\cite{kyutoku_st2014,moldenhauer_mjtb2014,haas_etal2016}. Although the
eccentricity reduction of binary black holes has long been performed
both in the excision \cite{pfeiffer_bklls2007} and puncture
\cite{purrer_hh2012} frameworks, it has been performed only in the
former for black hole--neutron star binaries
\cite{foucart_kpt2008}. Taking the robustness of the moving-puncture
simulations into account, our formulation may potentially be
advantageous for systematic exploration of a wide parameter space of
black hole--neutron star binaries (see
Refs.~\cite{kyutoku_st2010,kyutoku_ost2011} for our early effort).

Unsatisfactory aspects of our eccentricity reduction indicate future
directions for the improvement. The required number of iterative
corrections is generally larger than that for binary neutron stars,
particularly for the case that the black hole spin is high. In addition,
previous studies on binary black holes in the puncture framework suggest
that the gauge dynamics could degrade the accuracy in fitting the
orbital motion at $e \lesssim 0.001$ \cite{purrer_hh2012}. Although the
initial transition of the gauge condition could continue to be an
obstacle, these features may be improved by adopting a sophisticated
fitting procedure
\cite{buonanno_kmpt2011,ramosbuades_hp2019}. Differently from the case
of binary neutron stars \cite{kyutoku_st2014}, fitting formulas derived
from binary-black-hole simulations in the excision framework
\cite{mroue_pfeiffer2012} do not reduce the orbital eccentricity to a
satisfactory level of $e \approx 0.001$. It would be helpful to develop
phenomenological formulas tailored to black hole--neutron star binaries
in the puncture framework after performing systematic eccentricity
reduction. Regarding the low-eccentricity initial data themselves,
global quantities like the binding energy and the orbital angular
momentum do not agree very well with post-Newtonian predictions. Our
investigation suggests that a significant fraction of the deviation may
be ascribed to junk radiation, and formulation that suppresses it will
improve the accuracy of initial data and dynamical simulations (see also
Refs.~\cite{foucart_kpt2008,lovelace_opc2008} for issues related to the
high spin).

Detections of gravitational waves from black hole--neutron star binaries
are now becoming realistic. It is probable that we will detect their
coalescences with measurable matter effects, possibly associated with
electromagnetic counterparts (see, e.g.,
Refs.~\cite{kyutoku_is2013,kyutoku_iost2015}), in the foreseeable
future. We are now systematically performing long-term simulations of
inspiraling black hole--neutron star binaries, extending our previous
work on binary neutron stars \cite{kiuchi_kksst2017,kiuchi_kkss2020}, to
develop reliable gravitational-wave templates. We plan to report results
derived by these simulations elsewhere.

\begin{acknowledgments}
 Koutarou Kyutoku is grateful to Kota Hayashi for valuable
 discussions. Although the results shown in this paper are derived
 independently, we gain knowledge from numerical computations performed
 on Cray XC50 at CfCA of National Astronomical Observatory of Japan and
 Cray XC40 at Yukawa Institute for Theoretical Physics of Kyoto
 University. This work was supported by JSPS KAKENHI Grant-in-Aid
 (Grants No.~JP15H06857, No.~JP16H06342, No.~JP17H01131, No.~JP18H01213,
 No.~JP18H04595, No.~JP18H05236, No.~JP19K14720, and No.~JP20H00158).
\end{acknowledgments}

\appendix


\begin{thebibliography}{114}%
\makeatletter
\providecommand \@ifxundefined [1]{%
 \@ifx{#1\undefined}
}%
\providecommand \@ifnum [1]{%
 \ifnum #1\expandafter \@firstoftwo
 \else \expandafter \@secondoftwo
 \fi
}%
\providecommand \@ifx [1]{%
 \ifx #1\expandafter \@firstoftwo
 \else \expandafter \@secondoftwo
 \fi
}%
\providecommand \natexlab [1]{#1}%
\providecommand \enquote  [1]{``#1''}%
\providecommand \bibnamefont  [1]{#1}%
\providecommand \bibfnamefont [1]{#1}%
\providecommand \citenamefont [1]{#1}%
\providecommand \href@noop [0]{\@secondoftwo}%
\providecommand \href [0]{\begingroup \@sanitize@url \@href}%
\providecommand \@href[1]{\@@startlink{#1}\@@href}%
\providecommand \@@href[1]{\endgroup#1\@@endlink}%
\providecommand \@sanitize@url [0]{\catcode `\\12\catcode `\$12\catcode
  `\&12\catcode `\#12\catcode `\^12\catcode `\_12\catcode `\%12\relax}%
\providecommand \@@startlink[1]{}%
\providecommand \@@endlink[0]{}%
\providecommand \url  [0]{\begingroup\@sanitize@url \@url }%
\providecommand \@url [1]{\endgroup\@href {#1}{\urlprefix }}%
\providecommand \urlprefix  [0]{URL }%
\providecommand \Eprint [0]{\href }%
\providecommand \doibase [0]{http://dx.doi.org/}%
\providecommand \selectlanguage [0]{\@gobble}%
\providecommand \bibinfo  [0]{\@secondoftwo}%
\providecommand \bibfield  [0]{\@secondoftwo}%
\providecommand \translation [1]{[#1]}%
\providecommand \BibitemOpen [0]{}%
\providecommand \bibitemStop [0]{}%
\providecommand \bibitemNoStop [0]{.\EOS\space}%
\providecommand \EOS [0]{\spacefactor3000\relax}%
\providecommand \BibitemShut  [1]{\csname bibitem#1\endcsname}%
\let\auto@bib@innerbib\@empty
\bibitem [{\citenamefont {Shibata}\ and\ \citenamefont
  {Taniguchi}(2011)}]{shibata_taniguchi2011}%
  \BibitemOpen
  \bibfield  {author} {\bibinfo {author} {\bibfnamefont {M.}~\bibnamefont
  {Shibata}}\ and\ \bibinfo {author} {\bibfnamefont {K.}~\bibnamefont
  {Taniguchi}},\ }\href {\doibase 10.12942/lrr-2011-6} {\bibfield  {journal}
  {\bibinfo  {journal} {Living Reviews in Relativity}\ }\textbf {\bibinfo
  {volume} {14}},\ \bibinfo {pages} {6} (\bibinfo {year} {2011})}\BibitemShut
  {NoStop}%
\bibitem [{\citenamefont {{Abbott}}\ \emph
  {et~al.}(2020{\natexlab{a}})\citenamefont {{Abbott}}, \citenamefont
  {{Abbott}}, \citenamefont {{Abraham}}, \citenamefont {{Acernese}},
  \citenamefont {{Ackley}}, \citenamefont {{Adams}}, \citenamefont {{Adams}},
  \citenamefont {{Adhikari}}, \citenamefont {{Adya}}, \citenamefont
  {{Affeldt}},\ and\ \citenamefont {et~al.}}]{ligovirgo-gwtc2}%
  \BibitemOpen
  \bibfield  {author} {\bibinfo {author} {\bibfnamefont {R.}~\bibnamefont
  {{Abbott}}}, \bibinfo {author} {\bibfnamefont {T.~D.}\ \bibnamefont
  {{Abbott}}}, \bibinfo {author} {\bibfnamefont {S.}~\bibnamefont {{Abraham}}},
  \bibinfo {author} {\bibfnamefont {F.}~\bibnamefont {{Acernese}}}, \bibinfo
  {author} {\bibfnamefont {K.}~\bibnamefont {{Ackley}}}, \bibinfo {author}
  {\bibfnamefont {A.}~\bibnamefont {{Adams}}}, \bibinfo {author} {\bibfnamefont
  {C.}~\bibnamefont {{Adams}}}, \bibinfo {author} {\bibfnamefont {R.~X.}\
  \bibnamefont {{Adhikari}}}, \bibinfo {author} {\bibfnamefont {V.~B.}\
  \bibnamefont {{Adya}}}, \bibinfo {author} {\bibfnamefont {C.}~\bibnamefont
  {{Affeldt}}}, \ and\ \bibinfo {author} {\bibnamefont {et~al.}},\ }\href@noop
  {} {\bibfield  {journal} {\bibinfo  {journal} {arXiv:2010.14527}\ } (\bibinfo
  {year} {2020}{\natexlab{a}})}\BibitemShut {NoStop}%
\bibitem [{\citenamefont {{Abbott}}\ \emph
  {et~al.}(2020{\natexlab{b}})\citenamefont {{Abbott}}, \citenamefont
  {{Abbott}}, \citenamefont {{Abbott}}, \citenamefont {{Abraham}},
  \citenamefont {{Acernese}}, \citenamefont {{Ackley}}, \citenamefont
  {{Adams}}, \citenamefont {{Adhikari}}, \citenamefont {{Adya}}, \citenamefont
  {{Affeldt}},\ and\ \citenamefont {et~al.}}]{ligovirgo2020-2}%
  \BibitemOpen
  \bibfield  {author} {\bibinfo {author} {\bibfnamefont {B.~P.}\ \bibnamefont
  {{Abbott}}}, \bibinfo {author} {\bibfnamefont {R.}~\bibnamefont {{Abbott}}},
  \bibinfo {author} {\bibfnamefont {T.~D.}\ \bibnamefont {{Abbott}}}, \bibinfo
  {author} {\bibfnamefont {S.}~\bibnamefont {{Abraham}}}, \bibinfo {author}
  {\bibfnamefont {F.}~\bibnamefont {{Acernese}}}, \bibinfo {author}
  {\bibfnamefont {K.}~\bibnamefont {{Ackley}}}, \bibinfo {author}
  {\bibfnamefont {C.}~\bibnamefont {{Adams}}}, \bibinfo {author} {\bibfnamefont
  {R.~X.}\ \bibnamefont {{Adhikari}}}, \bibinfo {author} {\bibfnamefont
  {V.~B.}\ \bibnamefont {{Adya}}}, \bibinfo {author} {\bibfnamefont
  {C.}~\bibnamefont {{Affeldt}}}, \ and\ \bibinfo {author} {\bibnamefont
  {et~al.}},\ }\href {\doibase 10.3847/2041-8213/ab75f5} {\bibfield  {journal}
  {\bibinfo  {journal} {\apjl}\ }\textbf {\bibinfo {volume} {892}},\ \bibinfo
  {pages} {L3} (\bibinfo {year} {2020}{\natexlab{b}})}\BibitemShut {NoStop}%
\bibitem [{\citenamefont {{Abbott}}\ \emph
  {et~al.}(2020{\natexlab{c}})\citenamefont {{Abbott}}, \citenamefont
  {{Abbott}}, \citenamefont {{Abraham}}, \citenamefont {{Acernese}},
  \citenamefont {{Ackley}}, \citenamefont {{Adams}}, \citenamefont
  {{Adhikari}}, \citenamefont {{Adya}}, \citenamefont {{Affeldt}},
  \citenamefont {{Agathos}},\ and\ \citenamefont {et~al.}}]{ligovirgo2020-3}%
  \BibitemOpen
  \bibfield  {author} {\bibinfo {author} {\bibfnamefont {R.}~\bibnamefont
  {{Abbott}}}, \bibinfo {author} {\bibfnamefont {T.~D.}\ \bibnamefont
  {{Abbott}}}, \bibinfo {author} {\bibfnamefont {S.}~\bibnamefont {{Abraham}}},
  \bibinfo {author} {\bibfnamefont {F.}~\bibnamefont {{Acernese}}}, \bibinfo
  {author} {\bibfnamefont {K.}~\bibnamefont {{Ackley}}}, \bibinfo {author}
  {\bibfnamefont {C.}~\bibnamefont {{Adams}}}, \bibinfo {author} {\bibfnamefont
  {R.~X.}\ \bibnamefont {{Adhikari}}}, \bibinfo {author} {\bibfnamefont
  {V.~B.}\ \bibnamefont {{Adya}}}, \bibinfo {author} {\bibfnamefont
  {C.}~\bibnamefont {{Affeldt}}}, \bibinfo {author} {\bibfnamefont
  {M.}~\bibnamefont {{Agathos}}}, \ and\ \bibinfo {author} {\bibnamefont
  {et~al.}},\ }\href {\doibase 10.3847/2041-8213/ab960f} {\bibfield  {journal}
  {\bibinfo  {journal} {\apjl}\ }\textbf {\bibinfo {volume} {896}},\ \bibinfo
  {pages} {L44} (\bibinfo {year} {2020}{\natexlab{c}})}\BibitemShut {NoStop}%
\bibitem [{\citenamefont {Vallisneri}(2000)}]{vallisneri2000}%
  \BibitemOpen
  \bibfield  {author} {\bibinfo {author} {\bibfnamefont {M.}~\bibnamefont
  {Vallisneri}},\ }\href {\doibase 10.1103/PhysRevLett.84.3519} {\bibfield
  {journal} {\bibinfo  {journal} {\prl}\ }\textbf {\bibinfo {volume} {84}},\
  \bibinfo {pages} {3519} (\bibinfo {year} {2000})}\BibitemShut {NoStop}%
\bibitem [{\citenamefont {Shibata}\ \emph {et~al.}(2009)\citenamefont
  {Shibata}, \citenamefont {Kyutoku}, \citenamefont {Yamamoto},\ and\
  \citenamefont {Taniguchi}}]{shibata_kyt2009}%
  \BibitemOpen
  \bibfield  {author} {\bibinfo {author} {\bibfnamefont {M.}~\bibnamefont
  {Shibata}}, \bibinfo {author} {\bibfnamefont {K.}~\bibnamefont {Kyutoku}},
  \bibinfo {author} {\bibfnamefont {T.}~\bibnamefont {Yamamoto}}, \ and\
  \bibinfo {author} {\bibfnamefont {K.}~\bibnamefont {Taniguchi}},\ }\href
  {\doibase 10.1103/PhysRevD.79.044030} {\bibfield  {journal} {\bibinfo
  {journal} {\prd}\ }\textbf {\bibinfo {volume} {79}},\ \bibinfo {pages}
  {044030} (\bibinfo {year} {2009})}\BibitemShut {NoStop}%
\bibitem [{\citenamefont {Kyutoku}\ \emph {et~al.}(2010)\citenamefont
  {Kyutoku}, \citenamefont {Shibata},\ and\ \citenamefont
  {Taniguchi}}]{kyutoku_st2010}%
  \BibitemOpen
  \bibfield  {author} {\bibinfo {author} {\bibfnamefont {K.}~\bibnamefont
  {Kyutoku}}, \bibinfo {author} {\bibfnamefont {M.}~\bibnamefont {Shibata}}, \
  and\ \bibinfo {author} {\bibfnamefont {K.}~\bibnamefont {Taniguchi}},\ }\href
  {\doibase 10.1103/PhysRevD.82.044049} {\bibfield  {journal} {\bibinfo
  {journal} {\prd}\ }\textbf {\bibinfo {volume} {82}},\ \bibinfo {pages}
  {044049} (\bibinfo {year} {2010})}\BibitemShut {NoStop}%
\bibitem [{\citenamefont {Kyutoku}\ \emph {et~al.}(2011)\citenamefont
  {Kyutoku}, \citenamefont {Okawa}, \citenamefont {Shibata},\ and\
  \citenamefont {Taniguchi}}]{kyutoku_ost2011}%
  \BibitemOpen
  \bibfield  {author} {\bibinfo {author} {\bibfnamefont {K.}~\bibnamefont
  {Kyutoku}}, \bibinfo {author} {\bibfnamefont {H.}~\bibnamefont {Okawa}},
  \bibinfo {author} {\bibfnamefont {M.}~\bibnamefont {Shibata}}, \ and\
  \bibinfo {author} {\bibfnamefont {K.}~\bibnamefont {Taniguchi}},\ }\href
  {\doibase 10.1103/PhysRevD.84.064018} {\bibfield  {journal} {\bibinfo
  {journal} {\prd}\ }\textbf {\bibinfo {volume} {84}},\ \bibinfo {pages}
  {064018} (\bibinfo {year} {2011})}\BibitemShut {NoStop}%
\bibitem [{\citenamefont {Pannarale}\ \emph {et~al.}(2015)\citenamefont
  {Pannarale}, \citenamefont {Berti}, \citenamefont {Kyutoku}, \citenamefont
  {Lackey},\ and\ \citenamefont {Shibata}}]{pannarale_bkls2015}%
  \BibitemOpen
  \bibfield  {author} {\bibinfo {author} {\bibfnamefont {F.}~\bibnamefont
  {Pannarale}}, \bibinfo {author} {\bibfnamefont {E.}~\bibnamefont {Berti}},
  \bibinfo {author} {\bibfnamefont {K.}~\bibnamefont {Kyutoku}}, \bibinfo
  {author} {\bibfnamefont {B.~D.}\ \bibnamefont {Lackey}}, \ and\ \bibinfo
  {author} {\bibfnamefont {M.}~\bibnamefont {Shibata}},\ }\href {\doibase
  10.1103/PhysRevD.92.081504} {\bibfield  {journal} {\bibinfo  {journal}
  {\prd}\ }\textbf {\bibinfo {volume} {92}},\ \bibinfo {pages} {081504}
  (\bibinfo {year} {2015})}\BibitemShut {NoStop}%
\bibitem [{\citenamefont {Lackey}\ \emph {et~al.}(2012)\citenamefont {Lackey},
  \citenamefont {Kyutoku}, \citenamefont {Shibata}, \citenamefont {Brady},\
  and\ \citenamefont {Friedman}}]{lackey_ksbf2012}%
  \BibitemOpen
  \bibfield  {author} {\bibinfo {author} {\bibfnamefont {B.~D.}\ \bibnamefont
  {Lackey}}, \bibinfo {author} {\bibfnamefont {K.}~\bibnamefont {Kyutoku}},
  \bibinfo {author} {\bibfnamefont {M.}~\bibnamefont {Shibata}}, \bibinfo
  {author} {\bibfnamefont {P.~R.}\ \bibnamefont {Brady}}, \ and\ \bibinfo
  {author} {\bibfnamefont {J.~L.}\ \bibnamefont {Friedman}},\ }\href {\doibase
  10.1103/PhysRevD.85.044061} {\bibfield  {journal} {\bibinfo  {journal}
  {\prd}\ }\textbf {\bibinfo {volume} {85}},\ \bibinfo {pages} {044061}
  (\bibinfo {year} {2012})}\BibitemShut {NoStop}%
\bibitem [{\citenamefont {Lackey}\ \emph {et~al.}(2014)\citenamefont {Lackey},
  \citenamefont {Kyutoku}, \citenamefont {Shibata}, \citenamefont {Brady},\
  and\ \citenamefont {Friedman}}]{lackey_ksbf2014}%
  \BibitemOpen
  \bibfield  {author} {\bibinfo {author} {\bibfnamefont {B.~D.}\ \bibnamefont
  {Lackey}}, \bibinfo {author} {\bibfnamefont {K.}~\bibnamefont {Kyutoku}},
  \bibinfo {author} {\bibfnamefont {M.}~\bibnamefont {Shibata}}, \bibinfo
  {author} {\bibfnamefont {P.~R.}\ \bibnamefont {Brady}}, \ and\ \bibinfo
  {author} {\bibfnamefont {J.~L.}\ \bibnamefont {Friedman}},\ }\href {\doibase
  10.1103/PhysRevD.89.043009} {\bibfield  {journal} {\bibinfo  {journal}
  {\prd}\ }\textbf {\bibinfo {volume} {89}},\ \bibinfo {pages} {043009}
  (\bibinfo {year} {2014})}\BibitemShut {NoStop}%
\bibitem [{\citenamefont {Lattimer}\ and\ \citenamefont
  {Prakash}(2016)}]{lattimer_prakash2016}%
  \BibitemOpen
  \bibfield  {author} {\bibinfo {author} {\bibfnamefont {J.~M.}\ \bibnamefont
  {Lattimer}}\ and\ \bibinfo {author} {\bibfnamefont {M.}~\bibnamefont
  {Prakash}},\ }\href {\doibase 10.1016/j.physrep.2015.12.005} {\bibfield
  {journal} {\bibinfo  {journal} {\physrep}\ }\textbf {\bibinfo {volume}
  {621}},\ \bibinfo {pages} {127} (\bibinfo {year} {2016})}\BibitemShut
  {NoStop}%
\bibitem [{\citenamefont {Baym}\ \emph {et~al.}(2018)\citenamefont {Baym},
  \citenamefont {Hatsuda}, \citenamefont {Kojo}, \citenamefont {Powell},
  \citenamefont {Song},\ and\ \citenamefont {Takatsuka}}]{baym_hkpst2018}%
  \BibitemOpen
  \bibfield  {author} {\bibinfo {author} {\bibfnamefont {G.}~\bibnamefont
  {Baym}}, \bibinfo {author} {\bibfnamefont {T.}~\bibnamefont {Hatsuda}},
  \bibinfo {author} {\bibfnamefont {T.}~\bibnamefont {Kojo}}, \bibinfo {author}
  {\bibfnamefont {P.~D.}\ \bibnamefont {Powell}}, \bibinfo {author}
  {\bibfnamefont {Y.}~\bibnamefont {Song}}, \ and\ \bibinfo {author}
  {\bibfnamefont {T.}~\bibnamefont {Takatsuka}},\ }\href {\doibase
  10.1088/1361-6633/aaae14} {\bibfield  {journal} {\bibinfo  {journal} {Reports
  on Progress in Physics}\ }\textbf {\bibinfo {volume} {81}},\ \bibinfo {pages}
  {056902} (\bibinfo {year} {2018})}\BibitemShut {NoStop}%
\bibitem [{\citenamefont {{Abbott}}\ \emph
  {et~al.}(2018{\natexlab{a}})\citenamefont {{Abbott}}, \citenamefont
  {{Abbott}}, \citenamefont {{Abbott}}, \citenamefont {{Acernese}},
  \citenamefont {{Ackley}}, \citenamefont {{Adams}}, \citenamefont {{Adams}},
  \citenamefont {{Addesso}}, \citenamefont {{Adhikari}}, \citenamefont
  {{Adya}},\ and\ \citenamefont {et~al.}}]{ligovirgo2018}%
  \BibitemOpen
  \bibfield  {author} {\bibinfo {author} {\bibfnamefont {B.~P.}\ \bibnamefont
  {{Abbott}}}, \bibinfo {author} {\bibfnamefont {R.}~\bibnamefont {{Abbott}}},
  \bibinfo {author} {\bibfnamefont {T.~D.}\ \bibnamefont {{Abbott}}}, \bibinfo
  {author} {\bibfnamefont {F.}~\bibnamefont {{Acernese}}}, \bibinfo {author}
  {\bibfnamefont {K.}~\bibnamefont {{Ackley}}}, \bibinfo {author}
  {\bibfnamefont {C.}~\bibnamefont {{Adams}}}, \bibinfo {author} {\bibfnamefont
  {T.}~\bibnamefont {{Adams}}}, \bibinfo {author} {\bibfnamefont
  {P.}~\bibnamefont {{Addesso}}}, \bibinfo {author} {\bibfnamefont {R.~X.}\
  \bibnamefont {{Adhikari}}}, \bibinfo {author} {\bibfnamefont {V.~B.}\
  \bibnamefont {{Adya}}}, \ and\ \bibinfo {author} {\bibnamefont {et~al.}},\
  }\href {\doibase 10.1103/PhysRevLett.121.161101} {\bibfield  {journal}
  {\bibinfo  {journal} {\prl}\ }\textbf {\bibinfo {volume} {121}},\ \bibinfo
  {pages} {161101} (\bibinfo {year} {2018}{\natexlab{a}})}\BibitemShut
  {NoStop}%
\bibitem [{\citenamefont {{Abbott}}\ \emph
  {et~al.}(2019{\natexlab{a}})\citenamefont {{Abbott}}, \citenamefont
  {{Abbott}}, \citenamefont {{Abbott}}, \citenamefont {{Acernese}},
  \citenamefont {{Ackley}}, \citenamefont {{Adams}}, \citenamefont {{Adams}},
  \citenamefont {{Addesso}}, \citenamefont {{Adhikari}}, \citenamefont
  {{Adya}},\ and\ \citenamefont {et~al.}}]{ligovirgo2019}%
  \BibitemOpen
  \bibfield  {author} {\bibinfo {author} {\bibfnamefont {B.~P.}\ \bibnamefont
  {{Abbott}}}, \bibinfo {author} {\bibfnamefont {R.}~\bibnamefont {{Abbott}}},
  \bibinfo {author} {\bibfnamefont {T.~D.}\ \bibnamefont {{Abbott}}}, \bibinfo
  {author} {\bibfnamefont {F.}~\bibnamefont {{Acernese}}}, \bibinfo {author}
  {\bibfnamefont {K.}~\bibnamefont {{Ackley}}}, \bibinfo {author}
  {\bibfnamefont {C.}~\bibnamefont {{Adams}}}, \bibinfo {author} {\bibfnamefont
  {T.}~\bibnamefont {{Adams}}}, \bibinfo {author} {\bibfnamefont
  {P.}~\bibnamefont {{Addesso}}}, \bibinfo {author} {\bibfnamefont {R.~X.}\
  \bibnamefont {{Adhikari}}}, \bibinfo {author} {\bibfnamefont {V.~B.}\
  \bibnamefont {{Adya}}}, \ and\ \bibinfo {author} {\bibnamefont {et~al.}},\
  }\href {\doibase 10.1103/PhysRevX.9.011001} {\bibfield  {journal} {\bibinfo
  {journal} {\prx}\ }\textbf {\bibinfo {volume} {9}},\ \bibinfo {pages}
  {011001} (\bibinfo {year} {2019}{\natexlab{a}})}\BibitemShut {NoStop}%
\bibitem [{\citenamefont {{Abbott}}\ \emph {et~al.}(2016)\citenamefont
  {{Abbott}}, \citenamefont {{Abbott}}, \citenamefont {{Abbott}}, \citenamefont
  {{Abernathy}}, \citenamefont {{Acernese}}, \citenamefont {{Ackley}},
  \citenamefont {{Adams}}, \citenamefont {{Adams}}, \citenamefont {{Addesso}},
  \citenamefont {{Adhikari}},\ and\ \citenamefont {et~al.}}]{ligovirgo2016-5}%
  \BibitemOpen
  \bibfield  {author} {\bibinfo {author} {\bibfnamefont {B.~P.}\ \bibnamefont
  {{Abbott}}}, \bibinfo {author} {\bibfnamefont {R.}~\bibnamefont {{Abbott}}},
  \bibinfo {author} {\bibfnamefont {T.~D.}\ \bibnamefont {{Abbott}}}, \bibinfo
  {author} {\bibfnamefont {M.~R.}\ \bibnamefont {{Abernathy}}}, \bibinfo
  {author} {\bibfnamefont {F.}~\bibnamefont {{Acernese}}}, \bibinfo {author}
  {\bibfnamefont {K.}~\bibnamefont {{Ackley}}}, \bibinfo {author}
  {\bibfnamefont {C.}~\bibnamefont {{Adams}}}, \bibinfo {author} {\bibfnamefont
  {T.}~\bibnamefont {{Adams}}}, \bibinfo {author} {\bibfnamefont
  {P.}~\bibnamefont {{Addesso}}}, \bibinfo {author} {\bibfnamefont {R.~X.}\
  \bibnamefont {{Adhikari}}}, \ and\ \bibinfo {author} {\bibnamefont
  {et~al.}},\ }\href {\doibase 10.1103/PhysRevX.6.041015} {\bibfield  {journal}
  {\bibinfo  {journal} {Physical Review X}\ }\textbf {\bibinfo {volume} {6}},\
  \bibinfo {pages} {041015} (\bibinfo {year} {2016})}\BibitemShut {NoStop}%
\bibitem [{\citenamefont {{Abbott}}\ \emph
  {et~al.}(2019{\natexlab{b}})\citenamefont {{Abbott}}, \citenamefont
  {{Abbott}}, \citenamefont {{Abbott}}, \citenamefont {{Acernese}},
  \citenamefont {{Ackley}}, \citenamefont {{Adams}}, \citenamefont {{Adams}},
  \citenamefont {{Addesso}}, \citenamefont {{Adhikari}}, \citenamefont
  {{Adya}},\ and\ \citenamefont {et~al.}}]{ligovirgo2019-3}%
  \BibitemOpen
  \bibfield  {author} {\bibinfo {author} {\bibfnamefont {B.~P.}\ \bibnamefont
  {{Abbott}}}, \bibinfo {author} {\bibfnamefont {R.}~\bibnamefont {{Abbott}}},
  \bibinfo {author} {\bibfnamefont {T.~D.}\ \bibnamefont {{Abbott}}}, \bibinfo
  {author} {\bibfnamefont {F.}~\bibnamefont {{Acernese}}}, \bibinfo {author}
  {\bibfnamefont {K.}~\bibnamefont {{Ackley}}}, \bibinfo {author}
  {\bibfnamefont {C.}~\bibnamefont {{Adams}}}, \bibinfo {author} {\bibfnamefont
  {T.}~\bibnamefont {{Adams}}}, \bibinfo {author} {\bibfnamefont
  {P.}~\bibnamefont {{Addesso}}}, \bibinfo {author} {\bibfnamefont {R.~X.}\
  \bibnamefont {{Adhikari}}}, \bibinfo {author} {\bibfnamefont {V.~B.}\
  \bibnamefont {{Adya}}}, \ and\ \bibinfo {author} {\bibnamefont {et~al.}},\
  }\href {\doibase 10.1103/PhysRevX.9.031040} {\bibfield  {journal} {\bibinfo
  {journal} {\prx}\ }\textbf {\bibinfo {volume} {9}},\ \bibinfo {pages}
  {031040} (\bibinfo {year} {2019}{\natexlab{b}})}\BibitemShut {NoStop}%
\bibitem [{\citenamefont {Yagi}\ and\ \citenamefont
  {Yunes}(2014)}]{yagi_yunes2014}%
  \BibitemOpen
  \bibfield  {author} {\bibinfo {author} {\bibfnamefont {K.}~\bibnamefont
  {Yagi}}\ and\ \bibinfo {author} {\bibfnamefont {N.}~\bibnamefont {Yunes}},\
  }\href {\doibase 10.1103/PhysRevD.89.021303} {\bibfield  {journal} {\bibinfo
  {journal} {\prd}\ }\textbf {\bibinfo {volume} {89}},\ \bibinfo {pages}
  {021303} (\bibinfo {year} {2014})}\BibitemShut {NoStop}%
\bibitem [{\citenamefont {Favata}(2014)}]{favata2014}%
  \BibitemOpen
  \bibfield  {author} {\bibinfo {author} {\bibfnamefont {M.}~\bibnamefont
  {Favata}},\ }\href {\doibase 10.1103/PhysRevLett.112.101101} {\bibfield
  {journal} {\bibinfo  {journal} {\prl}\ }\textbf {\bibinfo {volume} {112}},\
  \bibinfo {pages} {101101} (\bibinfo {year} {2014})}\BibitemShut {NoStop}%
\bibitem [{\citenamefont {Wade}\ \emph {et~al.}(2014)\citenamefont {Wade},
  \citenamefont {Creighton}, \citenamefont {Ochsner}, \citenamefont {Lackey},
  \citenamefont {Farr}, \citenamefont {Littenberg},\ and\ \citenamefont
  {Raymond}}]{wade_colflr2014}%
  \BibitemOpen
  \bibfield  {author} {\bibinfo {author} {\bibfnamefont {L.}~\bibnamefont
  {Wade}}, \bibinfo {author} {\bibfnamefont {J.~D.~E.}\ \bibnamefont
  {Creighton}}, \bibinfo {author} {\bibfnamefont {E.}~\bibnamefont {Ochsner}},
  \bibinfo {author} {\bibfnamefont {B.~D.}\ \bibnamefont {Lackey}}, \bibinfo
  {author} {\bibfnamefont {B.~F.}\ \bibnamefont {Farr}}, \bibinfo {author}
  {\bibfnamefont {T.~B.}\ \bibnamefont {Littenberg}}, \ and\ \bibinfo {author}
  {\bibfnamefont {V.}~\bibnamefont {Raymond}},\ }\href {\doibase
  10.1103/PhysRevD.89.103012} {\bibfield  {journal} {\bibinfo  {journal}
  {\prd}\ }\textbf {\bibinfo {volume} {89}},\ \bibinfo {pages} {103012}
  (\bibinfo {year} {2014})}\BibitemShut {NoStop}%
\bibitem [{\citenamefont {{Abbott}}\ \emph
  {et~al.}(2020{\natexlab{d}})\citenamefont {{Abbott}}, \citenamefont
  {{Abbott}}, \citenamefont {{Abbott}}, \citenamefont {{Abraham}},
  \citenamefont {{Acernese}}, \citenamefont {{Ackley}}, \citenamefont
  {{Adams}}, \citenamefont {{Adya}}, \citenamefont {{Affeldt}}, \citenamefont
  {M.},\ and\ \citenamefont {et~al.}}]{ligovirgo2020}%
  \BibitemOpen
  \bibfield  {author} {\bibinfo {author} {\bibfnamefont {B.~P.}\ \bibnamefont
  {{Abbott}}}, \bibinfo {author} {\bibfnamefont {R.}~\bibnamefont {{Abbott}}},
  \bibinfo {author} {\bibfnamefont {T.~D.}\ \bibnamefont {{Abbott}}}, \bibinfo
  {author} {\bibfnamefont {S.}~\bibnamefont {{Abraham}}}, \bibinfo {author}
  {\bibfnamefont {F.}~\bibnamefont {{Acernese}}}, \bibinfo {author}
  {\bibfnamefont {K.}~\bibnamefont {{Ackley}}}, \bibinfo {author}
  {\bibfnamefont {C.}~\bibnamefont {{Adams}}}, \bibinfo {author} {\bibfnamefont
  {V.~B.}\ \bibnamefont {{Adya}}}, \bibinfo {author} {\bibfnamefont
  {C.}~\bibnamefont {{Affeldt}}}, \bibinfo {author} {\bibfnamefont
  {A.}~\bibnamefont {M.}}, \ and\ \bibinfo {author} {\bibnamefont {et~al.}},\
  }\href {\doibase 10.1088/1361-6382/ab5f7c} {\bibfield  {journal} {\bibinfo
  {journal} {Classical and Quantum Gravity}\ }\textbf {\bibinfo {volume}
  {37}},\ \bibinfo {pages} {045006} (\bibinfo {year}
  {2020}{\natexlab{d}})}\BibitemShut {NoStop}%
\bibitem [{\citenamefont {Narikawa}\ \emph {et~al.}(2020)\citenamefont
  {Narikawa}, \citenamefont {Uchikata}, \citenamefont {Kawaguchi},
  \citenamefont {Kiuchi}, \citenamefont {Kyutoku}, \citenamefont {Shibata},\
  and\ \citenamefont {Tagoshi}}]{narikawa_ukkkst2020}%
  \BibitemOpen
  \bibfield  {author} {\bibinfo {author} {\bibfnamefont {T.}~\bibnamefont
  {Narikawa}}, \bibinfo {author} {\bibfnamefont {N.}~\bibnamefont {Uchikata}},
  \bibinfo {author} {\bibfnamefont {K.}~\bibnamefont {Kawaguchi}}, \bibinfo
  {author} {\bibfnamefont {K.}~\bibnamefont {Kiuchi}}, \bibinfo {author}
  {\bibfnamefont {K.}~\bibnamefont {Kyutoku}}, \bibinfo {author} {\bibfnamefont
  {M.}~\bibnamefont {Shibata}}, \ and\ \bibinfo {author} {\bibfnamefont
  {H.}~\bibnamefont {Tagoshi}},\ }\href {\doibase
  10.1103/PhysRevResearch.2.043039} {\bibfield  {journal} {\bibinfo  {journal}
  {Physical Review Research}\ }\textbf {\bibinfo {volume} {2}},\ \bibinfo
  {pages} {043039} (\bibinfo {year} {2020})}\BibitemShut {NoStop}%
\bibitem [{\citenamefont {{Abbott}}\ \emph
  {et~al.}(2018{\natexlab{b}})\citenamefont {{Abbott}}, \citenamefont
  {{Abbott}}, \citenamefont {{Abbott}}, \citenamefont {{Abernathy}},
  \citenamefont {{Acernese}}, \citenamefont {{Ackley}}, \citenamefont
  {{Adams}}, \citenamefont {{Adams}}, \citenamefont {{Addesso}}, \citenamefont
  {{Adhikari}},\ and\ \citenamefont {et~al.}}]{lvk2018}%
  \BibitemOpen
  \bibfield  {author} {\bibinfo {author} {\bibfnamefont {B.~P.}\ \bibnamefont
  {{Abbott}}}, \bibinfo {author} {\bibfnamefont {R.}~\bibnamefont {{Abbott}}},
  \bibinfo {author} {\bibfnamefont {T.~D.}\ \bibnamefont {{Abbott}}}, \bibinfo
  {author} {\bibfnamefont {M.~R.}\ \bibnamefont {{Abernathy}}}, \bibinfo
  {author} {\bibfnamefont {F.}~\bibnamefont {{Acernese}}}, \bibinfo {author}
  {\bibfnamefont {K.}~\bibnamefont {{Ackley}}}, \bibinfo {author}
  {\bibfnamefont {C.}~\bibnamefont {{Adams}}}, \bibinfo {author} {\bibfnamefont
  {T.}~\bibnamefont {{Adams}}}, \bibinfo {author} {\bibfnamefont
  {P.}~\bibnamefont {{Addesso}}}, \bibinfo {author} {\bibfnamefont {R.~X.}\
  \bibnamefont {{Adhikari}}}, \ and\ \bibinfo {author} {\bibnamefont
  {et~al.}},\ }\href {\doibase 10.1007/s4114-018-0012-9} {\bibfield  {journal}
  {\bibinfo  {journal} {Living Reviews in Relativity}\ }\textbf {\bibinfo
  {volume} {21}},\ \bibinfo {pages} {3} (\bibinfo {year}
  {2018}{\natexlab{b}})}\BibitemShut {NoStop}%
\bibitem [{\citenamefont {Kawaguchi}\ \emph {et~al.}(2018)\citenamefont
  {Kawaguchi}, \citenamefont {Kiuchi}, \citenamefont {Kyutoku}, \citenamefont
  {Sekiguchi}, \citenamefont {Shibata},\ and\ \citenamefont
  {Taniguchi}}]{kawaguchi_kksst2018}%
  \BibitemOpen
  \bibfield  {author} {\bibinfo {author} {\bibfnamefont {K.}~\bibnamefont
  {Kawaguchi}}, \bibinfo {author} {\bibfnamefont {K.}~\bibnamefont {Kiuchi}},
  \bibinfo {author} {\bibfnamefont {K.}~\bibnamefont {Kyutoku}}, \bibinfo
  {author} {\bibfnamefont {Y.}~\bibnamefont {Sekiguchi}}, \bibinfo {author}
  {\bibfnamefont {M.}~\bibnamefont {Shibata}}, \ and\ \bibinfo {author}
  {\bibfnamefont {K.}~\bibnamefont {Taniguchi}},\ }\href {\doibase
  10.1103/PhysRevD.97.044044} {\bibfield  {journal} {\bibinfo  {journal}
  {\prd}\ }\textbf {\bibinfo {volume} {97}},\ \bibinfo {pages} {044044}
  (\bibinfo {year} {2018})}\BibitemShut {NoStop}%
\bibitem [{\citenamefont {Dudi}\ \emph {et~al.}(2018)\citenamefont {Dudi},
  \citenamefont {Pannarale}, \citenamefont {Dietrich}, \citenamefont {Hannam},
  \citenamefont {Bernuzzi}, \citenamefont {Ohme},\ and\ \citenamefont
  {Br{\"u}gmann}}]{dudi_pdhbob2018}%
  \BibitemOpen
  \bibfield  {author} {\bibinfo {author} {\bibfnamefont {R.}~\bibnamefont
  {Dudi}}, \bibinfo {author} {\bibfnamefont {F.}~\bibnamefont {Pannarale}},
  \bibinfo {author} {\bibfnamefont {T.}~\bibnamefont {Dietrich}}, \bibinfo
  {author} {\bibfnamefont {M.}~\bibnamefont {Hannam}}, \bibinfo {author}
  {\bibfnamefont {S.}~\bibnamefont {Bernuzzi}}, \bibinfo {author}
  {\bibfnamefont {F.}~\bibnamefont {Ohme}}, \ and\ \bibinfo {author}
  {\bibfnamefont {B.}~\bibnamefont {Br{\"u}gmann}},\ }\href {\doibase
  10.1103/PhysRevD.98.084061} {\bibfield  {journal} {\bibinfo  {journal}
  {\prd}\ }\textbf {\bibinfo {volume} {98}},\ \bibinfo {pages} {084061}
  (\bibinfo {year} {2018})}\BibitemShut {NoStop}%
\bibitem [{\citenamefont {{Chakravarti}}\ \emph {et~al.}(2019)\citenamefont
  {{Chakravarti}}, \citenamefont {{Gupta}}, \citenamefont {{Bose}},
  \citenamefont {{Duez}}, \citenamefont {{Caro}}, \citenamefont {{Brege}},
  \citenamefont {{Foucart}}, \citenamefont {{Ghosh}}, \citenamefont
  {{Kyutoku}}, \citenamefont {{Lackey}},\ and\ \citenamefont
  {et~al.}}]{chakravarti_etal2019}%
  \BibitemOpen
  \bibfield  {author} {\bibinfo {author} {\bibfnamefont {K.}~\bibnamefont
  {{Chakravarti}}}, \bibinfo {author} {\bibfnamefont {A.}~\bibnamefont
  {{Gupta}}}, \bibinfo {author} {\bibfnamefont {S.}~\bibnamefont {{Bose}}},
  \bibinfo {author} {\bibfnamefont {M.~D.}\ \bibnamefont {{Duez}}}, \bibinfo
  {author} {\bibfnamefont {J.}~\bibnamefont {{Caro}}}, \bibinfo {author}
  {\bibfnamefont {W.}~\bibnamefont {{Brege}}}, \bibinfo {author} {\bibfnamefont
  {F.}~\bibnamefont {{Foucart}}}, \bibinfo {author} {\bibfnamefont
  {S.}~\bibnamefont {{Ghosh}}}, \bibinfo {author} {\bibfnamefont
  {K.}~\bibnamefont {{Kyutoku}}}, \bibinfo {author} {\bibfnamefont {B.~D.}\
  \bibnamefont {{Lackey}}}, \ and\ \bibinfo {author} {\bibnamefont {et~al.}},\
  }\href {\doibase 10.1103/PhysRevD.99.024049} {\bibfield  {journal} {\bibinfo
  {journal} {\prd}\ }\textbf {\bibinfo {volume} {99}},\ \bibinfo {pages}
  {024049} (\bibinfo {year} {2019})}\BibitemShut {NoStop}%
\bibitem [{\citenamefont {Huang}\ \emph {et~al.}(2020)\citenamefont {Huang},
  \citenamefont {Haster}, \citenamefont {Vitale}, \citenamefont {Varma},
  \citenamefont {Foucart},\ and\ \citenamefont {Biscoveanu}}]{huang_hvvfb2020}%
  \BibitemOpen
  \bibfield  {author} {\bibinfo {author} {\bibfnamefont {Y.}~\bibnamefont
  {Huang}}, \bibinfo {author} {\bibfnamefont {C.-J.}\ \bibnamefont {Haster}},
  \bibinfo {author} {\bibfnamefont {S.}~\bibnamefont {Vitale}}, \bibinfo
  {author} {\bibfnamefont {V.}~\bibnamefont {Varma}}, \bibinfo {author}
  {\bibfnamefont {F.}~\bibnamefont {Foucart}}, \ and\ \bibinfo {author}
  {\bibfnamefont {S.}~\bibnamefont {Biscoveanu}},\ }\href@noop {} {\  (\bibinfo
  {year} {2020})},\ \Eprint {http://arxiv.org/abs/arXiv:2005.11850}
  {arXiv:2005.11850} \BibitemShut {NoStop}%
\bibitem [{\citenamefont {Thompson}\ \emph {et~al.}(2020)\citenamefont
  {Thompson}, \citenamefont {Fauchon-Jones}, \citenamefont {Khan},
  \citenamefont {Nitoglia}, \citenamefont {Pannarale}, \citenamefont
  {Dietrich},\ and\ \citenamefont {Hannam}}]{thompson_fknpdh2020}%
  \BibitemOpen
  \bibfield  {author} {\bibinfo {author} {\bibfnamefont {J.~E.}\ \bibnamefont
  {Thompson}}, \bibinfo {author} {\bibfnamefont {E.}~\bibnamefont
  {Fauchon-Jones}}, \bibinfo {author} {\bibfnamefont {S.}~\bibnamefont {Khan}},
  \bibinfo {author} {\bibfnamefont {E.}~\bibnamefont {Nitoglia}}, \bibinfo
  {author} {\bibfnamefont {F.}~\bibnamefont {Pannarale}}, \bibinfo {author}
  {\bibfnamefont {T.}~\bibnamefont {Dietrich}}, \ and\ \bibinfo {author}
  {\bibfnamefont {M.}~\bibnamefont {Hannam}},\ }\href {\doibase
  10.1103/PhysRevD.101.124059} {\bibfield  {journal} {\bibinfo  {journal}
  {\prd}\ }\textbf {\bibinfo {volume} {101}},\ \bibinfo {pages} {124059}
  (\bibinfo {year} {2020})}\BibitemShut {NoStop}%
\bibitem [{\citenamefont {Matas}\ \emph {et~al.}(2020)\citenamefont {Matas},
  \citenamefont {Dietrich}, \citenamefont {Buonanno}, \citenamefont {Hinderer},
  \citenamefont {P{\"u}rrer}, \citenamefont {Foucart}, \citenamefont {Boyle},
  \citenamefont {Duez}, \citenamefont {Kidder}, \citenamefont {Pfeiffer},\ and\
  \citenamefont {Scheel}}]{matas_etal2020}%
  \BibitemOpen
  \bibfield  {author} {\bibinfo {author} {\bibfnamefont {A.}~\bibnamefont
  {Matas}}, \bibinfo {author} {\bibfnamefont {T.}~\bibnamefont {Dietrich}},
  \bibinfo {author} {\bibfnamefont {A.}~\bibnamefont {Buonanno}}, \bibinfo
  {author} {\bibfnamefont {T.}~\bibnamefont {Hinderer}}, \bibinfo {author}
  {\bibfnamefont {M.}~\bibnamefont {P{\"u}rrer}}, \bibinfo {author}
  {\bibfnamefont {F.}~\bibnamefont {Foucart}}, \bibinfo {author} {\bibfnamefont
  {M.}~\bibnamefont {Boyle}}, \bibinfo {author} {\bibfnamefont {M.~D.}\
  \bibnamefont {Duez}}, \bibinfo {author} {\bibfnamefont {L.~E.}\ \bibnamefont
  {Kidder}}, \bibinfo {author} {\bibfnamefont {H.~P.}\ \bibnamefont
  {Pfeiffer}}, \ and\ \bibinfo {author} {\bibfnamefont {M.~A.}\ \bibnamefont
  {Scheel}},\ }\href {\doibase 10.1103/PhysRevD.102.043023} {\bibfield
  {journal} {\bibinfo  {journal} {\prd}\ }\textbf {\bibinfo {volume} {102}},\
  \bibinfo {pages} {043023} (\bibinfo {year} {2020})}\BibitemShut {NoStop}%
\bibitem [{\citenamefont {Chawla}\ \emph {et~al.}(2010)\citenamefont {Chawla},
  \citenamefont {Anderson}, \citenamefont {Besselman}, \citenamefont {Lehner},
  \citenamefont {Liebling}, \citenamefont {Motl},\ and\ \citenamefont
  {Neilsen}}]{chawla_abllmn2010}%
  \BibitemOpen
  \bibfield  {author} {\bibinfo {author} {\bibfnamefont {S.}~\bibnamefont
  {Chawla}}, \bibinfo {author} {\bibfnamefont {M.}~\bibnamefont {Anderson}},
  \bibinfo {author} {\bibfnamefont {M.}~\bibnamefont {Besselman}}, \bibinfo
  {author} {\bibfnamefont {L.}~\bibnamefont {Lehner}}, \bibinfo {author}
  {\bibfnamefont {S.~L.}\ \bibnamefont {Liebling}}, \bibinfo {author}
  {\bibfnamefont {P.~M.}\ \bibnamefont {Motl}}, \ and\ \bibinfo {author}
  {\bibfnamefont {D.}~\bibnamefont {Neilsen}},\ }\href {\doibase
  10.1103/PhysRevLett.105.111101} {\bibfield  {journal} {\bibinfo  {journal}
  {\prl}\ }\textbf {\bibinfo {volume} {105}},\ \bibinfo {pages} {111101}
  (\bibinfo {year} {2010})}\BibitemShut {NoStop}%
\bibitem [{\citenamefont {Foucart}\ \emph
  {et~al.}(2013{\natexlab{a}})\citenamefont {Foucart}, \citenamefont {Deaton},
  \citenamefont {Duez}, \citenamefont {Kidder}, \citenamefont {MacDonald},
  \citenamefont {Ott}, \citenamefont {Pfeiffer}, \citenamefont {Scheel},
  \citenamefont {Szilagyi},\ and\ \citenamefont
  {Teukolsky}}]{foucart_ddkmopsst2013}%
  \BibitemOpen
  \bibfield  {author} {\bibinfo {author} {\bibfnamefont {F.}~\bibnamefont
  {Foucart}}, \bibinfo {author} {\bibfnamefont {M.~B.}\ \bibnamefont {Deaton}},
  \bibinfo {author} {\bibfnamefont {M.~D.}\ \bibnamefont {Duez}}, \bibinfo
  {author} {\bibfnamefont {L.~E.}\ \bibnamefont {Kidder}}, \bibinfo {author}
  {\bibfnamefont {I.}~\bibnamefont {MacDonald}}, \bibinfo {author}
  {\bibfnamefont {C.~D.}\ \bibnamefont {Ott}}, \bibinfo {author} {\bibfnamefont
  {H.~P.}\ \bibnamefont {Pfeiffer}}, \bibinfo {author} {\bibfnamefont {M.~A.}\
  \bibnamefont {Scheel}}, \bibinfo {author} {\bibfnamefont {B.}~\bibnamefont
  {Szilagyi}}, \ and\ \bibinfo {author} {\bibfnamefont {S.~A.}\ \bibnamefont
  {Teukolsky}},\ }\href {\doibase 10.1103/PhysRevD.87.084006} {\bibfield
  {journal} {\bibinfo  {journal} {\prd}\ }\textbf {\bibinfo {volume} {87}},\
  \bibinfo {pages} {084006} (\bibinfo {year} {2013}{\natexlab{a}})}\BibitemShut
  {NoStop}%
\bibitem [{\citenamefont {Kyutoku}\ \emph {et~al.}(2013)\citenamefont
  {Kyutoku}, \citenamefont {Ioka},\ and\ \citenamefont
  {Shibata}}]{kyutoku_is2013}%
  \BibitemOpen
  \bibfield  {author} {\bibinfo {author} {\bibfnamefont {K.}~\bibnamefont
  {Kyutoku}}, \bibinfo {author} {\bibfnamefont {K.}~\bibnamefont {Ioka}}, \
  and\ \bibinfo {author} {\bibfnamefont {M.}~\bibnamefont {Shibata}},\ }\href
  {\doibase 10.1103/PhysRevD.88.041503} {\bibfield  {journal} {\bibinfo
  {journal} {\prd}\ }\textbf {\bibinfo {volume} {88}},\ \bibinfo {pages}
  {041503} (\bibinfo {year} {2013})}\BibitemShut {NoStop}%
\bibitem [{\citenamefont {Foucart}\ \emph {et~al.}(2014)\citenamefont
  {Foucart}, \citenamefont {Deaton}, \citenamefont {Duez}, \citenamefont
  {O'Connor}, \citenamefont {Ott}, \citenamefont {Haas}, \citenamefont
  {Kidder}, \citenamefont {Pfeiffer}, \citenamefont {Scheel},\ and\
  \citenamefont {Szilagyi}}]{foucart_etal2014}%
  \BibitemOpen
  \bibfield  {author} {\bibinfo {author} {\bibfnamefont {F.}~\bibnamefont
  {Foucart}}, \bibinfo {author} {\bibfnamefont {M.~B.}\ \bibnamefont {Deaton}},
  \bibinfo {author} {\bibfnamefont {M.~D.}\ \bibnamefont {Duez}}, \bibinfo
  {author} {\bibfnamefont {E.}~\bibnamefont {O'Connor}}, \bibinfo {author}
  {\bibfnamefont {C.~D.}\ \bibnamefont {Ott}}, \bibinfo {author} {\bibfnamefont
  {R.}~\bibnamefont {Haas}}, \bibinfo {author} {\bibfnamefont {L.~E.}\
  \bibnamefont {Kidder}}, \bibinfo {author} {\bibfnamefont {H.~P.}\
  \bibnamefont {Pfeiffer}}, \bibinfo {author} {\bibfnamefont {M.~A.}\
  \bibnamefont {Scheel}}, \ and\ \bibinfo {author} {\bibfnamefont
  {B.}~\bibnamefont {Szilagyi}},\ }\href {\doibase 10.1103/PhysRevD.90.024026}
  {\bibfield  {journal} {\bibinfo  {journal} {\prd}\ }\textbf {\bibinfo
  {volume} {90}},\ \bibinfo {pages} {024026} (\bibinfo {year}
  {2014})}\BibitemShut {NoStop}%
\bibitem [{\citenamefont {Kawaguchi}\ \emph {et~al.}(2015)\citenamefont
  {Kawaguchi}, \citenamefont {Kyutoku}, \citenamefont {Nakano}, \citenamefont
  {Okawa}, \citenamefont {Shibata},\ and\ \citenamefont
  {Taniguchi}}]{kawaguchi_knost2015}%
  \BibitemOpen
  \bibfield  {author} {\bibinfo {author} {\bibfnamefont {K.}~\bibnamefont
  {Kawaguchi}}, \bibinfo {author} {\bibfnamefont {K.}~\bibnamefont {Kyutoku}},
  \bibinfo {author} {\bibfnamefont {H.}~\bibnamefont {Nakano}}, \bibinfo
  {author} {\bibfnamefont {H.}~\bibnamefont {Okawa}}, \bibinfo {author}
  {\bibfnamefont {M.}~\bibnamefont {Shibata}}, \ and\ \bibinfo {author}
  {\bibfnamefont {K.}~\bibnamefont {Taniguchi}},\ }\href {\doibase
  10.1103/PhysRevD.92.024014} {\bibfield  {journal} {\bibinfo  {journal}
  {\prd}\ }\textbf {\bibinfo {volume} {92}},\ \bibinfo {pages} {024014}
  (\bibinfo {year} {2015})}\BibitemShut {NoStop}%
\bibitem [{\citenamefont {Kyutoku}\ \emph {et~al.}(2015)\citenamefont
  {Kyutoku}, \citenamefont {Ioka}, \citenamefont {Okawa}, \citenamefont
  {Shibata},\ and\ \citenamefont {Taniguchi}}]{kyutoku_iost2015}%
  \BibitemOpen
  \bibfield  {author} {\bibinfo {author} {\bibfnamefont {K.}~\bibnamefont
  {Kyutoku}}, \bibinfo {author} {\bibfnamefont {K.}~\bibnamefont {Ioka}},
  \bibinfo {author} {\bibfnamefont {H.}~\bibnamefont {Okawa}}, \bibinfo
  {author} {\bibfnamefont {M.}~\bibnamefont {Shibata}}, \ and\ \bibinfo
  {author} {\bibfnamefont {K.}~\bibnamefont {Taniguchi}},\ }\href {\doibase
  10.1103/PhysRevD.92.044028} {\bibfield  {journal} {\bibinfo  {journal}
  {\prd}\ }\textbf {\bibinfo {volume} {92}},\ \bibinfo {pages} {044028}
  (\bibinfo {year} {2015})}\BibitemShut {NoStop}%
\bibitem [{\citenamefont {Foucart}\ \emph {et~al.}(2017)\citenamefont
  {Foucart}, \citenamefont {Desai}, \citenamefont {Brege}, \citenamefont
  {Duez}, \citenamefont {Kasen}, \citenamefont {Hemberger}, \citenamefont
  {Kidder}, \citenamefont {Pfeiffer},\ and\ \citenamefont
  {Scheel}}]{foucart_dbdkhkps2017}%
  \BibitemOpen
  \bibfield  {author} {\bibinfo {author} {\bibfnamefont {F.}~\bibnamefont
  {Foucart}}, \bibinfo {author} {\bibfnamefont {D.}~\bibnamefont {Desai}},
  \bibinfo {author} {\bibfnamefont {W.}~\bibnamefont {Brege}}, \bibinfo
  {author} {\bibfnamefont {M.~D.}\ \bibnamefont {Duez}}, \bibinfo {author}
  {\bibfnamefont {D.}~\bibnamefont {Kasen}}, \bibinfo {author} {\bibfnamefont
  {D.~A.}\ \bibnamefont {Hemberger}}, \bibinfo {author} {\bibfnamefont {L.~E.}\
  \bibnamefont {Kidder}}, \bibinfo {author} {\bibfnamefont {H.~P.}\
  \bibnamefont {Pfeiffer}}, \ and\ \bibinfo {author} {\bibfnamefont {M.~A.}\
  \bibnamefont {Scheel}},\ }\href {\doibase 10.1088/1361-6382/aa573b}
  {\bibfield  {journal} {\bibinfo  {journal} {Classical and Quantum Gravity}\
  }\textbf {\bibinfo {volume} {34}},\ \bibinfo {pages} {044002} (\bibinfo
  {year} {2017})}\BibitemShut {NoStop}%
\bibitem [{\citenamefont {Kyutoku}\ \emph {et~al.}(2018)\citenamefont
  {Kyutoku}, \citenamefont {Kiuchi}, \citenamefont {Sekiguchi}, \citenamefont
  {Shibata},\ and\ \citenamefont {Taniguchi}}]{kyutoku_ksst2018}%
  \BibitemOpen
  \bibfield  {author} {\bibinfo {author} {\bibfnamefont {K.}~\bibnamefont
  {Kyutoku}}, \bibinfo {author} {\bibfnamefont {K.}~\bibnamefont {Kiuchi}},
  \bibinfo {author} {\bibfnamefont {Y.}~\bibnamefont {Sekiguchi}}, \bibinfo
  {author} {\bibfnamefont {M.}~\bibnamefont {Shibata}}, \ and\ \bibinfo
  {author} {\bibfnamefont {K.}~\bibnamefont {Taniguchi}},\ }\href {\doibase
  10.1103/PhysRevD.97.023009} {\bibfield  {journal} {\bibinfo  {journal}
  {\prd}\ }\textbf {\bibinfo {volume} {97}},\ \bibinfo {pages} {023009}
  (\bibinfo {year} {2018})}\BibitemShut {NoStop}%
\bibitem [{\citenamefont {Brege}\ \emph {et~al.}(2018)\citenamefont {Brege},
  \citenamefont {Duez}, \citenamefont {Foucart}, \citenamefont {Deaton},
  \citenamefont {Caro}, \citenamefont {Hemberger}, \citenamefont {Kidder},
  \citenamefont {O'Conner}, \citenamefont {Pfeiffer},\ and\ \citenamefont
  {Scheel}}]{brege_dfdchkops2018}%
  \BibitemOpen
  \bibfield  {author} {\bibinfo {author} {\bibfnamefont {W.}~\bibnamefont
  {Brege}}, \bibinfo {author} {\bibfnamefont {M.~D.}\ \bibnamefont {Duez}},
  \bibinfo {author} {\bibfnamefont {F.}~\bibnamefont {Foucart}}, \bibinfo
  {author} {\bibfnamefont {M.~B.}\ \bibnamefont {Deaton}}, \bibinfo {author}
  {\bibfnamefont {J.}~\bibnamefont {Caro}}, \bibinfo {author} {\bibfnamefont
  {D.~A.}\ \bibnamefont {Hemberger}}, \bibinfo {author} {\bibfnamefont {L.~E.}\
  \bibnamefont {Kidder}}, \bibinfo {author} {\bibfnamefont {E.}~\bibnamefont
  {O'Conner}}, \bibinfo {author} {\bibfnamefont {H.~P.}\ \bibnamefont
  {Pfeiffer}}, \ and\ \bibinfo {author} {\bibfnamefont {M.~A.}\ \bibnamefont
  {Scheel}},\ }\href {\doibase 10.1103/PhysRevD.98.063009} {\bibfield
  {journal} {\bibinfo  {journal} {\prd}\ }\textbf {\bibinfo {volume} {98}},\
  \bibinfo {pages} {063009} (\bibinfo {year} {2018})}\BibitemShut {NoStop}%
\bibitem [{\citenamefont {Foucart}\ \emph
  {et~al.}(2019{\natexlab{a}})\citenamefont {Foucart}, \citenamefont {Duez},
  \citenamefont {Kidder}, \citenamefont {Nissanke}, \citenamefont {Pfeiffer},\
  and\ \citenamefont {Scheel}}]{foucart_dknps2019}%
  \BibitemOpen
  \bibfield  {author} {\bibinfo {author} {\bibfnamefont {F.}~\bibnamefont
  {Foucart}}, \bibinfo {author} {\bibfnamefont {M.~D.}\ \bibnamefont {Duez}},
  \bibinfo {author} {\bibfnamefont {L.~E.}\ \bibnamefont {Kidder}}, \bibinfo
  {author} {\bibfnamefont {S.}~\bibnamefont {Nissanke}}, \bibinfo {author}
  {\bibfnamefont {H.~P.}\ \bibnamefont {Pfeiffer}}, \ and\ \bibinfo {author}
  {\bibfnamefont {M.~A.}\ \bibnamefont {Scheel}},\ }\href {\doibase
  10.1103/PhysRevD.99.103025} {\bibfield  {journal} {\bibinfo  {journal}
  {\prd}\ }\textbf {\bibinfo {volume} {99}},\ \bibinfo {pages} {103025}
  (\bibinfo {year} {2019}{\natexlab{a}})}\BibitemShut {NoStop}%
\bibitem [{\citenamefont {Foucart}\ \emph
  {et~al.}(2013{\natexlab{b}})\citenamefont {Foucart}, \citenamefont {Buchman},
  \citenamefont {Duez}, \citenamefont {Grudich}, \citenamefont {Kidder},
  \citenamefont {MacDonald}, \citenamefont {Mroue}, \citenamefont {Pfeiffer},
  \citenamefont {Scheel},\ and\ \citenamefont
  {Szilagyi}}]{foucart_bdgkmmpss2013}%
  \BibitemOpen
  \bibfield  {author} {\bibinfo {author} {\bibfnamefont {F.}~\bibnamefont
  {Foucart}}, \bibinfo {author} {\bibfnamefont {L.}~\bibnamefont {Buchman}},
  \bibinfo {author} {\bibfnamefont {M.~D.}\ \bibnamefont {Duez}}, \bibinfo
  {author} {\bibfnamefont {M.}~\bibnamefont {Grudich}}, \bibinfo {author}
  {\bibfnamefont {L.~E.}\ \bibnamefont {Kidder}}, \bibinfo {author}
  {\bibfnamefont {I.}~\bibnamefont {MacDonald}}, \bibinfo {author}
  {\bibfnamefont {A.}~\bibnamefont {Mroue}}, \bibinfo {author} {\bibfnamefont
  {H.~P.}\ \bibnamefont {Pfeiffer}}, \bibinfo {author} {\bibfnamefont {M.~A.}\
  \bibnamefont {Scheel}}, \ and\ \bibinfo {author} {\bibfnamefont
  {B.}~\bibnamefont {Szilagyi}},\ }\href {\doibase 10.1103/PhysRevD.88.064017}
  {\bibfield  {journal} {\bibinfo  {journal} {\prd}\ }\textbf {\bibinfo
  {volume} {88}},\ \bibinfo {pages} {064017} (\bibinfo {year}
  {2013}{\natexlab{b}})}\BibitemShut {NoStop}%
\bibitem [{\citenamefont {Foucart}\ \emph
  {et~al.}(2019{\natexlab{b}})\citenamefont {Foucart}, \citenamefont {Duez},
  \citenamefont {Hinderer}, \citenamefont {Caro}, \citenamefont {Williamson},
  \citenamefont {Boyle}, \citenamefont {Buonanno}, \citenamefont {Haas},
  \citenamefont {Hemberger}, \citenamefont {Kidder}, \citenamefont {Pfeiffer},\
  and\ \citenamefont {Scheel}}]{foucart_etal2019}%
  \BibitemOpen
  \bibfield  {author} {\bibinfo {author} {\bibfnamefont {F.}~\bibnamefont
  {Foucart}}, \bibinfo {author} {\bibfnamefont {M.~D.}\ \bibnamefont {Duez}},
  \bibinfo {author} {\bibfnamefont {T.}~\bibnamefont {Hinderer}}, \bibinfo
  {author} {\bibfnamefont {J.}~\bibnamefont {Caro}}, \bibinfo {author}
  {\bibfnamefont {A.~R.}\ \bibnamefont {Williamson}}, \bibinfo {author}
  {\bibfnamefont {M.}~\bibnamefont {Boyle}}, \bibinfo {author} {\bibfnamefont
  {A.}~\bibnamefont {Buonanno}}, \bibinfo {author} {\bibfnamefont
  {R.}~\bibnamefont {Haas}}, \bibinfo {author} {\bibfnamefont {D.~A.}\
  \bibnamefont {Hemberger}}, \bibinfo {author} {\bibfnamefont {L.~E.}\
  \bibnamefont {Kidder}}, \bibinfo {author} {\bibfnamefont {H.~P.}\
  \bibnamefont {Pfeiffer}}, \ and\ \bibinfo {author} {\bibfnamefont {M.~A.}\
  \bibnamefont {Scheel}},\ }\href {\doibase 10.1103/PhysRevD.99.044008}
  {\bibfield  {journal} {\bibinfo  {journal} {\prd}\ }\textbf {\bibinfo
  {volume} {99}},\ \bibinfo {pages} {044008} (\bibinfo {year}
  {2019}{\natexlab{b}})}\BibitemShut {NoStop}%
\bibitem [{\citenamefont {Peters}\ and\ \citenamefont
  {Mathews}(1963)}]{peters_mathews1963}%
  \BibitemOpen
  \bibfield  {author} {\bibinfo {author} {\bibfnamefont {P.~C.}\ \bibnamefont
  {Peters}}\ and\ \bibinfo {author} {\bibfnamefont {J.}~\bibnamefont
  {Mathews}},\ }\href {\doibase 10.1103/PhysRev.131.435} {\bibfield  {journal}
  {\bibinfo  {journal} {Physical Review}\ }\textbf {\bibinfo {volume} {131}},\
  \bibinfo {pages} {435} (\bibinfo {year} {1963})}\BibitemShut {NoStop}%
\bibitem [{\citenamefont {Peters}(1964)}]{peters1964}%
  \BibitemOpen
  \bibfield  {author} {\bibinfo {author} {\bibfnamefont {P.~C.}\ \bibnamefont
  {Peters}},\ }\href {\doibase 10.1103/PhysRev.136.B1224} {\bibfield  {journal}
  {\bibinfo  {journal} {Physical Review}\ }\textbf {\bibinfo {volume} {136}},\
  \bibinfo {pages} {B1224} (\bibinfo {year} {1964})}\BibitemShut {NoStop}%
\bibitem [{\citenamefont {Foucart}\ \emph {et~al.}(2008)\citenamefont
  {Foucart}, \citenamefont {Kidder}, \citenamefont {Pfeiffer},\ and\
  \citenamefont {Teukolsky}}]{foucart_kpt2008}%
  \BibitemOpen
  \bibfield  {author} {\bibinfo {author} {\bibfnamefont {F.}~\bibnamefont
  {Foucart}}, \bibinfo {author} {\bibfnamefont {L.~E.}\ \bibnamefont {Kidder}},
  \bibinfo {author} {\bibfnamefont {H.~P.}\ \bibnamefont {Pfeiffer}}, \ and\
  \bibinfo {author} {\bibfnamefont {S.~A.}\ \bibnamefont {Teukolsky}},\ }\href
  {\doibase 10.1103/PhysRevD.77.124051} {\bibfield  {journal} {\bibinfo
  {journal} {\prd}\ }\textbf {\bibinfo {volume} {77}},\ \bibinfo {pages}
  {124051} (\bibinfo {year} {2008})}\BibitemShut {NoStop}%
\bibitem [{\citenamefont {Shibata}\ and\ \citenamefont
  {Ury{\=u}}(2006)}]{shibata_uryu2006}%
  \BibitemOpen
  \bibfield  {author} {\bibinfo {author} {\bibfnamefont {M.}~\bibnamefont
  {Shibata}}\ and\ \bibinfo {author} {\bibfnamefont {K.}~\bibnamefont
  {Ury{\=u}}},\ }\href {\doibase 10.1103/PhysRevD.74.121503} {\bibfield
  {journal} {\bibinfo  {journal} {\prd}\ }\textbf {\bibinfo {volume} {74}},\
  \bibinfo {pages} {121503} (\bibinfo {year} {2006})}\BibitemShut {NoStop}%
\bibitem [{\citenamefont {Shibata}\ and\ \citenamefont
  {Ury{\=u}}(2007)}]{shibata_uryu2007}%
  \BibitemOpen
  \bibfield  {author} {\bibinfo {author} {\bibfnamefont {M.}~\bibnamefont
  {Shibata}}\ and\ \bibinfo {author} {\bibfnamefont {K.}~\bibnamefont
  {Ury{\=u}}},\ }\href {\doibase 10.1088/0264-9381/24/12/S09} {\bibfield
  {journal} {\bibinfo  {journal} {Classical and Quantum Gravity}\ }\textbf
  {\bibinfo {volume} {24}},\ \bibinfo {pages} {S125} (\bibinfo {year}
  {2007})}\BibitemShut {NoStop}%
\bibitem [{\citenamefont {Kyutoku}\ \emph
  {et~al.}(2014{\natexlab{a}})\citenamefont {Kyutoku}, \citenamefont
  {Shibata},\ and\ \citenamefont {Taniguchi}}]{kyutoku_st2014}%
  \BibitemOpen
  \bibfield  {author} {\bibinfo {author} {\bibfnamefont {K.}~\bibnamefont
  {Kyutoku}}, \bibinfo {author} {\bibfnamefont {M.}~\bibnamefont {Shibata}}, \
  and\ \bibinfo {author} {\bibfnamefont {K.}~\bibnamefont {Taniguchi}},\ }\href
  {\doibase 10.1103/PhysRevD.90.064006} {\bibfield  {journal} {\bibinfo
  {journal} {\prd}\ }\textbf {\bibinfo {volume} {90}},\ \bibinfo {pages}
  {064006} (\bibinfo {year} {2014}{\natexlab{a}})}\BibitemShut {NoStop}%
\bibitem [{\citenamefont {Husa}\ \emph {et~al.}(2008)\citenamefont {Husa},
  \citenamefont {Hannam}, \citenamefont {Gonz{\'a}lez}, \citenamefont
  {Sperhake},\ and\ \citenamefont {Br{\"u}gmann}}]{husa_hgsb2008}%
  \BibitemOpen
  \bibfield  {author} {\bibinfo {author} {\bibfnamefont {S.}~\bibnamefont
  {Husa}}, \bibinfo {author} {\bibfnamefont {M.}~\bibnamefont {Hannam}},
  \bibinfo {author} {\bibfnamefont {J.~A.}\ \bibnamefont {Gonz{\'a}lez}},
  \bibinfo {author} {\bibfnamefont {U.}~\bibnamefont {Sperhake}}, \ and\
  \bibinfo {author} {\bibfnamefont {B.}~\bibnamefont {Br{\"u}gmann}},\ }\href
  {\doibase 10.1103/PhysRevD.77.044037} {\bibfield  {journal} {\bibinfo
  {journal} {\prd}\ }\textbf {\bibinfo {volume} {77}},\ \bibinfo {pages}
  {044037} (\bibinfo {year} {2008})}\BibitemShut {NoStop}%
\bibitem [{\citenamefont {Kyutoku}\ \emph {et~al.}(2009)\citenamefont
  {Kyutoku}, \citenamefont {Shibata},\ and\ \citenamefont
  {Taniguchi}}]{kyutoku_st2009}%
  \BibitemOpen
  \bibfield  {author} {\bibinfo {author} {\bibfnamefont {K.}~\bibnamefont
  {Kyutoku}}, \bibinfo {author} {\bibfnamefont {M.}~\bibnamefont {Shibata}}, \
  and\ \bibinfo {author} {\bibfnamefont {K.}~\bibnamefont {Taniguchi}},\ }\href
  {\doibase 10.1103/PhysRevD.79.124018} {\bibfield  {journal} {\bibinfo
  {journal} {\prd}\ }\textbf {\bibinfo {volume} {79}},\ \bibinfo {pages}
  {124018} (\bibinfo {year} {2009})}\BibitemShut {NoStop}%
\bibitem [{\citenamefont {Brandt}\ and\ \citenamefont
  {Br{\"u}gmann}(1997)}]{brandt_brugmann1997}%
  \BibitemOpen
  \bibfield  {author} {\bibinfo {author} {\bibfnamefont {S.}~\bibnamefont
  {Brandt}}\ and\ \bibinfo {author} {\bibfnamefont {B.}~\bibnamefont
  {Br{\"u}gmann}},\ }\href {\doibase 10.1103/PhysRevLett.78.3606} {\bibfield
  {journal} {\bibinfo  {journal} {\prl}\ }\textbf {\bibinfo {volume} {78}},\
  \bibinfo {pages} {3606} (\bibinfo {year} {1997})}\BibitemShut {NoStop}%
\bibitem [{\citenamefont {Shibata}\ and\ \citenamefont
  {Taniguchi}(2008)}]{shibata_taniguchi2008}%
  \BibitemOpen
  \bibfield  {author} {\bibinfo {author} {\bibfnamefont {M.}~\bibnamefont
  {Shibata}}\ and\ \bibinfo {author} {\bibfnamefont {K.}~\bibnamefont
  {Taniguchi}},\ }\href {\doibase 10.1103/PhysRevD.77.084015} {\bibfield
  {journal} {\bibinfo  {journal} {\prd}\ }\textbf {\bibinfo {volume} {77}},\
  \bibinfo {pages} {084015} (\bibinfo {year} {2008})}\BibitemShut {NoStop}%
\bibitem [{\citenamefont {York}(1999)}]{york1999}%
  \BibitemOpen
  \bibfield  {author} {\bibinfo {author} {\bibfnamefont {J.~W.}\ \bibnamefont
  {York}},\ }\href {\doibase 10.1103/PhysRevLett.82.1350} {\bibfield  {journal}
  {\bibinfo  {journal} {\prl}\ }\textbf {\bibinfo {volume} {82}},\ \bibinfo
  {pages} {1350} (\bibinfo {year} {1999})}\BibitemShut {NoStop}%
\bibitem [{\citenamefont {Pfeiffer}\ and\ \citenamefont
  {York}(2003)}]{pfeiffer_york2003}%
  \BibitemOpen
  \bibfield  {author} {\bibinfo {author} {\bibfnamefont {H.~P.}\ \bibnamefont
  {Pfeiffer}}\ and\ \bibinfo {author} {\bibfnamefont {J.~W.}\ \bibnamefont
  {York}},\ }\href {\doibase 10.1103/PhysRevD.67.044022} {\bibfield  {journal}
  {\bibinfo  {journal} {\prd}\ }\textbf {\bibinfo {volume} {67}},\ \bibinfo
  {pages} {044022} (\bibinfo {year} {2003})}\BibitemShut {NoStop}%
\bibitem [{\citenamefont {York}(1979)}]{york1979}%
  \BibitemOpen
  \bibfield  {author} {\bibinfo {author} {\bibfnamefont {J.~W.}\ \bibnamefont
  {York}},\ }in\ \href@noop {} {\emph {\bibinfo {booktitle} {Gravitational
  Radiation}}},\ \bibinfo {editor} {edited by\ \bibinfo {editor} {\bibfnamefont
  {L.}~\bibnamefont {Smarr}}}\ (\bibinfo {year} {1979})\BibitemShut {NoStop}%
\bibitem [{\citenamefont {Bonazzola}\ \emph {et~al.}(1997)\citenamefont
  {Bonazzola}, \citenamefont {Gourgoulhon},\ and\ \citenamefont
  {Marck}}]{bonazzola_gm1997}%
  \BibitemOpen
  \bibfield  {author} {\bibinfo {author} {\bibfnamefont {S.}~\bibnamefont
  {Bonazzola}}, \bibinfo {author} {\bibfnamefont {E.}~\bibnamefont
  {Gourgoulhon}}, \ and\ \bibinfo {author} {\bibfnamefont {J.-A.}\ \bibnamefont
  {Marck}},\ }\href {\doibase 10.1103/PhysRevD.56.7740} {\bibfield  {journal}
  {\bibinfo  {journal} {\prd}\ }\textbf {\bibinfo {volume} {56}},\ \bibinfo
  {pages} {7740} (\bibinfo {year} {1997})}\BibitemShut {NoStop}%
\bibitem [{\citenamefont {Asada}(1998)}]{asada1998}%
  \BibitemOpen
  \bibfield  {author} {\bibinfo {author} {\bibfnamefont {H.}~\bibnamefont
  {Asada}},\ }\href {\doibase 10.1103/PhysRevD.57.7292} {\bibfield  {journal}
  {\bibinfo  {journal} {\prd}\ }\textbf {\bibinfo {volume} {57}},\ \bibinfo
  {pages} {7292} (\bibinfo {year} {1998})}\BibitemShut {NoStop}%
\bibitem [{\citenamefont {Shibata}(1998)}]{shibata1998}%
  \BibitemOpen
  \bibfield  {author} {\bibinfo {author} {\bibfnamefont {M.}~\bibnamefont
  {Shibata}},\ }\href {\doibase 10.1103/PhysRevD.58.024012} {\bibfield
  {journal} {\bibinfo  {journal} {\prd}\ }\textbf {\bibinfo {volume} {58}},\
  \bibinfo {pages} {024012} (\bibinfo {year} {1998})}\BibitemShut {NoStop}%
\bibitem [{\citenamefont {Teukolsky}(1998)}]{teukolsky1998}%
  \BibitemOpen
  \bibfield  {author} {\bibinfo {author} {\bibfnamefont {S.~A.}\ \bibnamefont
  {Teukolsky}},\ }\href {\doibase 10.1086/306082} {\bibfield  {journal}
  {\bibinfo  {journal} {\apj}\ }\textbf {\bibinfo {volume} {504}},\ \bibinfo
  {pages} {442} (\bibinfo {year} {1998})}\BibitemShut {NoStop}%
\bibitem [{\citenamefont {Gourgoulhon}\ \emph {et~al.}(2001)\citenamefont
  {Gourgoulhon}, \citenamefont {Grandcl{\'e}ment}, \citenamefont {Taniguchi},
  \citenamefont {Marck},\ and\ \citenamefont
  {Bonazzola}}]{gourgoulhon_gtmb2001}%
  \BibitemOpen
  \bibfield  {author} {\bibinfo {author} {\bibfnamefont {E.}~\bibnamefont
  {Gourgoulhon}}, \bibinfo {author} {\bibfnamefont {P.}~\bibnamefont
  {Grandcl{\'e}ment}}, \bibinfo {author} {\bibfnamefont {K.}~\bibnamefont
  {Taniguchi}}, \bibinfo {author} {\bibfnamefont {J.-A.}\ \bibnamefont
  {Marck}}, \ and\ \bibinfo {author} {\bibfnamefont {S.}~\bibnamefont
  {Bonazzola}},\ }\href {\doibase 10.1103/PhysRevD.63.064029} {\bibfield
  {journal} {\bibinfo  {journal} {\prd}\ }\textbf {\bibinfo {volume} {63}},\
  \bibinfo {pages} {064029} (\bibinfo {year} {2001})}\BibitemShut {NoStop}%
\bibitem [{\citenamefont {{LORENE website}}()}]{LORENE}%
  \BibitemOpen
  \bibfield  {author} {\bibinfo {author} {\bibnamefont {{LORENE website}}},\
  }\href@noop {} {}\bibinfo {note} {{http://www.lorene.obspm.fr/}}\BibitemShut
  {NoStop}%
\bibitem [{\citenamefont {Wilson}\ and\ \citenamefont
  {Mathews}(1995)}]{wilson_mathews1995}%
  \BibitemOpen
  \bibfield  {author} {\bibinfo {author} {\bibfnamefont {J.~R.}\ \bibnamefont
  {Wilson}}\ and\ \bibinfo {author} {\bibfnamefont {G.~J.}\ \bibnamefont
  {Mathews}},\ }\href {\doibase 10.1103/PhysRevLett.75.4161} {\bibfield
  {journal} {\bibinfo  {journal} {\prl}\ }\textbf {\bibinfo {volume} {75}},\
  \bibinfo {pages} {4161} (\bibinfo {year} {1995})}\BibitemShut {NoStop}%
\bibitem [{\citenamefont {Wilson}\ \emph {et~al.}(1996)\citenamefont {Wilson},
  \citenamefont {Mathews},\ and\ \citenamefont {Marronetti}}]{wilson_mm1996}%
  \BibitemOpen
  \bibfield  {author} {\bibinfo {author} {\bibfnamefont {J.~R.}\ \bibnamefont
  {Wilson}}, \bibinfo {author} {\bibfnamefont {G.~J.}\ \bibnamefont {Mathews}},
  \ and\ \bibinfo {author} {\bibfnamefont {P.}~\bibnamefont {Marronetti}},\
  }\href {\doibase 10.1103/PhysRevD.54.1317} {\bibfield  {journal} {\bibinfo
  {journal} {\prd}\ }\textbf {\bibinfo {volume} {54}},\ \bibinfo {pages} {1317}
  (\bibinfo {year} {1996})}\BibitemShut {NoStop}%
\bibitem [{\citenamefont {Bowen}\ and\ \citenamefont
  {York}(1980)}]{bowen_york1980}%
  \BibitemOpen
  \bibfield  {author} {\bibinfo {author} {\bibfnamefont {J.~M.}\ \bibnamefont
  {Bowen}}\ and\ \bibinfo {author} {\bibfnamefont {J.~W.}\ \bibnamefont
  {York}},\ }\href {\doibase 10.1103/PhysRevD.21.2047} {\bibfield  {journal}
  {\bibinfo  {journal} {\prd}\ }\textbf {\bibinfo {volume} {21}},\ \bibinfo
  {pages} {2047} (\bibinfo {year} {1980})}\BibitemShut {NoStop}%
\bibitem [{\citenamefont {Pfeiffer}\ \emph {et~al.}(2007)\citenamefont
  {Pfeiffer}, \citenamefont {Brown}, \citenamefont {Kidder}, \citenamefont
  {Lindblom}, \citenamefont {Lovelace},\ and\ \citenamefont
  {Scheel}}]{pfeiffer_bklls2007}%
  \BibitemOpen
  \bibfield  {author} {\bibinfo {author} {\bibfnamefont {H.~P.}\ \bibnamefont
  {Pfeiffer}}, \bibinfo {author} {\bibfnamefont {D.~A.}\ \bibnamefont {Brown}},
  \bibinfo {author} {\bibfnamefont {L.~E.}\ \bibnamefont {Kidder}}, \bibinfo
  {author} {\bibfnamefont {L.}~\bibnamefont {Lindblom}}, \bibinfo {author}
  {\bibfnamefont {G.}~\bibnamefont {Lovelace}}, \ and\ \bibinfo {author}
  {\bibfnamefont {M.~A.}\ \bibnamefont {Scheel}},\ }\href {\doibase
  10.1088/0264-9381/24/12/S06} {\bibfield  {journal} {\bibinfo  {journal}
  {Classical and Quantum Gravity}\ }\textbf {\bibinfo {volume} {24}},\ \bibinfo
  {pages} {S59} (\bibinfo {year} {2007})}\BibitemShut {NoStop}%
\bibitem [{\citenamefont {Buonanno}\ \emph {et~al.}(2011)\citenamefont
  {Buonanno}, \citenamefont {Kidder}, \citenamefont {Mrou{\'e}}, \citenamefont
  {Pfeiffer},\ and\ \citenamefont {Taracchini}}]{buonanno_kmpt2011}%
  \BibitemOpen
  \bibfield  {author} {\bibinfo {author} {\bibfnamefont {A.}~\bibnamefont
  {Buonanno}}, \bibinfo {author} {\bibfnamefont {L.~E.}\ \bibnamefont
  {Kidder}}, \bibinfo {author} {\bibfnamefont {A.~H.}\ \bibnamefont
  {Mrou{\'e}}}, \bibinfo {author} {\bibfnamefont {H.~P.}\ \bibnamefont
  {Pfeiffer}}, \ and\ \bibinfo {author} {\bibfnamefont {A.}~\bibnamefont
  {Taracchini}},\ }\href {\doibase 10.1103/PhysRevD.83.104034} {\bibfield
  {journal} {\bibinfo  {journal} {\prd}\ }\textbf {\bibinfo {volume} {83}},\
  \bibinfo {pages} {104034} (\bibinfo {year} {2011})}\BibitemShut {NoStop}%
\bibitem [{\citenamefont {Kyutoku}\ \emph
  {et~al.}(2014{\natexlab{b}})\citenamefont {Kyutoku}, \citenamefont {Ioka},\
  and\ \citenamefont {Shibata}}]{kyutoku_is2014}%
  \BibitemOpen
  \bibfield  {author} {\bibinfo {author} {\bibfnamefont {K.}~\bibnamefont
  {Kyutoku}}, \bibinfo {author} {\bibfnamefont {K.}~\bibnamefont {Ioka}}, \
  and\ \bibinfo {author} {\bibfnamefont {M.}~\bibnamefont {Shibata}},\ }\href
  {\doibase 10.1093/mnrasl/slt128} {\bibfield  {journal} {\bibinfo  {journal}
  {\mnras}\ }\textbf {\bibinfo {volume} {437}},\ \bibinfo {pages} {L6}
  (\bibinfo {year} {2014}{\natexlab{b}})}\BibitemShut {NoStop}%
\bibitem [{\citenamefont {Gourgoulhon}\ and\ \citenamefont
  {Jaramillo}(2006)}]{gourgoulhon_jaramillo2006}%
  \BibitemOpen
  \bibfield  {author} {\bibinfo {author} {\bibfnamefont {E.}~\bibnamefont
  {Gourgoulhon}}\ and\ \bibinfo {author} {\bibfnamefont {J.~L.}\ \bibnamefont
  {Jaramillo}},\ }\href {\doibase 10.1016/j.physrep.2005.10.005} {\bibfield
  {journal} {\bibinfo  {journal} {\physrep}\ }\textbf {\bibinfo {volume}
  {423}},\ \bibinfo {pages} {159} (\bibinfo {year} {2006})}\BibitemShut
  {NoStop}%
\bibitem [{\citenamefont {Cook}\ and\ \citenamefont
  {Whiting}(2007)}]{cook_whiting2007}%
  \BibitemOpen
  \bibfield  {author} {\bibinfo {author} {\bibfnamefont {G.~B.}\ \bibnamefont
  {Cook}}\ and\ \bibinfo {author} {\bibfnamefont {B.~F.}\ \bibnamefont
  {Whiting}},\ }\href {\doibase 10.1103/PhysRevD.76.041501} {\bibfield
  {journal} {\bibinfo  {journal} {\prd}\ }\textbf {\bibinfo {volume} {76}},\
  \bibinfo {pages} {041501} (\bibinfo {year} {2007})}\BibitemShut {NoStop}%
\bibitem [{\citenamefont {P{\"u}rrer}\ \emph {et~al.}(2012)\citenamefont
  {P{\"u}rrer}, \citenamefont {Husa},\ and\ \citenamefont
  {Hannam}}]{purrer_hh2012}%
  \BibitemOpen
  \bibfield  {author} {\bibinfo {author} {\bibfnamefont {M.}~\bibnamefont
  {P{\"u}rrer}}, \bibinfo {author} {\bibfnamefont {S.}~\bibnamefont {Husa}}, \
  and\ \bibinfo {author} {\bibfnamefont {M.}~\bibnamefont {Hannam}},\ }\href
  {\doibase 10.1103/PhysRevD.85.124051} {\bibfield  {journal} {\bibinfo
  {journal} {\prd}\ }\textbf {\bibinfo {volume} {85}},\ \bibinfo {pages}
  {124051} (\bibinfo {year} {2012})}\BibitemShut {NoStop}%
\bibitem [{\citenamefont {Ramos-Buades}\ \emph {et~al.}(2019)\citenamefont
  {Ramos-Buades}, \citenamefont {Husa},\ and\ \citenamefont
  {Pratten}}]{ramosbuades_hp2019}%
  \BibitemOpen
  \bibfield  {author} {\bibinfo {author} {\bibfnamefont {A.}~\bibnamefont
  {Ramos-Buades}}, \bibinfo {author} {\bibfnamefont {S.}~\bibnamefont {Husa}},
  \ and\ \bibinfo {author} {\bibfnamefont {G.}~\bibnamefont {Pratten}},\ }\href
  {\doibase 10.1103/PhysRevD.99.023003} {\bibfield  {journal} {\bibinfo
  {journal} {\prd}\ }\textbf {\bibinfo {volume} {99}},\ \bibinfo {pages}
  {023003} (\bibinfo {year} {2019})}\BibitemShut {NoStop}%
\bibitem [{\citenamefont {Beig}(1978)}]{beig1978}%
  \BibitemOpen
  \bibfield  {author} {\bibinfo {author} {\bibfnamefont {R.}~\bibnamefont
  {Beig}},\ }\href {\doibase 10.1016/0375-9601(78)90198-6} {\bibfield
  {journal} {\bibinfo  {journal} {Physics Letters A}\ }\textbf {\bibinfo
  {volume} {69}},\ \bibinfo {pages} {153} (\bibinfo {year} {1978})}\BibitemShut
  {NoStop}%
\bibitem [{\citenamefont {Ashtekar}\ and\ \citenamefont
  {Magnon-Ashtekar}(1979)}]{ashtekar_magnonashtekar1979}%
  \BibitemOpen
  \bibfield  {author} {\bibinfo {author} {\bibfnamefont {A.}~\bibnamefont
  {Ashtekar}}\ and\ \bibinfo {author} {\bibfnamefont {A.}~\bibnamefont
  {Magnon-Ashtekar}},\ }\href {\doibase 10.1063/1.524151} {\bibfield  {journal}
  {\bibinfo  {journal} {Journal of Mathematical Physics}\ }\textbf {\bibinfo
  {volume} {20}},\ \bibinfo {pages} {793} (\bibinfo {year} {1979})}\BibitemShut
  {NoStop}%
\bibitem [{\citenamefont {Friedman}\ \emph {et~al.}(2002)\citenamefont
  {Friedman}, \citenamefont {Ury{\=u}},\ and\ \citenamefont
  {Shibata}}]{friedman_us2002}%
  \BibitemOpen
  \bibfield  {author} {\bibinfo {author} {\bibfnamefont {J.~L.}\ \bibnamefont
  {Friedman}}, \bibinfo {author} {\bibfnamefont {K.}~\bibnamefont {Ury{\=u}}},
  \ and\ \bibinfo {author} {\bibfnamefont {M.}~\bibnamefont {Shibata}},\ }\href
  {\doibase 10.1103/PhysRevD.65.064035} {\bibfield  {journal} {\bibinfo
  {journal} {\prd}\ }\textbf {\bibinfo {volume} {65}},\ \bibinfo {pages}
  {064035} (\bibinfo {year} {2002})}\BibitemShut {NoStop}%
\bibitem [{\citenamefont {Shibata}\ \emph {et~al.}(2004)\citenamefont
  {Shibata}, \citenamefont {Ury{\=u}},\ and\ \citenamefont
  {Friedman}}]{shibata_uf2004}%
  \BibitemOpen
  \bibfield  {author} {\bibinfo {author} {\bibfnamefont {M.}~\bibnamefont
  {Shibata}}, \bibinfo {author} {\bibfnamefont {K.}~\bibnamefont {Ury{\=u}}}, \
  and\ \bibinfo {author} {\bibfnamefont {J.~L.}\ \bibnamefont {Friedman}},\
  }\href {\doibase 10.1103/PhysRevD.70.044044} {\bibfield  {journal} {\bibinfo
  {journal} {\prd}\ }\textbf {\bibinfo {volume} {70}},\ \bibinfo {pages}
  {044044} (\bibinfo {year} {2004})}\BibitemShut {NoStop}%
\bibitem [{\citenamefont {Gourgoulhon}\ \emph {et~al.}(2002)\citenamefont
  {Gourgoulhon}, \citenamefont {Grandcl{\'e}ment},\ and\ \citenamefont
  {Bonazzola}}]{gourgoulhon_gb2002}%
  \BibitemOpen
  \bibfield  {author} {\bibinfo {author} {\bibfnamefont {E.}~\bibnamefont
  {Gourgoulhon}}, \bibinfo {author} {\bibfnamefont {P.}~\bibnamefont
  {Grandcl{\'e}ment}}, \ and\ \bibinfo {author} {\bibfnamefont
  {S.}~\bibnamefont {Bonazzola}},\ }\href {\doibase 10.1103/PhysRevD.65.044020}
  {\bibfield  {journal} {\bibinfo  {journal} {\prd}\ }\textbf {\bibinfo
  {volume} {65}},\ \bibinfo {pages} {044020} (\bibinfo {year}
  {2002})}\BibitemShut {NoStop}%
\bibitem [{\citenamefont {Grandcl{\'e}ment}\ \emph {et~al.}(2002)\citenamefont
  {Grandcl{\'e}ment}, \citenamefont {gourgoulhon},\ and\ \citenamefont
  {Bonazzola}}]{grandclement_gb2002}%
  \BibitemOpen
  \bibfield  {author} {\bibinfo {author} {\bibfnamefont {P.}~\bibnamefont
  {Grandcl{\'e}ment}}, \bibinfo {author} {\bibfnamefont {E.}~\bibnamefont
  {gourgoulhon}}, \ and\ \bibinfo {author} {\bibfnamefont {S.}~\bibnamefont
  {Bonazzola}},\ }\href {\doibase 10.1103/PhysRevD.65.044021} {\bibfield
  {journal} {\bibinfo  {journal} {\prd}\ }\textbf {\bibinfo {volume} {65}},\
  \bibinfo {pages} {044021} (\bibinfo {year} {2002})}\BibitemShut {NoStop}%
\bibitem [{\citenamefont {Caudill}\ \emph {et~al.}(2006)\citenamefont
  {Caudill}, \citenamefont {Cook}, \citenamefont {Grigsby},\ and\ \citenamefont
  {Pfeiffer}}]{caudill_cgp2006}%
  \BibitemOpen
  \bibfield  {author} {\bibinfo {author} {\bibfnamefont {M.}~\bibnamefont
  {Caudill}}, \bibinfo {author} {\bibfnamefont {G.~B.}\ \bibnamefont {Cook}},
  \bibinfo {author} {\bibfnamefont {J.~D.}\ \bibnamefont {Grigsby}}, \ and\
  \bibinfo {author} {\bibfnamefont {H.~P.}\ \bibnamefont {Pfeiffer}},\ }\href
  {\doibase 10.1103/PhysRevD.74.064011} {\bibfield  {journal} {\bibinfo
  {journal} {\prd}\ }\textbf {\bibinfo {volume} {74}},\ \bibinfo {pages}
  {064011} (\bibinfo {year} {2006})}\BibitemShut {NoStop}%
\bibitem [{\citenamefont {Taniguchi}\ \emph {et~al.}(2006)\citenamefont
  {Taniguchi}, \citenamefont {Baumgarte}, \citenamefont {Faber},\ and\
  \citenamefont {Shapiro}}]{taniguchi_bfs2006}%
  \BibitemOpen
  \bibfield  {author} {\bibinfo {author} {\bibfnamefont {K.}~\bibnamefont
  {Taniguchi}}, \bibinfo {author} {\bibfnamefont {T.~W.}\ \bibnamefont
  {Baumgarte}}, \bibinfo {author} {\bibfnamefont {J.~A.}\ \bibnamefont
  {Faber}}, \ and\ \bibinfo {author} {\bibfnamefont {S.~L.}\ \bibnamefont
  {Shapiro}},\ }\href {\doibase 10.1103/PhysRevD.74.041502} {\bibfield
  {journal} {\bibinfo  {journal} {\prd}\ }\textbf {\bibinfo {volume} {74}},\
  \bibinfo {pages} {041502} (\bibinfo {year} {2006})}\BibitemShut {NoStop}%
\bibitem [{\citenamefont {Taniguchi}\ \emph {et~al.}(2007)\citenamefont
  {Taniguchi}, \citenamefont {Baumgarte}, \citenamefont {Faber},\ and\
  \citenamefont {Shapiro}}]{taniguchi_bfs2007}%
  \BibitemOpen
  \bibfield  {author} {\bibinfo {author} {\bibfnamefont {K.}~\bibnamefont
  {Taniguchi}}, \bibinfo {author} {\bibfnamefont {T.~W.}\ \bibnamefont
  {Baumgarte}}, \bibinfo {author} {\bibfnamefont {J.~A.}\ \bibnamefont
  {Faber}}, \ and\ \bibinfo {author} {\bibfnamefont {S.~L.}\ \bibnamefont
  {Shapiro}},\ }\href {\doibase 10.1103/PhysRevD.75.084005} {\bibfield
  {journal} {\bibinfo  {journal} {\prd}\ }\textbf {\bibinfo {volume} {75}},\
  \bibinfo {pages} {084005} (\bibinfo {year} {2007})}\BibitemShut {NoStop}%
\bibitem [{\citenamefont {Taniguchi}\ \emph {et~al.}(2008)\citenamefont
  {Taniguchi}, \citenamefont {Baumgarte}, \citenamefont {Faber},\ and\
  \citenamefont {Shapiro}}]{taniguchi_bfs2008}%
  \BibitemOpen
  \bibfield  {author} {\bibinfo {author} {\bibfnamefont {K.}~\bibnamefont
  {Taniguchi}}, \bibinfo {author} {\bibfnamefont {T.~W.}\ \bibnamefont
  {Baumgarte}}, \bibinfo {author} {\bibfnamefont {J.~A.}\ \bibnamefont
  {Faber}}, \ and\ \bibinfo {author} {\bibfnamefont {S.~L.}\ \bibnamefont
  {Shapiro}},\ }\href {\doibase 10.1103/PhysRevD.77.044003} {\bibfield
  {journal} {\bibinfo  {journal} {\prd}\ }\textbf {\bibinfo {volume} {77}},\
  \bibinfo {pages} {044003} (\bibinfo {year} {2008})}\BibitemShut {NoStop}%
\bibitem [{\citenamefont {Foucart}\ \emph {et~al.}(2011)\citenamefont
  {Foucart}, \citenamefont {Duez}, \citenamefont {Kidder},\ and\ \citenamefont
  {Teukolsky}}]{foucart_dkt2011}%
  \BibitemOpen
  \bibfield  {author} {\bibinfo {author} {\bibfnamefont {F.}~\bibnamefont
  {Foucart}}, \bibinfo {author} {\bibfnamefont {M.~D.}\ \bibnamefont {Duez}},
  \bibinfo {author} {\bibfnamefont {L.~E.}\ \bibnamefont {Kidder}}, \ and\
  \bibinfo {author} {\bibfnamefont {S.~A.}\ \bibnamefont {Teukolsky}},\ }\href
  {\doibase 10.1103/PhysRevD.83.024005} {\bibfield  {journal} {\bibinfo
  {journal} {\prd}\ }\textbf {\bibinfo {volume} {83}},\ \bibinfo {pages}
  {024005} (\bibinfo {year} {2011})}\BibitemShut {NoStop}%
\bibitem [{\citenamefont {Moldenhauer}\ \emph {et~al.}(2014)\citenamefont
  {Moldenhauer}, \citenamefont {Markakis}, \citenamefont {Johnson-McDaniel},
  \citenamefont {Tichy},\ and\ \citenamefont
  {Br{\"u}gmann}}]{moldenhauer_mjtb2014}%
  \BibitemOpen
  \bibfield  {author} {\bibinfo {author} {\bibfnamefont {N.}~\bibnamefont
  {Moldenhauer}}, \bibinfo {author} {\bibfnamefont {C.~M.}\ \bibnamefont
  {Markakis}}, \bibinfo {author} {\bibfnamefont {N.~K.}\ \bibnamefont
  {Johnson-McDaniel}}, \bibinfo {author} {\bibfnamefont {W.}~\bibnamefont
  {Tichy}}, \ and\ \bibinfo {author} {\bibfnamefont {B.}~\bibnamefont
  {Br{\"u}gmann}},\ }\href {\doibase 10.1103/PhysRevD.90.084043} {\bibfield
  {journal} {\bibinfo  {journal} {\prd}\ }\textbf {\bibinfo {volume} {90}},\
  \bibinfo {pages} {084043} (\bibinfo {year} {2014})}\BibitemShut {NoStop}%
\bibitem [{\citenamefont {Haas}\ \emph {et~al.}(2016)\citenamefont {Haas},
  \citenamefont {Ott}, \citenamefont {Szilagyi}, \citenamefont {Kaplan},
  \citenamefont {Lippuner}, \citenamefont {Scheel}, \citenamefont {Barkett},
  \citenamefont {Muhlberger}, \citenamefont {Dietrich}, \citenamefont {Duez},
  \citenamefont {Foucart}, \citenamefont {Pfeiffer}, \citenamefont {Kidder},\
  and\ \citenamefont {Teukolsky}}]{haas_etal2016}%
  \BibitemOpen
  \bibfield  {author} {\bibinfo {author} {\bibfnamefont {R.}~\bibnamefont
  {Haas}}, \bibinfo {author} {\bibfnamefont {C.~D.}\ \bibnamefont {Ott}},
  \bibinfo {author} {\bibfnamefont {B.}~\bibnamefont {Szilagyi}}, \bibinfo
  {author} {\bibfnamefont {J.~D.}\ \bibnamefont {Kaplan}}, \bibinfo {author}
  {\bibfnamefont {J.}~\bibnamefont {Lippuner}}, \bibinfo {author}
  {\bibfnamefont {M.~A.}\ \bibnamefont {Scheel}}, \bibinfo {author}
  {\bibfnamefont {K.}~\bibnamefont {Barkett}}, \bibinfo {author} {\bibfnamefont
  {C.~D.}\ \bibnamefont {Muhlberger}}, \bibinfo {author} {\bibfnamefont
  {T.}~\bibnamefont {Dietrich}}, \bibinfo {author} {\bibfnamefont {M.~D.}\
  \bibnamefont {Duez}}, \bibinfo {author} {\bibfnamefont {F.}~\bibnamefont
  {Foucart}}, \bibinfo {author} {\bibfnamefont {H.~P.}\ \bibnamefont
  {Pfeiffer}}, \bibinfo {author} {\bibfnamefont {L.~E.}\ \bibnamefont
  {Kidder}}, \ and\ \bibinfo {author} {\bibfnamefont {S.~A.}\ \bibnamefont
  {Teukolsky}},\ }\href {\doibase 10.1103/PhysRevD.93.124062} {\bibfield
  {journal} {\bibinfo  {journal} {\prd}\ }\textbf {\bibinfo {volume} {93}},\
  \bibinfo {pages} {124062} (\bibinfo {year} {2016})}\BibitemShut {NoStop}%
\bibitem [{\citenamefont {Yamamoto}\ \emph {et~al.}(2008)\citenamefont
  {Yamamoto}, \citenamefont {Shibata},\ and\ \citenamefont
  {Taniguchi}}]{yamamoto_st2008}%
  \BibitemOpen
  \bibfield  {author} {\bibinfo {author} {\bibfnamefont {T.}~\bibnamefont
  {Yamamoto}}, \bibinfo {author} {\bibfnamefont {M.}~\bibnamefont {Shibata}}, \
  and\ \bibinfo {author} {\bibfnamefont {K.}~\bibnamefont {Taniguchi}},\ }\href
  {\doibase 10.1103/PhysRevD.78.064054} {\bibfield  {journal} {\bibinfo
  {journal} {\prd}\ }\textbf {\bibinfo {volume} {78}},\ \bibinfo {pages}
  {064054} (\bibinfo {year} {2008})}\BibitemShut {NoStop}%
\bibitem [{\citenamefont {Kiuchi}\ \emph {et~al.}(2017)\citenamefont {Kiuchi},
  \citenamefont {Kawaguchi}, \citenamefont {Kyutoku}, \citenamefont
  {Sekiguchi}, \citenamefont {Shibata},\ and\ \citenamefont
  {Taniguchi}}]{kiuchi_kksst2017}%
  \BibitemOpen
  \bibfield  {author} {\bibinfo {author} {\bibfnamefont {K.}~\bibnamefont
  {Kiuchi}}, \bibinfo {author} {\bibfnamefont {K.}~\bibnamefont {Kawaguchi}},
  \bibinfo {author} {\bibfnamefont {K.}~\bibnamefont {Kyutoku}}, \bibinfo
  {author} {\bibfnamefont {Y.}~\bibnamefont {Sekiguchi}}, \bibinfo {author}
  {\bibfnamefont {M.}~\bibnamefont {Shibata}}, \ and\ \bibinfo {author}
  {\bibfnamefont {K.}~\bibnamefont {Taniguchi}},\ }\href {\doibase
  10.1103/PhysRevD.96.084060} {\bibfield  {journal} {\bibinfo  {journal}
  {\prd}\ }\textbf {\bibinfo {volume} {96}},\ \bibinfo {pages} {084060}
  (\bibinfo {year} {2017})}\BibitemShut {NoStop}%
\bibitem [{\citenamefont {Campanelli}\ \emph {et~al.}(2006)\citenamefont
  {Campanelli}, \citenamefont {Lousto}, \citenamefont {Marronetti},\ and\
  \citenamefont {Zlochower}}]{campanelli_lmz2006}%
  \BibitemOpen
  \bibfield  {author} {\bibinfo {author} {\bibfnamefont {M.}~\bibnamefont
  {Campanelli}}, \bibinfo {author} {\bibfnamefont {C.~O.}\ \bibnamefont
  {Lousto}}, \bibinfo {author} {\bibfnamefont {P.}~\bibnamefont {Marronetti}},
  \ and\ \bibinfo {author} {\bibfnamefont {Y.}~\bibnamefont {Zlochower}},\
  }\href {\doibase 10.1103/PhysRevLett.96.111101} {\bibfield  {journal}
  {\bibinfo  {journal} {\prl}\ }\textbf {\bibinfo {volume} {96}},\ \bibinfo
  {pages} {111101} (\bibinfo {year} {2006})}\BibitemShut {NoStop}%
\bibitem [{\citenamefont {Br{\"u}gmann}\ \emph {et~al.}(2008)\citenamefont
  {Br{\"u}gmann}, \citenamefont {Gonz{\'a}lez}, \citenamefont {Hannam},
  \citenamefont {Husa}, \citenamefont {Sperhake},\ and\ \citenamefont
  {Tichy}}]{brugmann_ghhst2008}%
  \BibitemOpen
  \bibfield  {author} {\bibinfo {author} {\bibfnamefont {B.}~\bibnamefont
  {Br{\"u}gmann}}, \bibinfo {author} {\bibfnamefont {J.~A.}\ \bibnamefont
  {Gonz{\'a}lez}}, \bibinfo {author} {\bibfnamefont {M.}~\bibnamefont
  {Hannam}}, \bibinfo {author} {\bibfnamefont {S.}~\bibnamefont {Husa}},
  \bibinfo {author} {\bibfnamefont {U.}~\bibnamefont {Sperhake}}, \ and\
  \bibinfo {author} {\bibfnamefont {W.}~\bibnamefont {Tichy}},\ }\href
  {\doibase 10.1103/PhysRevD.77.024027} {\bibfield  {journal} {\bibinfo
  {journal} {\prd}\ }\textbf {\bibinfo {volume} {77}},\ \bibinfo {pages}
  {024027} (\bibinfo {year} {2008})}\BibitemShut {NoStop}%
\bibitem [{\citenamefont {Boyle}(2013)}]{Boyle2013}%
  \BibitemOpen
  \bibfield  {author} {\bibinfo {author} {\bibfnamefont {M.}~\bibnamefont
  {Boyle}},\ }\href {\doibase 10.1103/PhysRevD.87.104006} {\bibfield  {journal}
  {\bibinfo  {journal} {\prd}\ }\textbf {\bibinfo {volume} {87}},\ \bibinfo
  {pages} {104006} (\bibinfo {year} {2013})}\BibitemShut {NoStop}%
\bibitem [{\citenamefont {Boyle}\ \emph {et~al.}(2007)\citenamefont {Boyle},
  \citenamefont {Brown}, \citenamefont {Kidder}, \citenamefont {Mrou{\'e}},
  \citenamefont {Pfeiffer}, \citenamefont {Scheel}, \citenamefont {Cook},\ and\
  \citenamefont {Teukolsky}}]{boyle_bkmpsct2007}%
  \BibitemOpen
  \bibfield  {author} {\bibinfo {author} {\bibfnamefont {M.}~\bibnamefont
  {Boyle}}, \bibinfo {author} {\bibfnamefont {D.~A.}\ \bibnamefont {Brown}},
  \bibinfo {author} {\bibfnamefont {L.~E.}\ \bibnamefont {Kidder}}, \bibinfo
  {author} {\bibfnamefont {A.~H.}\ \bibnamefont {Mrou{\'e}}}, \bibinfo {author}
  {\bibfnamefont {H.~P.}\ \bibnamefont {Pfeiffer}}, \bibinfo {author}
  {\bibfnamefont {M.~A.}\ \bibnamefont {Scheel}}, \bibinfo {author}
  {\bibfnamefont {G.~B.}\ \bibnamefont {Cook}}, \ and\ \bibinfo {author}
  {\bibfnamefont {S.~A.}\ \bibnamefont {Teukolsky}},\ }\href {\doibase
  10.1103/PhysRevD.76.124038} {\bibfield  {journal} {\bibinfo  {journal}
  {\prd}\ }\textbf {\bibinfo {volume} {76}},\ \bibinfo {pages} {124038}
  (\bibinfo {year} {2007})}\BibitemShut {NoStop}%
\bibitem [{\citenamefont {Read}\ \emph {et~al.}(2009)\citenamefont {Read},
  \citenamefont {Lackey}, \citenamefont {Owen},\ and\ \citenamefont
  {Friedman}}]{read_lof2009}%
  \BibitemOpen
  \bibfield  {author} {\bibinfo {author} {\bibfnamefont {J.~S.}\ \bibnamefont
  {Read}}, \bibinfo {author} {\bibfnamefont {B.~D.}\ \bibnamefont {Lackey}},
  \bibinfo {author} {\bibfnamefont {B.~J.}\ \bibnamefont {Owen}}, \ and\
  \bibinfo {author} {\bibfnamefont {J.~L.}\ \bibnamefont {Friedman}},\ }\href
  {\doibase 10.1103/PhysRevD.79.124032} {\bibfield  {journal} {\bibinfo
  {journal} {\prd}\ }\textbf {\bibinfo {volume} {79}},\ \bibinfo {pages}
  {124032} (\bibinfo {year} {2009})}\BibitemShut {NoStop}%
\bibitem [{\citenamefont {Akmal}\ \emph {et~al.}(1998)\citenamefont {Akmal},
  \citenamefont {Pandharipande},\ and\ \citenamefont
  {Ravenhall}}]{akmal_pr1998}%
  \BibitemOpen
  \bibfield  {author} {\bibinfo {author} {\bibfnamefont {A.}~\bibnamefont
  {Akmal}}, \bibinfo {author} {\bibfnamefont {V.~R.}\ \bibnamefont
  {Pandharipande}}, \ and\ \bibinfo {author} {\bibfnamefont {D.~G.}\
  \bibnamefont {Ravenhall}},\ }\href {\doibase 10.1103/PhysRevC.58.1804}
  {\bibfield  {journal} {\bibinfo  {journal} {\prc}\ }\textbf {\bibinfo
  {volume} {58}},\ \bibinfo {pages} {1804} (\bibinfo {year}
  {1998})}\BibitemShut {NoStop}%
\bibitem [{\citenamefont {De}\ \emph {et~al.}(2018)\citenamefont {De},
  \citenamefont {Finstad}, \citenamefont {Lattimer}, \citenamefont {Brown},
  \citenamefont {Berger},\ and\ \citenamefont {Biwer}}]{de_flbbb2018}%
  \BibitemOpen
  \bibfield  {author} {\bibinfo {author} {\bibfnamefont {S.}~\bibnamefont
  {De}}, \bibinfo {author} {\bibfnamefont {D.}~\bibnamefont {Finstad}},
  \bibinfo {author} {\bibfnamefont {J.~M.}\ \bibnamefont {Lattimer}}, \bibinfo
  {author} {\bibfnamefont {D.~A.}\ \bibnamefont {Brown}}, \bibinfo {author}
  {\bibfnamefont {E.}~\bibnamefont {Berger}}, \ and\ \bibinfo {author}
  {\bibfnamefont {C.~M.}\ \bibnamefont {Biwer}},\ }\href {\doibase
  10.1103/PhysRevLett.121.091102} {\bibfield  {journal} {\bibinfo  {journal}
  {\prl}\ }\textbf {\bibinfo {volume} {121}},\ \bibinfo {pages} {091102}
  (\bibinfo {year} {2018})}\BibitemShut {NoStop}%
\bibitem [{\citenamefont {Mrou{\'e}}\ and\ \citenamefont
  {Pfeiffer}(2012)}]{mroue_pfeiffer2012}%
  \BibitemOpen
  \bibfield  {author} {\bibinfo {author} {\bibfnamefont {A.~H.}\ \bibnamefont
  {Mrou{\'e}}}\ and\ \bibinfo {author} {\bibfnamefont {H.~P.}\ \bibnamefont
  {Pfeiffer}},\ }\href@noop {} {\bibfield  {journal} {\bibinfo  {journal}
  {arXiv:1210.2958}\ } (\bibinfo {year} {2012})}\BibitemShut {NoStop}%
\bibitem [{\citenamefont {Hotokezaka}\ \emph {et~al.}(2015)\citenamefont
  {Hotokezaka}, \citenamefont {Kyutoku}, \citenamefont {Okawa},\ and\
  \citenamefont {Shibata}}]{hotokezaka_kos2015}%
  \BibitemOpen
  \bibfield  {author} {\bibinfo {author} {\bibfnamefont {K.}~\bibnamefont
  {Hotokezaka}}, \bibinfo {author} {\bibfnamefont {K.}~\bibnamefont {Kyutoku}},
  \bibinfo {author} {\bibfnamefont {H.}~\bibnamefont {Okawa}}, \ and\ \bibinfo
  {author} {\bibfnamefont {M.}~\bibnamefont {Shibata}},\ }\href {\doibase
  10.1103/PhysRevD.91.064060} {\bibfield  {journal} {\bibinfo  {journal}
  {\prd}\ }\textbf {\bibinfo {volume} {91}},\ \bibinfo {pages} {064060}
  (\bibinfo {year} {2015})}\BibitemShut {NoStop}%
\bibitem [{\citenamefont {Hotokezaka}\ \emph {et~al.}(2016)\citenamefont
  {Hotokezaka}, \citenamefont {Kyutoku}, \citenamefont {Sekiguchi},\ and\
  \citenamefont {Shibata}}]{hotokezaka_kss2016}%
  \BibitemOpen
  \bibfield  {author} {\bibinfo {author} {\bibfnamefont {K.}~\bibnamefont
  {Hotokezaka}}, \bibinfo {author} {\bibfnamefont {K.}~\bibnamefont {Kyutoku}},
  \bibinfo {author} {\bibfnamefont {Y.-I.}\ \bibnamefont {Sekiguchi}}, \ and\
  \bibinfo {author} {\bibfnamefont {M.}~\bibnamefont {Shibata}},\ }\href
  {\doibase 10.1103/PhysRevD.93.064082} {\bibfield  {journal} {\bibinfo
  {journal} {\prd}\ }\textbf {\bibinfo {volume} {93}},\ \bibinfo {pages}
  {064082} (\bibinfo {year} {2016})}\BibitemShut {NoStop}%
\bibitem [{\citenamefont {Kiuchi}\ \emph {et~al.}(2020)\citenamefont {Kiuchi},
  \citenamefont {Kawaguchi}, \citenamefont {Kyutoku}, \citenamefont
  {Sekiguchi},\ and\ \citenamefont {Shibata}}]{kiuchi_kkss2020}%
  \BibitemOpen
  \bibfield  {author} {\bibinfo {author} {\bibfnamefont {K.}~\bibnamefont
  {Kiuchi}}, \bibinfo {author} {\bibfnamefont {K.}~\bibnamefont {Kawaguchi}},
  \bibinfo {author} {\bibfnamefont {K.}~\bibnamefont {Kyutoku}}, \bibinfo
  {author} {\bibfnamefont {Y.}~\bibnamefont {Sekiguchi}}, \ and\ \bibinfo
  {author} {\bibfnamefont {M.}~\bibnamefont {Shibata}},\ }\href {\doibase
  10.1103/PhysRevD.101.084006} {\bibfield  {journal} {\bibinfo  {journal}
  {\prd}\ }\textbf {\bibinfo {volume} {101}},\ \bibinfo {pages} {084006}
  (\bibinfo {year} {2020})}\BibitemShut {NoStop}%
\bibitem [{\citenamefont {{Abbott}}\ \emph
  {et~al.}(2020{\natexlab{e}})\citenamefont {{Abbott}}, \citenamefont
  {{Abbott}}, \citenamefont {{Abraham}}, \citenamefont {{Acernese}},
  \citenamefont {{Ackley}}, \citenamefont {{Adams}}, \citenamefont
  {{Adhikari}}, \citenamefont {{Adya}}, \citenamefont {{Affeldt}},
  \citenamefont {{Agathos}},\ and\ \citenamefont {et~al.}}]{ligovirgo2020-4}%
  \BibitemOpen
  \bibfield  {author} {\bibinfo {author} {\bibfnamefont {R.}~\bibnamefont
  {{Abbott}}}, \bibinfo {author} {\bibfnamefont {T.~D.}\ \bibnamefont
  {{Abbott}}}, \bibinfo {author} {\bibfnamefont {S.}~\bibnamefont {{Abraham}}},
  \bibinfo {author} {\bibfnamefont {F.}~\bibnamefont {{Acernese}}}, \bibinfo
  {author} {\bibfnamefont {K.}~\bibnamefont {{Ackley}}}, \bibinfo {author}
  {\bibfnamefont {C.}~\bibnamefont {{Adams}}}, \bibinfo {author} {\bibfnamefont
  {R.~X.}\ \bibnamefont {{Adhikari}}}, \bibinfo {author} {\bibfnamefont
  {V.~B.}\ \bibnamefont {{Adya}}}, \bibinfo {author} {\bibfnamefont
  {C.}~\bibnamefont {{Affeldt}}}, \bibinfo {author} {\bibfnamefont
  {M.}~\bibnamefont {{Agathos}}}, \ and\ \bibinfo {author} {\bibnamefont
  {et~al.}},\ }\href {\doibase 10.1103/PhysRevD.102.043015} {\bibfield
  {journal} {\bibinfo  {journal} {\prd}\ }\textbf {\bibinfo {volume} {102}},\
  \bibinfo {pages} {043015} (\bibinfo {year} {2020}{\natexlab{e}})}\BibitemShut
  {NoStop}%
\bibitem [{\citenamefont {Blanchet}(2014)}]{blanchet2014}%
  \BibitemOpen
  \bibfield  {author} {\bibinfo {author} {\bibfnamefont {L.}~\bibnamefont
  {Blanchet}},\ }\href {\doibase 10.12942/lrr-2014-2} {\bibfield  {journal}
  {\bibinfo  {journal} {Living Reviews in Relativity}\ }\textbf {\bibinfo
  {volume} {17}},\ \bibinfo {pages} {2} (\bibinfo {year} {2014})}\BibitemShut
  {NoStop}%
\bibitem [{\citenamefont {Vines}\ and\ \citenamefont
  {Flanagan}(2013)}]{vines_flanagan2013}%
  \BibitemOpen
  \bibfield  {author} {\bibinfo {author} {\bibfnamefont {J.~E.}\ \bibnamefont
  {Vines}}\ and\ \bibinfo {author} {\bibfnamefont {{\'E}.~{\'E}.}\ \bibnamefont
  {Flanagan}},\ }\href {\doibase 10.1103/PhysRevD.88.024046} {\bibfield
  {journal} {\bibinfo  {journal} {\prd}\ }\textbf {\bibinfo {volume} {88}},\
  \bibinfo {pages} {024046} (\bibinfo {year} {2013})}\BibitemShut {NoStop}%
\bibitem [{\citenamefont {Boh{\'e}}\ \emph {et~al.}(2013)\citenamefont
  {Boh{\'e}}, \citenamefont {Marsat}, \citenamefont {Faye},\ and\ \citenamefont
  {Blancet}}]{bohe_mfb2013}%
  \BibitemOpen
  \bibfield  {author} {\bibinfo {author} {\bibfnamefont {A.}~\bibnamefont
  {Boh{\'e}}}, \bibinfo {author} {\bibfnamefont {S.}~\bibnamefont {Marsat}},
  \bibinfo {author} {\bibfnamefont {G.}~\bibnamefont {Faye}}, \ and\ \bibinfo
  {author} {\bibfnamefont {L.}~\bibnamefont {Blancet}},\ }\href {\doibase
  10.1088/0264-9381/30/7/075017} {\bibfield  {journal} {\bibinfo  {journal}
  {Classical and Quantum Gravity}\ }\textbf {\bibinfo {volume} {30}},\ \bibinfo
  {pages} {075017} (\bibinfo {year} {2013})}\BibitemShut {NoStop}%
\bibitem [{\citenamefont {Barker}\ and\ \citenamefont
  {O'Connell}(1975)}]{barker_oconnel1975}%
  \BibitemOpen
  \bibfield  {author} {\bibinfo {author} {\bibfnamefont {B.~M.}\ \bibnamefont
  {Barker}}\ and\ \bibinfo {author} {\bibfnamefont {R.~F.}\ \bibnamefont
  {O'Connell}},\ }\href {\doibase 10.1103/PhysRevD.12.329} {\bibfield
  {journal} {\bibinfo  {journal} {\prd}\ }\textbf {\bibinfo {volume} {12}},\
  \bibinfo {pages} {329} (\bibinfo {year} {1975})}\BibitemShut {NoStop}%
\bibitem [{\citenamefont {Apostolatos}\ \emph {et~al.}(1994)\citenamefont
  {Apostolatos}, \citenamefont {Cutler}, \citenamefont {Sussman},\ and\
  \citenamefont {Thorne}}]{apostolatos_cst1994}%
  \BibitemOpen
  \bibfield  {author} {\bibinfo {author} {\bibfnamefont {T.~A.}\ \bibnamefont
  {Apostolatos}}, \bibinfo {author} {\bibfnamefont {C.}~\bibnamefont {Cutler}},
  \bibinfo {author} {\bibfnamefont {G.~J.}\ \bibnamefont {Sussman}}, \ and\
  \bibinfo {author} {\bibfnamefont {K.~S.}\ \bibnamefont {Thorne}},\ }\href
  {\doibase 10.1103/PhysRevD.49.6274} {\bibfield  {journal} {\bibinfo
  {journal} {\prd}\ }\textbf {\bibinfo {volume} {49}},\ \bibinfo {pages} {6274}
  (\bibinfo {year} {1994})}\BibitemShut {NoStop}%
\bibitem [{\citenamefont {Racine}(2008)}]{racine2008}%
  \BibitemOpen
  \bibfield  {author} {\bibinfo {author} {\bibfnamefont {{\'E}.}~\bibnamefont
  {Racine}},\ }\href {\doibase 10.1103/PhysRevD.78.044021} {\bibfield
  {journal} {\bibinfo  {journal} {\prd}\ }\textbf {\bibinfo {volume} {78}},\
  \bibinfo {pages} {044021} (\bibinfo {year} {2008})}\BibitemShut {NoStop}%
\bibitem [{\citenamefont {Kidder}(1995)}]{kidder1995}%
  \BibitemOpen
  \bibfield  {author} {\bibinfo {author} {\bibfnamefont {L.~E.}\ \bibnamefont
  {Kidder}},\ }\href {\doibase 10.1103/PhysRevD.52.821} {\bibfield  {journal}
  {\bibinfo  {journal} {\prd}\ }\textbf {\bibinfo {volume} {52}},\ \bibinfo
  {pages} {821} (\bibinfo {year} {1995})}\BibitemShut {NoStop}%
\bibitem [{\citenamefont {O'Shaughnessy}\ \emph {et~al.}(2011)\citenamefont
  {O'Shaughnessy}, \citenamefont {Vaishnav}, \citenamefont {Healy},
  \citenamefont {Zachary},\ and\ \citenamefont
  {Shoemaker}}]{oshaughnessy_vhms2011}%
  \BibitemOpen
  \bibfield  {author} {\bibinfo {author} {\bibfnamefont {R.}~\bibnamefont
  {O'Shaughnessy}}, \bibinfo {author} {\bibfnamefont {B.}~\bibnamefont
  {Vaishnav}}, \bibinfo {author} {\bibfnamefont {J.}~\bibnamefont {Healy}},
  \bibinfo {author} {\bibnamefont {Zachary}}, \ and\ \bibinfo {author}
  {\bibfnamefont {D.}~\bibnamefont {Shoemaker}},\ }\href {\doibase
  10.1103/PhysRevD.84.124002} {\bibfield  {journal} {\bibinfo  {journal}
  {\prd}\ }\textbf {\bibinfo {volume} {84}},\ \bibinfo {pages} {124002}
  (\bibinfo {year} {2011})}\BibitemShut {NoStop}%
\bibitem [{\citenamefont {Schmidt}\ \emph {et~al.}(2012)\citenamefont
  {Schmidt}, \citenamefont {Hannam},\ and\ \citenamefont
  {Husa}}]{schmidt_hh2012}%
  \BibitemOpen
  \bibfield  {author} {\bibinfo {author} {\bibfnamefont {P.}~\bibnamefont
  {Schmidt}}, \bibinfo {author} {\bibfnamefont {M.}~\bibnamefont {Hannam}}, \
  and\ \bibinfo {author} {\bibfnamefont {S.}~\bibnamefont {Husa}},\ }\href
  {\doibase 10.1103/PhysRevD.86.104063} {\bibfield  {journal} {\bibinfo
  {journal} {\prd}\ }\textbf {\bibinfo {volume} {86}},\ \bibinfo {pages}
  {104063} (\bibinfo {year} {2012})}\BibitemShut {NoStop}%
\bibitem [{\citenamefont {Boyle}\ \emph {et~al.}(2011)\citenamefont {Boyle},
  \citenamefont {Owen},\ and\ \citenamefont {Pfeiffer}}]{boyle_op2011}%
  \BibitemOpen
  \bibfield  {author} {\bibinfo {author} {\bibfnamefont {M.}~\bibnamefont
  {Boyle}}, \bibinfo {author} {\bibfnamefont {R.}~\bibnamefont {Owen}}, \ and\
  \bibinfo {author} {\bibfnamefont {H.~P.}\ \bibnamefont {Pfeiffer}},\ }\href
  {\doibase 10.1103/PhysRevD.84.124011} {\bibfield  {journal} {\bibinfo
  {journal} {\prd}\ }\textbf {\bibinfo {volume} {84}},\ \bibinfo {pages}
  {124011} (\bibinfo {year} {2011})}\BibitemShut {NoStop}%
\bibitem [{\citenamefont {Ochsner}\ and\ \citenamefont
  {O'Shaughnessy}(2012)}]{ochsner_oshaughnessy2012}%
  \BibitemOpen
  \bibfield  {author} {\bibinfo {author} {\bibfnamefont {E.}~\bibnamefont
  {Ochsner}}\ and\ \bibinfo {author} {\bibfnamefont {R.}~\bibnamefont
  {O'Shaughnessy}},\ }\href {\doibase 10.1103/PhysRevD.86.104037} {\bibfield
  {journal} {\bibinfo  {journal} {\prd}\ }\textbf {\bibinfo {volume} {86}},\
  \bibinfo {pages} {104037} (\bibinfo {year} {2012})}\BibitemShut {NoStop}%
\bibitem [{\citenamefont {Kawaguchi}\ \emph {et~al.}(2017)\citenamefont
  {Kawaguchi}, \citenamefont {Kyutoku}, \citenamefont {Nakano},\ and\
  \citenamefont {Shibata}}]{kawaguchi_kns2017}%
  \BibitemOpen
  \bibfield  {author} {\bibinfo {author} {\bibfnamefont {K.}~\bibnamefont
  {Kawaguchi}}, \bibinfo {author} {\bibfnamefont {K.}~\bibnamefont {Kyutoku}},
  \bibinfo {author} {\bibfnamefont {H.}~\bibnamefont {Nakano}}, \ and\ \bibinfo
  {author} {\bibfnamefont {M.}~\bibnamefont {Shibata}},\ }\href@noop {} {\
  (\bibinfo {year} {2017})},\ \Eprint {http://arxiv.org/abs/arXiv:1709.02754}
  {arXiv:1709.02754} \BibitemShut {NoStop}%
\bibitem [{\citenamefont {Schmidt}\ \emph {et~al.}(2011)\citenamefont
  {Schmidt}, \citenamefont {Hannam}, \citenamefont {Husa},\ and\ \citenamefont
  {Ajith}}]{schmidt_hha2011}%
  \BibitemOpen
  \bibfield  {author} {\bibinfo {author} {\bibfnamefont {P.}~\bibnamefont
  {Schmidt}}, \bibinfo {author} {\bibfnamefont {M.}~\bibnamefont {Hannam}},
  \bibinfo {author} {\bibfnamefont {S.}~\bibnamefont {Husa}}, \ and\ \bibinfo
  {author} {\bibfnamefont {P.}~\bibnamefont {Ajith}},\ }\href {\doibase
  10.1103/PhysRevD.84.024046} {\bibfield  {journal} {\bibinfo  {journal}
  {\prd}\ }\textbf {\bibinfo {volume} {84}},\ \bibinfo {pages} {024046}
  (\bibinfo {year} {2011})}\BibitemShut {NoStop}%
\bibitem [{\citenamefont {Pekowsky}\ \emph {et~al.}(2013)\citenamefont
  {Pekowsky}, \citenamefont {O'Shaughnessy}, \citenamefont {Healy},\ and\
  \citenamefont {Shoemaker}}]{pekowsky_ohs2013}%
  \BibitemOpen
  \bibfield  {author} {\bibinfo {author} {\bibfnamefont {L.}~\bibnamefont
  {Pekowsky}}, \bibinfo {author} {\bibfnamefont {R.}~\bibnamefont
  {O'Shaughnessy}}, \bibinfo {author} {\bibfnamefont {J.}~\bibnamefont
  {Healy}}, \ and\ \bibinfo {author} {\bibfnamefont {D.}~\bibnamefont
  {Shoemaker}},\ }\href {\doibase 10.1103/PhysRevD.88.024040} {\bibfield
  {journal} {\bibinfo  {journal} {\prd}\ }\textbf {\bibinfo {volume} {88}},\
  \bibinfo {pages} {024040} (\bibinfo {year} {2013})}\BibitemShut {NoStop}%
\bibitem [{\citenamefont {Boyle}\ \emph {et~al.}(2014)\citenamefont {Boyle},
  \citenamefont {Kidder}, \citenamefont {Ossikine},\ and\ \citenamefont
  {Pfeiffer}}]{boyle_kop2014}%
  \BibitemOpen
  \bibfield  {author} {\bibinfo {author} {\bibfnamefont {M.}~\bibnamefont
  {Boyle}}, \bibinfo {author} {\bibfnamefont {L.~E.}\ \bibnamefont {Kidder}},
  \bibinfo {author} {\bibfnamefont {S.}~\bibnamefont {Ossikine}}, \ and\
  \bibinfo {author} {\bibfnamefont {H.~P.}\ \bibnamefont {Pfeiffer}},\
  }\href@noop {} {\  (\bibinfo {year} {2014})},\ \Eprint
  {http://arxiv.org/abs/arXiv:1409.4431} {arXiv:1409.4431} \BibitemShut
  {NoStop}%
\bibitem [{\citenamefont {Boyle}\ \emph {et~al.}(2008)\citenamefont {Boyle},
  \citenamefont {Buonanno}, \citenamefont {Kidder}, \citenamefont {Mrou{\'e}},
  \citenamefont {Pan}, \citenamefont {Pfeiffer},\ and\ \citenamefont
  {Scheel}}]{boyle_bkmpps2008}%
  \BibitemOpen
  \bibfield  {author} {\bibinfo {author} {\bibfnamefont {M.}~\bibnamefont
  {Boyle}}, \bibinfo {author} {\bibfnamefont {A.}~\bibnamefont {Buonanno}},
  \bibinfo {author} {\bibfnamefont {L.~E.}\ \bibnamefont {Kidder}}, \bibinfo
  {author} {\bibfnamefont {A.~H.}\ \bibnamefont {Mrou{\'e}}}, \bibinfo {author}
  {\bibfnamefont {Y.}~\bibnamefont {Pan}}, \bibinfo {author} {\bibfnamefont
  {H.~P.}\ \bibnamefont {Pfeiffer}}, \ and\ \bibinfo {author} {\bibfnamefont
  {M.~A.}\ \bibnamefont {Scheel}},\ }\href {\doibase
  10.1103/PhysRevD.78.104020} {\bibfield  {journal} {\bibinfo  {journal}
  {\prd}\ }\textbf {\bibinfo {volume} {78}},\ \bibinfo {pages} {104020}
  (\bibinfo {year} {2008})}\BibitemShut {NoStop}%
\bibitem [{\citenamefont {Lovelace}\ \emph {et~al.}(2008)\citenamefont
  {Lovelace}, \citenamefont {Owen}, \citenamefont {Pfeiffer},\ and\
  \citenamefont {Chu}}]{lovelace_opc2008}%
  \BibitemOpen
  \bibfield  {author} {\bibinfo {author} {\bibfnamefont {G.}~\bibnamefont
  {Lovelace}}, \bibinfo {author} {\bibfnamefont {R.}~\bibnamefont {Owen}},
  \bibinfo {author} {\bibfnamefont {H.~P.}\ \bibnamefont {Pfeiffer}}, \ and\
  \bibinfo {author} {\bibfnamefont {T.}~\bibnamefont {Chu}},\ }\href {\doibase
  10.1103/PhysRevD.78.084017} {\bibfield  {journal} {\bibinfo  {journal}
  {\prd}\ }\textbf {\bibinfo {volume} {78}},\ \bibinfo {pages} {084017}
  (\bibinfo {year} {2008})}\BibitemShut {NoStop}%
\end{thebibliography}
%

\end{document}